\documentclass[11pt,preprint]{aastex}
\usepackage{natbib}
\received{}
\accepted{}
\shorttitle{Cepheid Calibrations of Modern SNe Ia}
\shortauthors{Riess et al.}

\newcommand{\bq}{\begin{equation}} 
\newcommand{\eq}{\end{equation}}

\newcommand{\beq}{\begin{equation}}
\newcommand{\eeq}{\end{equation}}
\newcommand{\beqa}{\begin{eqnarray}}
\newcommand{\eeqa}{\end{eqnarray}}

\newcommand{\PL}{$P\hbox{-}L \ $}
\newcommand{\nd}{\multicolumn{1}{c}{$\dots$}}
\newcommand{\mc}{\multicolumn{2}{l}}

\begin{document} 

\title{Cepheid Calibrations of Modern Type Ia Supernovae:
Implications for the Hubble Constant\altaffilmark{1}}

\author{Adam G. Riess\altaffilmark{2,3}, Lucas Macri\altaffilmark{4}, Weidong Li\altaffilmark{5}, Hubert Lampeitl\altaffilmark{3,8}, Stefano Casertano\altaffilmark{3},
Henry C. Ferguson\altaffilmark{3}, Alexei V. Filippenko\altaffilmark{5},
Saurabh W. Jha\altaffilmark{6}, , 
Ryan Chornock\altaffilmark{5}, Lincoln Greenhill\altaffilmark{7}, Max Mutchler\altaffilmark{3}, Mohan Ganeshalingham\altaffilmark{5}}

\altaffiltext{1}{Based on observations with the NASA/ESA {\it Hubble Space 
Telescope}, obtained at the Space Telescope Science Institute, which is 
operated by AURA, Inc., under NASA contract NAS 5-26555.}
\altaffiltext{2}{Department of Physics and Astronomy, Johns Hopkins 
University, Baltimore, MD 21218.}
\altaffiltext{3}{Space Telescope Science Institute, 3700 San Martin 
Drive, Baltimore, MD 21218; ariess@stsci.edu .}
\altaffiltext{4}{George P. and Cynthia W. Mitchell Institute for Fundamental Physics and Astronomy, 
Department of Physics, Texas A\&M University, 4242 
TAMU, College Station, TX 77843-4242.}
\altaffiltext{5}{Department of Astronomy, University of California,
Berkeley, CA 94720-3411.}
\altaffiltext{6}{Department of Physics and Astronomy, Rutgers University,
136 Frelinghuysen Road, Piscataway, NJ 08854.}
\altaffiltext{7}{Smithsonian Astrophysical Observatory, Cambridge, MA}
\altaffiltext{8}{ Institute of Cosmology and Gravitation, University of Portsmouth, Portsmouth, PO1 3FX, UK}

\begin{abstract} 

\noindent
This is the first of two papers reporting measurements from a program to
determine the Hubble constant to $\sim$5\% precision from a
refurbished distance ladder.  We present new observations of 110
Cepheid variables in the host galaxies of two recent Type Ia
supernovae (SNe~Ia), NGC 1309 and NGC 3021, using the Advanced Camera
for Surveys on the {\it Hubble Space Telescope (HST)}.  We also
present new observations of the hosts previously observed with {\it
HST} whose SNe Ia provide the most precise luminosity calibrations: SN
1994ae in NGC 3370, SN 1998aq in NGC 3982, SN 1990N in NGC 4639, and
SN 1981B in NGC 4536, as well as the maser host, NGC 4258.  Increasing
the interval between observations enabled the discovery of new, longer-period
Cepheids, including 57 with $P > 60$~days, which extend these
period-luminosity (\PL) relations.  We present 93 measurements of the
metallicity parameter, 12 + log[O/H], measured from H~II regions in
the vicinity of the Cepheids and show these
are consistent with solar metallicity.  We find the slope of the
seven dereddened \PL relations to be consistent with that of the Large
Magellanic Cloud Cepheids and with parallax measurements of Galactic
Cepheids, and we address the implications for the Hubble constant.  We
also present multi-band light curves of SN 2002fk (in NGC 1309) and SN
1995al (in NGC 3021) which may be used to calibrate their luminosities.
In the second paper we present observations of
the Cepheids in the $H$ band obtained with the Near Infrared Camera
and Multi-Object Spectrometer on {\it HST}, further mitigating
systematic errors along the distance ladder resulting from dust and
chemical variations. The quality and homogeneity of these SN and
Cepheid data provide the basis for a more precise determination of the
Hubble constant.

\end{abstract} 


\section{Introduction} 

The most accurate method to measure the Hubble
constant, $H_0$, in the low-redshift Universe 
has been to meld the Type Ia supernova (SN~Ia) and
Cepheid distance scales through {\it Hubble Space Telescope (HST)}
observations of Cepheid variables in SN~Ia host galaxies within $\sim$
20 Mpc.  Having extremely high and moderately uniform luminosities,
SNe~Ia are the most precise distance indicators known for sampling the
present expansion rate of the Universe.  Methods which use the
relationship between SN~Ia light-curve shape and luminosity
\citep{phillips93,hamuy95,hamuy96,riess95,riess96a,riess98,perlmutter97,saha01}
and supernova color to constrain absorption by dust
\citep{riess96b,phillips99,guy05,wang09} yield distances with {\it
relative} precision approaching 5\% {\it if applied to modern
photometry}.  The $\sim$100 high-quality, modern SN~Ia light curves
currently published in the redshift range $0.01 < z < 0.1$ establish
the {\it relative} expansion rate to an unprecedented uncertainty of
$<$1\% (i.e., internal statistical error; Jha, Riess, Kirshner 2005).  Yet because SNe~Ia are a
secondary distance indicator, the measurement of the {\it true}
expansion rate is seriously limited by the few opportunities to
calibrate their peak luminosity.

The SN~Ia {\it HST} Calibration Program \citep{saha01,sandage06} and
the {\it HST} Key Project \citep{freedman01} calibrated
$H_0$ via Cepheids and SNe~Ia using Wide Field and Planetary Camera 2
(WFPC2).  These measurements resolved decades of extreme
uncertainty about the scale and age of the Universe.  However, despite
a great amount of careful work, the final estimates of $H_0$ by the
two teams differ by 20\% and the overall uncertainty in each estimate
has proved difficult to reduce to 10\% or less.  The bulk of
this can be traced to systematic uncertainties among the 
many steps on the path to determining the luminosity of SNe Ia and to the known
inaccuracy of the SN Ia calibration sample.

The use of the LMC as the anchor in the Key Project 
 distance ladder presents
specific challenges which result in several sources of systematic
uncertainty if used for calibrating SNe Ia with Cepheids.  The LMC
distance is known to only 5\% to 10\% and its Cepheids, all observed
from the ground, are of shorter mean period ($<P>$ $\approx 5$~d) and
lower metallicity than those found in the spiral galaxies that host
SNe Ia beyond the Local Group.  These mismatches propagate unwanted
uncertainties to the measurement of the Hubble constant from the use
of imperfectly known relations between Cepheid luminosity,
metallicity, and period.  The use of distance estimates to Galactic Cepheids as anchors by the HST Calibration Program also presents stiff challenges as discussed in \S 5.

Additional systematic uncertainties in both teams` measurements of $H_0$
result from the photometric idiosyncracies of the WFPC2 camera and the
unreliability of those nearby SNe Ia which were {\it photographically}
observed, highly reddened, atypical, or discovered after maximum
brightness.  Only three SNe~Ia (SNe 1990N, 1981B, and 1998aq) were free from
these aforementioned shortcomings, making ÒidealÓ SNe a minority of the set used for calibrating $H_0$ with WFPC2 (see Table 1).  The use of many unreliable
calibrators was necessitated by the limited reach of WFPC2, defining a
volume within which Nature provides a suitable SN~Ia only once every
decade.

The installation of the Advanced Camera for Surveys \citep[ACS;
][]{ford03} on board {\it HST} provided an important improvement in
the optical imaging capabilities of the observatory.  In regards to the
distance ladder, the improved resolution and sensitivity of ACS
 extended the reach of {\it HST} Cepheid observations to
about 35~Mpc, tripling the enclosed volume and the available number of
potential calibrators of SNe Ia since the WFPC2 era.  In Cycle 11 we
\citep{riess05} used ACS to measure Cepheid variables in the host of
an ideal calibrator, SN 1994ae, situated in this extended volume.  Two
more opportunities to augment the very small, modern sample of
calibrated SNe~Ia are presented by SN 1995al in NGC 3021 and SN 2002fk
in NGC 1309.  These SNe Ia were discovered well before their maxima
and were observed with the same $UBVRI$ passbands and equipment we
employed to measure the SNe~Ia that help define the Hubble flow.
These SNe Ia present two of the most complete photometric records ever
of any SNe~Ia (see \S 4).  Calibration of $H_0$ from SN 1995al and SN
2002fk would add only the fifth and sixth reliable SN Ia to Table 1
and, more importantly, only the second and third to use the more
reliable and advantageous photometric system of ACS.

By replacing previous anchors of the distance scale 
with the ``maser galaxy'' NGC 4258 we can also
wring additional precision for the Hubble constant.  
First, the distance to the masers has been measured geometrically to unprecedented precision for an extragalactic source (Herrnstein et al. 1999, Humphreys et al. 2005, 2008,2009).
NGC 4258 also has the largest extragalactic set of long-period
Cepheids observed with HST (Macri et al. 2006)
Furthermore, we could bypass uncertainties in the determination of the
photometric system zeropoints from the ground and space, and in the
functional form of the period-luminosity (\PL) relation, because
we would only need to determine the relative magnitude offset between
each galaxy's set of Cepheids.  To maximize this advantage we may also
reobserve the hosts of nearby SNe Ia already observed with WFPC2 to
place their Cepheids on the same photometric system and to find
additional Cepheids whose periods would have been greater than the
length of the WFPC2 campaigns.  Lastly, the metallicity of the spiral
hosts of Cepheid-calibratable SNe~Ia is much more similar to that of
NGC 4258 than to that of the LMC, reducing the overall dependence of
$H_0$ on metallicity corrections.  Ultimately, the improved precision
in $H_0$ gained from a reduction in rungs on the distance ladder using
NGC 4258 will be limited by the number of SN Ia hosts calibrated by
ACS (due to the intrinsic scatter of SN Ia magnitudes).

In \S 2 we present the observations of (a) Cepheids in the new SN
hosts, (b) new, longer period Cepheids in the previously observed
hosts, and (c) measurements of the metallicities in the vicinity of
the Cepheids.  The light curves of the SNe~Ia are presented in \S 3.
In \S 4 we describe the value of these data for the distance scale, a
task undertaken in a companion paper \citep{riess09}.

\begin{table}[h]
\begin{small}
\begin{center}
\vspace{0.4cm}
\begin{tabular}{cccccc}
\multicolumn{6}{c}{Table 1: SN~Ia Light Curves with $HST$ Cepheid
Calibration} \\
\hline
\hline
SN Ia & Modern photometry?$^e$ & Low reddening?$^a$ & Observed before max? & Normal & Ideal? \\
\hline
1895B$^b$  & No & Unknown & No & Unknown & \\
1937C$^b$ & No & Yes & Yes & Yes & \\
1960F$^b$ & No & No & Yes & ? & \\
1972E$^b$ & Yes & Yes & No & Yes & \\
1974G & No & No & Yes & ? & \\
1981B$^b$ & Yes & Yes & Yes & Yes & $\surd$ \\
1989B$^b$ & Yes & No & Yes & Yes & \\
1990N$^b$ & Yes & Yes & Yes & Yes & $\surd$ \\
1991T$^b$ & Yes & Yes? & Yes & No & \\
1994ae$^c$ & Yes & Yes & Yes & Yes & $\surd$ \\
1995al$^d$ & Yes & Yes & Yes & Yes & $\surd$ \\
1998aq$^c$ & Yes & Yes & Yes & Yes & $\surd$ \\  
1998bu & Yes & No & Yes & Yes & \\
1999by & Yes & Yes & Yes & No & \\
2002fk$^d$ & Yes & Yes & Yes & Yes & $\surd$ \\

\hline
\hline
\multicolumn{6}{l}{$^aA_V < 0.5$ mag.} \\
\multicolumn{6}{l}{$^b$Calibrated by the Sandage/Tammann/Saha
collaboration.} \\
\multicolumn{6}{l}{$^c$Calibration presented by \citet{riess05}.} \\
\multicolumn{6}{l}{$^d$Calibration first presented in this paper.} \\
\multicolumn{6}{l}{$^e$CCD or photoelectric, not photographic.} \\
\end{tabular}
\end{center}
\end{small}
\end{table}

\section{Cepheid Observations}

The average of a small sample may be significantly impacted by a moderate systematic error among one of its members.
Therefore, {\it it is essential that each SN~Ia in the
calibration sample have a reliable photometric record which is
accurate and comparable to those of the SNe~Ia used to measure the
Hubble flow}.  To define a reliable calibration sample we use the
criteria for inclusion of a SN Ia given by \citet{riess05}: (a) modern
data (i.e., photoelectric or CCD), (b) observed before maximum
brightness, (c) low reddening, and (d) spectroscopically typical.  In
addition, their hosts must be suitable targets for observations by
{\it HST} of their Cepheids, which requires a relatively face-on,
late-type host within $\sim$ 35 Mpc.

\begin{table}[h]
\begin{small}
\begin{center}
\vspace{0.4cm}
\begin{tabular}{cccc}
\multicolumn{4}{c}{Table 2: Optical Observations of SN Ia Calibration
Sample} \\
\hline
\hline
Host & SN Ia &  Initial campaign & Reobservation   \\
\hline
NGC 4536 & SN 1981B &  WFPC2  & WFPC2  \\
NGC 4639 & SN 1990N &  WFPC2 & ACS \\
NGC 3982 & SN 1998aq &  WFPC2 & ACS  \\
NGC 3370 & SN 1994ae &  ACS &  ACS  \\
NGC 3021 & SN 1995al &   ACS &  ACS  \\
NGC 1309 & SN 2002fk &  ACS &  ACS  \\
\hline
\hline
\end{tabular}
\end{center}
\end{small} 
\end{table}

The recent SN 1995al in NGC 3021 and SN 2002fk in NGC 1309 provide two
valuable additions to the small calibration set (shown in Table 2).
The SN data are described in \S 4, but both are ideal candidates
by the previous criteria and their magnitudes indicate distances of
30--35 Mpc.  The host galaxies are moderately sized (1.5$^\prime
\times 1.5^\prime$ and 2.0$^\prime \times 2.2^\prime$, respectively),
face-on Sbc galaxies, each fitting well within the ACS WFC field of
view as shown in Figures 1 and 2.

In Cycle 14 (2005--2006, GO-10497) we observed NGC 1309 and NGC 3021
for 12 epochs of 4800~s each with ACS WFC $F555W$ using a power-law
spacing of intervals designed to reduce period aliasing up to the full
monitoring duration of 52~d.  We also observed each host at 5
epochs with $F814W$ for 4800~s each.  The telescope position and
orientation were fixed for each host and a four-position dither
(noninteger shift of 4.5 pixels in each detector coordinate) was
performed to better sample the point-spread function (PSF) in the 
median image.

In Cycle 15 (2006--2007, GO-10802) we reobserved NGC 1309 and NGC
3021, as well as the other four hosts in the calibration sample of
Table 2 at two epochs using ACS and $F555W$ (each $F555W$ observation
was separated by $\sim$ 10~d) to constrain the phase of the Cepheid
light curves for subsequent infrared (IR) observations with the Near
Infrared Camera and Multi-Object Spectrometer (NICMOS; Riess et
al. 2009) and to aid in the identification of Cepheids with $P > 60$~d
which were beyond the interval of the initial, contiguous
campaigns.\footnote{The Cycle 15 observations of NGC 4536 were
obtained with WFPC2 due to the failure of ACS in February, 2007.}
Logs of the exposures for NGC 3021 and NGC 1309 are given in Tables 3 and 4, respectively.

\tabletypesize{\normalsize}
\tablewidth{0pt}
\begin{deluxetable}{lll}
\tablenum{3}
\tablecaption{Log of Observations for NGC 3021}
\tablehead{\colhead{Epoch} &  \multicolumn{2}{c}{MJD at mid exposure$^*$} \\
\colhead{\#} & \colhead{V} & \colhead{I}}
\startdata
01 & 3686.0407 & 3686.1735 \\
02 & 3686.1066 & 3686.2398 \\
03 & 3696.4242 & \nd       \\
04 & 3696.5566 & \nd       \\
05 & 3704.6240 & 3704.7597 \\
06 & 3704.6929 & 3704.8261 \\
07 & 3708.7942 & \nd       \\
08 & 3708.8564 & \nd       \\
09 & 3712.7275 & \nd       \\
10 & 3712.7894 & \nd       \\
11 & 3715.7923 & \nd       \\
12 & 3715.8559 & \nd       \\
13 & 3716.7919 & 3716.9227 \\
14 & 3716.8555 & 3717.0197 \\
15 & 3718.8569 & \nd       \\
16 & 3718.9211 & \nd       \\
17 & 3721.9208 & \nd       \\
18 & 3722.0161 & \nd       \\
19 & 3725.7835 & 3725.9159 \\
20 & 3725.8488 & 3726.0127 \\
21 & 3730.5123 & \nd       \\
22 & 3730.5778 & \nd       \\
23 & 3737.5724 & 3737.7050 \\
24 & 3737.6380 & 3737.7712 \\
25 & 4058.9217 & \nd       \\
26 & 4066.7134 & \nd       \\
\tableline
\multicolumn{3}{l}{$*$: JD-2450000.0} \\
\enddata
\end{deluxetable}

\tabletypesize{\normalsize}
\tablewidth{0pt}
\begin{deluxetable}{lll}
\tablenum{4}
\tablecaption{Log of Observations for NGC 1309}
\tablehead{\colhead{Epoch} &  \multicolumn{2}{c}{MJD at mid exposure$^*$} \\
\colhead{\#} & \colhead{V} & \colhead{I}}
\startdata
01 & 3588.6856 & 3588.8192 \\
02 & 3588.7523 & 3588.8856 \\
03 & 3599.9153 & \nd       \\
04 & 3599.9797 & \nd       \\
05 & 3606.7768 & 3606.9082 \\
06 & 3606.8411 & 3606.9744 \\
07 & 3609.9078 & \nd       \\
08 & 3609.9723 & \nd       \\
09 & 3615.5304 & \nd       \\
10 & 3615.6030 & \nd       \\
11 & 3616.9033 & \nd       \\
12 & 3617.0008 & \nd       \\
13 & 3618.5021 & 3618.6347 \\
14 & 3618.5676 & 3618.7009 \\
15 & 3620.9007 & \nd       \\
16 & 3620.9992 & \nd       \\
17 & 3623.9338 & \nd       \\
18 & 3624.0679 & \nd       \\
19 & 3629.7599 & 3629.9226 \\
20 & 3629.8252 & 3630.0626 \\
21 & 3633.3574 & \nd       \\
22 & 3633.4228 & \nd       \\
23 & 3640.4216 & 3640.5530 \\
24 & 3640.4859 & 3640.6192 \\
25 & 4015.3368 & \nd       \\
26 & 4032.5923 & \nd       \\
\tableline
\multicolumn{3}{l}{$*$: JD-2450000.0} \\
\enddata
\end{deluxetable}

\subsection{ACS Photometry}

For the four hosts with extensive ACS measurements NGC 1309, NGC 3021,  NGC 3370 and NGC 4258,
the data from programs GO-9810 (P.I. Greenhill), GO-9351 (P.I. Riess),
GO-10497 (P.I. Riess), and GO-10802 (P.I. Riess) were retrieved
from the {\it HST} archive.  While ACS data for NGC 3370
\citep{riess05} and NGC 4258 \citep{macri06} were previously analyzed
to identify Cepheids, the acquisition of the two new epochs of ACS
imaging in program GO-10802, better calibration data, and the need for
a homogeneous reduction process for the comparison of Cepheid
magnitudes necessitated new data reductions undertaken here.

The data retrieval from the Space Telescope Science Institute (STScI)
MAST archive made use of the most up-to-date calibration as applied by
the software suite {\it calacs} in ``pyraf.''  Epoch-by-epoch
photometry for all stellar sources identified in a master, composite
image was measured for the images obtained in $F555W$ and $F814W$
using the DAOPHOT\footnote{Past comparisons of photometry with the
popular PSF fitting tools DoPHOT, DAOPHOT \citep{stetson87}, and HSTphot by
\citet{gibson00}, \citet{leonard03}, and \citet{saha01} show that the
mean difference in {\it HST} photometry obtained with these packages
is not significant.} set of routines following the procedures given by
\citet{stetson94}, \citet{stetson98}, \citet{riess05}, and
\citet{macri06}.  Magnitudes from PSF fitting were initially measured
to include the charge within 0.5$\arcsec$ radius apertures.  These
natural-system magnitudes for each candidate Cepheid at each observed
epoch are given in Tables 5 and 6, evaluated as 2.5\, log($e$/s)
$-$ CTE + 25.0, where $e$/s is the measured flux (electrons per
second) and CTE (charge transfer efficiency) is the magnitude loss
given by \citet{chiaberge09} for a star as a function of observation
date, pixel position, brightness, and background.  The median uncertainty
for an individual Cepheid magnitude in an epoch was 0.09 mag.

We selected Cepheids following the methodology outlined by
\citet{macri06}. Initially, Cepheid candidates were identified from
the time-sampled data by selecting all stellar sources with a modified
Welch/Stetson variability index $J_S$ \citep{stetson96} in excess of
0.75 and that were detected in at least 12 of the 14 epochs of $F555W$
data.  These candidates were then subjected to a ``minimum
string-length analysis,'' which identified as likely periods those
that minimized the sum of magnitude variations for observations at
similar trial phases \citep{stetson98}.  Robust least-squares fits were then
performed in the two bands, comparing the single-epoch magnitudes to
template Cepheid light curves from \citet{stetson98}, where six
parameters representing (1) period, (2) $V$-band amplitude, (3)
$I$-band amplitude, (4) epoch of zero phase, (5) mean magnitude in
$V$, and (6) mean magnitude in $I$ were free, but the shape of the
light curve was a unique function of the assumed period. Finally, one
of us (L.M.M.) applied experienced judgement in a visual comparison of
the data to the best-fit Cepheid model, rejecting unconvincing
candidates (approximately 10\%).  The criteria used to settle on the final set were (a) how
well the light-curve sampling was distributed in phase, (b) the
appearance of the characteristic rapid rise and slow fall of the light
curve, and (c) the proximity of the amplitudes and colors to the
expected range.  This same partially subjective process was compared
to a purely numerical selection by \citet{riess05} for the Cepheids in
NGC 3370, resulting in small differences in Cepheid-list membership,
and less than 0.04 mag difference in the intercept of the \PL
relation. When the optimum fit had been achieved, the fitted light
curves were converted to flux units and numerically integrated over a
cycle to achieve flux-weighted mean apparent brightnesses in each
bandpass; these were then converted back to magnitude units.

The photometric zeropoints (the magnitudes resulting in 1 electron
s$^{-1}$ in an infinite aperture) for our ACS magnitudes were obtained
using the Vegamag system (i.e., Vega $\equiv 0$ mag in all passbands).
The official STScI ACS zeropoints were revised from those given by
\citet{sirianni05}, and as of the beginning of 2009 have values of
$F555W=25.744$ and $F814W=25.536$ for data obtained before 2006 July 4
when ACS operated at $-77$~C, and $F555W=25.727$ and $F814W=25.520$
after that date when ACS operated at $-81$~C. These are based on an
empirical measurement of the system throughput at all wavelengths
using the known spectral energy distribution of five white dwarfs.
This approach is similar to that of \citet{holtzman95} for
WFPC2.\footnote{The Key Project zeropoint is based on matching {\it
HST} photometry of globular clusters (including 47 Tuc) to
ground-based data obtained by \citet{stetson00} on 22 nights from 12
distinct observing runs.  \citet{riess05} found that the ACS
photometric zeropoints of \citet{sirianni05}, $F555W=25.704$ mag and
$F814W=25.492$ mag, and those based on the \citet{stetson00} system
are quite consistent, with a mean difference of 0.015 mag in $V$ and
0.026 mag in $I$ for 250 calibrating stars in 47 Tuc.}  The PSF
magnitudes in Tables 5 and 6 were extended from the 0.5$\arcsec$
radius aperture to infinity using the aperture corrections given by
\citet{sirianni05} of 0.092 and 0.087 mag in $F555W$ and $F814W$,
respectively.  The natural-system magnitudes were converted to the
Johnson system (for ease of comparison to non-{\it HST} Cepheid data)
using the formulae in \citet{sirianni05}.  Thus, the natural-system
magnitudes in Tables 5 and 6 with zeropoint 25.0 are converted to the
Johnson system using the transformations

\bq V = F555W + 0.744 - 0.092 - 0.054(V-I), \eq
\bq I = F814W + 0.536 - 0.087 - 0.002(V-I). \eq

\noindent
This transformation is performed during the simultaneous fitting of
the $V$ and $I$ light-curve templates to interpolate the $V-I$
color at epochs when only $F555W$ was measured.

\tabletypesize{\scriptsize}
\tablewidth{0pt}
\begin{deluxetable}{lllllllll}
\tablenum{5}
\tablecaption{Cepheid photometry in NGC 3021$^a$}
\tablehead{\colhead{\#} & \multicolumn{1}{c}{F555W} & \multicolumn{1}{c}{F814W} & \multicolumn{1}{c}{F555W} & \multicolumn{1}{c}{F814W} & \multicolumn{1}{c}{F555W} & \multicolumn{1}{c}{F814W} & \multicolumn{1}{c}{F555W} & \multicolumn{1}{c}{F814W}}
\startdata
  &\mc{\bf 31556\boldmath $P=11.18$~d}&\mc{\bf 30672\boldmath $P=13.93$~d}&\mc{\bf 08621\boldmath $P=15.38$~d}&\mc{\bf 08102\boldmath $P=18.71$~d}\\
 1 & 26.841(189)  & 26.382(185)* & 27.381(323)  & 26.677(182)  & 26.195(166)  & 25.881(145)  & 26.313(174)  & 25.904(148) \\
 2 & 26.822(218)  & 26.807(199)  & 27.218(199)  & 26.799(206)  & 26.390(115)  & 25.942(135)  & 26.398(108)  & 26.110(122) \\
 3 & 27.399(285)  &     \nd      & 26.600(165)  &     \nd      & 27.078(237)  &     \nd      &     \nd      &     \nd     \\
 4 & 26.981(188)  &     \nd      & 26.750(152)  &     \nd      & 26.742(190)  &     \nd      & 26.845(168)  &     \nd     \\
 5 & 26.576(137)  & 26.543(274)  & 27.297(236)  & 26.820(181)  & 26.614(183)  & 25.719(136)  & 26.625(171)  & 26.056(203) \\
 6 & 26.708(207)  & 27.024(332)* & 27.298(282)  & 26.854(185)  & 26.592(178)  & 26.011(124)  & 26.144(149)  & 26.189(117) \\
 7 & 27.227(371)  &     \nd      & 26.593(132)  &     \nd      & 27.004(190)  &     \nd      & 26.213(248)  &     \nd     \\
\enddata
\end{deluxetable}


\tabletypesize{\scriptsize}
\tablewidth{0pt}
\begin{deluxetable}{lllllllll}
\tablenum{6}
\tablecaption{Cepheid photometry in NGC 1309 $^a$}
\tablehead{\colhead{\#} & \multicolumn{1}{c}{F555W} & \multicolumn{1}{c}{F814W} & \multicolumn{1}{c}{F555W} & \multicolumn{1}{c}{F814W} & \multicolumn{1}{c}{F555W} & \multicolumn{1}{c}{F814W} & \multicolumn{1}{c}{F555W} & \multicolumn{1}{c}{F814W}}
\startdata
  &\mc{\bf 21599\boldmath $P=20.93$~d}&\mc{\bf 09778\boldmath $P=21.98$~d}&\mc{\bf 08610\boldmath $P=23.23$~d}&\mc{\bf 06631\boldmath $P=24.82$~d}\\
 1 & 27.127(234)  & 26.249(129)  & 27.137(249)  & 27.083(303)* & 27.039(291)  & 26.456(166)  & 25.882(101)  & 25.646(127) \\
 2 & 26.627(221)  & 26.238(114)  & 27.535(393)  & 26.796(228)  & 26.965(368)  & 26.012(148)* & 25.574(167)  & 25.750(085) \\
 3 & 26.833(225)  &     \nd      & 26.316(252)  &     \nd      & 26.751(269)  &     \nd      & 26.543(262)  &     \nd     \\
 4 & 27.354(217)  &     \nd      & 26.378(245)  &     \nd      & 25.896(130)  &     \nd      & 26.628(223)  &     \nd     \\
 5 & 27.289(283)  & 26.931(180)* &     \nd      & 26.242(181)  & 26.792(209)  & 26.160(105)  & 26.636(135)  & 26.135(154) \\
 6 &     \nd      & 26.529(128)  & 26.999(280)  & 26.442(153)  & 26.709(249)  & 25.796(142)* &     \nd      & 26.241(139) \\
 7 & 27.012(101)  &     \nd      & 26.937(181)  &     \nd      & 26.954(277)  &     \nd      & 26.483(107)  &     \nd     \\
\enddata
\end{deluxetable}

However we note that our determination of $H_0$ in Riess et al. (2009) is insensitive to the value of the optical zeropoints and aperture corrections since we will make use of {\it the difference} in the photometry of Cepheids in NGC 4258 and the SN hosts.

The impact of blending and crowding on Cepheid magnitudes in optical
{\it HST} data has been addressed through Monte Carlo ``artificial
star'' experiments.  \citet{ferrarese00} have shown that the impact of
crowding on the measured magnitudes is largely eliminated by
application of the previously described selection criteria, whose
effect is to reject Cepheids that are significantly contaminated by
a close companion.  The presence of significant contamination will
alter the shape parameters of the Cepheid light curves, reducing the
amplitude of variation and flattening the ``sawtooth'' near minimum
light (by contributing a greater fraction to the total flux when the
Cepheid is faint).  Alternatively, a partial blend will result in a
poor PSF fit and a large reported uncertainty, rendering the apparent
variability less significant, and also causing a Cepheid candidate to
fail one or more of the previous criteria.  \citet{ferrarese00} found
that for multi-epoch data, the net crowding bias on the distance
modulus is only $\sim$1\%. \citet{riess05} similarly found that
candidates in ``crowded'' environments (defined here as having an
additional source within at least $0.1\arcsec$ which contributes at
least $\sim$10\% of the peak flux of the variable candidate) usually
failed one or more of the previously discussed selection criteria.

The {\it net} effect of even modest crowding and blending on the
distance scale is further reduced by the use NGC
4258 (instead of the LMC or the Galaxy) as an anchor of the distance
scale.  As shown in \citet{riess09}, the effect of crowding is
reduced to the {\it difference} in crowding between the SN hosts and
NGC 4258, which is negligible as determined from artificial-star
tests.

\subsection{Cepheids in NGC 3021, NGC 1309, and NGC 3370}

Here we present the first identification of Cepheids in NGC 1309 and
NGC 3021.  Each host yielded a sufficient number of Cepheids to provide a mean distance precision which is greater than the SN it hosts.

For NGC 1309 we identified 79 Cepheids, all with $P > 20$~d, providing
one of the largest sets of extragalactic Cepheids observed by {\it HST}. 
NGC 3021, a third the size of NGC 1309, not surprisingly yielded
fewer Cepheids, a total of 31 with 27 at $P> 20$~d.  Their light
curves are shown in Figures 3 and 4, and the parameters of the Cepheids
are given in Tables 7 and 8.  The \PL relations in $V$ and $I$ are
shown in Figures 5 and 6.

Cepheid samples in a magnitude-limited survey may suffer selection
bias at the short-period end due to the loss 
of Cepheids faint for their period (e.g., Ferrarese et al. 2006, Leonard et al. 2003).   Such Cepheids may fall below the detection limit or their light curves may be dominated by blending, reducing the significance of their variability. \citet{riess05} found this bias to apply for those Cepheids with $P < 20$~d for NGC 3370.  This limit
applies to NGC 3021 which has similar Cepheid magnitudes at a given period as NGC 3370.
As seen in Figure 6, the few Cepheids found with $P < 20$~d tend to be
brighter than expected, though no such bias appears for the Wesenheit
reddening-free magnitudes (defined in \S 3; see \citealt{madore82} and Figure 8).  
For NGC 1309, whose Cepheids indicate a greater distance than NGC 3370 or NGC 3021, the periods with apparent bias rises to $P < 38$~d (see
Fig.~5).  Again, the Wesenheit magnitudes do not show this bias at
shorter periods (see Figure 8).  In both cases this would imply that the Cepheids with periods shorter than the bias limit suffer a modest amount of blending from
bluer sources which is largely removed by the color correction.  While the Wesenheit magnitudes appear useful at lower
periods, it is safer to restrict the use of Cepheids to those with periods
greater than the range where their selection appears biased in the individual passbands.

Our additional imaging of NGC 3370 in Cycle 15 allowed us to identify
new Cepheids with periods in excess of the original 60 day campaign
as well as a few
more at shorter periods.  For NGC 3370, we have detected 127 Cepheids
of which 110 have $P > 20$~d, nearly double the sample found
by \citet{riess05} and reducing the mean wesenheit magnitudes by 0.035 mag.  These are shown in Figure 7 and their parameters
are given in Table 9 (where $L_V$ is the $V$-band variability index).

\begin{deluxetable}{cccccccccc}
\footnotesize
\tablenum{7}
\tablecaption{Cepheid Candidates in NGC 3021}
\tablehead{\colhead{id}&\colhead{$\alpha$}&\colhead{$\delta$}&\colhead{period}&\colhead{$<V>$}&\colhead{$<I>$}&\colhead{Amp$_V$} &\colhead{Amp$_I$}& \colhead{$L_V$}&\colhead{$t_0$} \\
\colhead{  }&\colhead{(J2000)}&\colhead{(J2000)}&\colhead{(days)}&\colhead{(mag)}&\colhead{(mag)}&\colhead{(mag)} &\colhead{(mag)}& \colhead{ }&\colhead{(2400000+)}} 
\startdata
31556 & 147.72782 & 33.55528 & 11.17 & 27.418(0.243) & 27.052(0.072) &0.600 & 0.322 & 0.804 & 54071.48 \nl 
30672 & 147.72778 & 33.54702 & 13.92 & 27.489(0.224) & 26.996(0.137) &0.618 & 0.429 & 1.187 & 54082.60 \nl 
8621. & 147.74838 & 33.55002 & 15.37 & 27.225(0.209) & 26.479(0.111) &0.470 & 0.206 & 0.908 & 54083.98 \nl 
8102. & 147.74935 & 33.55170 & 18.71 & 27.391(0.251) & 26.760(0.121) &0.536 & 0.270 & 0.991 & 54077.91 \nl 
20774 & 147.73750 & 33.55041 & 20.38 & 26.770(0.144) & 26.179(0.174) &0.447 & 0.268 & 1.313 & 54077.16 \nl 
10786 & 147.74693 & 33.55663 & 21.16 & 27.242(0.176) & 26.349(0.152) &0.465 & 0.243 & 1.256 & 54081.52 \nl 
47390 & 147.73410 & 33.55873 & 21.84 & 27.186(0.165) & 26.373(0.151) &0.455 & 0.212 & 1.300 & 54080.90 \nl 
33607 & 147.72083 & 33.55514 & 23.13 & 27.328(0.160) & 26.626(0.050) &0.450 & 0.268 & 1.532 & 54084.59 \nl 
32375 & 147.72586 & 33.55581 & 24.01 & 27.280(0.210) & 26.461(0.148) &0.552 & 0.301 & 1.530 & 54092.85 \nl 
8636. & 147.74871 & 33.55237 & 24.36 & 26.848(0.085) & 26.116(0.102) &0.451 & 0.288 & 1.588 & 54083.61 \nl 
32380 & 147.72645 & 33.56000 & 25.18 & 26.769(0.095) & 26.040(0.134) &0.560 & 0.346 & 2.752 & 54096.04 \nl 
32088 & 147.72678 & 33.55614 & 25.77 & 27.199(0.228) & 26.220(0.103) &0.552 & 0.245 & 1.733 & 54080.94 \nl 
26946 & 147.73211 & 33.54878 & 26.84 & 26.816(0.152) & 25.974(0.149) &0.587 & 0.170 & 2.380 & 54102.95 \nl 
9028. & 147.74791 & 33.55032 & 31.89 & 26.356(0.095) & 25.642(0.071) &0.516 & 0.277 & 3.677 & 54099.86 \nl 
23149 & 147.73688 & 33.55930 & 32.53 & 26.633(0.108) & 25.713(0.157) &0.555 & 0.367 & 2.523 & 54104.49 \nl 
30428 & 147.72812 & 33.54750 & 32.60 & 26.636(0.123) & 25.849(0.088) &0.576 & 0.244 & 3.384 & 54097.22 \nl 
26126 & 147.73336 & 33.55230 & 34.87 & 26.319(0.111) & 25.721(0.150) &0.328 & 0.176 & 1.092 & 54086.75 \nl 
12135 & 147.74553 & 33.55600 & 36.50 & 26.743(0.146) & 25.721(0.127) &0.426 & 0.303 & 1.492 & 54092.92 \nl 
31803 & 147.72789 & 33.55893 & 37.27 & 27.039(0.114) & 26.152(0.156) &0.486 & 0.353 & 1.601 & 54088.95 \nl 
45787 & 147.73632 & 33.55657 & 37.31 & 26.023(0.122) & 25.330(0.059) &0.282 & 0.125 & 0.709 & 54095.88 \nl 
26545 & 147.73249 & 33.54885 & 39.57 & 26.327(0.139) & 25.461(0.140) &0.590 & 0.323 & 3.045 & 54089.38 \nl 
9402. & 147.74757 & 33.55109 & 39.78 & 26.885(0.162) & 25.729(0.104) &0.623 & 0.317 & 1.558 & 54100.07 \nl 
25375 & 147.73388 & 33.55151 & 39.95 & 26.359(0.150) & 25.850(0.118) &0.614 & 0.560 & 3.372 & 54083.58 \nl 
9611. & 147.74740 & 33.55142 & 40.49 & 26.295(0.117) & 25.663(0.101) &0.400 & 0.396 & 0.963 & 54106.75 \nl 
20415 & 147.73892 & 33.55806 & 51.51 & 26.551(0.163) & 25.467(0.124) &0.476 & 0.220 & 1.712 & 54088.40 \nl 
12778 & 147.74489 & 33.55619 & 63.19 & 25.613(0.070) & 25.097(0.095) &0.278 & 0.198 & 1.265 & 54113.26 \nl 
19817 & 147.73982 & 33.56093 & 68.61 & 26.322(0.089) & 25.253(0.057) &0.370 & 0.259 & 2.211 & 54146.16 \nl 
7098. & 147.75116 & 33.55414 & 82.66 & 25.913(0.129) & 25.182(0.138) &0.344 & 0.186 & 1.127 & 54131.46 \nl 
9558. & 147.74734 & 33.55075 & 88.18 & 26.884(0.152) & 25.435(0.085) &0.435 & 0.262 & 0.932 & 54164.64 \nl 
12013 & 147.74545 & 33.55484 & 90.73 & 25.735(0.065) & 24.819(0.094) &0.237 & 0.119 & 0.757 & 54172.44 \nl 
10203 & 147.74683 & 33.55170 & 95.91 & 25.756(0.063) & 24.909(0.049) &0.170 & 0.020 & 0.908 & 54135.11 \nl 
\enddata
\end{deluxetable}
\begin{deluxetable}{cccccccccc}
\footnotesize
\tablenum{8}
\tablecaption{Cepheid Candidates in NGC 1309}
\tablehead{\colhead{id}&\colhead{$\alpha$}&\colhead{$\delta$}&\colhead{period}&\colhead{$<V>$}&\colhead{$<I>$}&\colhead{Amp$_V$} &\colhead{Amp$_I$}& \colhead{$L_V$}&\colhead{$t_0$} \\
\colhead{  }&\colhead{(J2000)}&\colhead{(J2000)}&\colhead{(days)}&\colhead{(mag)}&\colhead{(mag)}&\colhead{(mag)} &\colhead{(mag)}& \colhead{ }&\colhead{(2400000+)}} 
\startdata
21599 & 50.52930 & -15.41687 & 20.93 & 27.434(0.201) & 26.744(0.187) &0.555 & 0.162 & 1.631 & 54053.33 \nl 
9778. & 50.53294 & -15.38609 & 21.98 & 27.453(0.199) & 26.869(0.194) &0.513 & 0.279 & 1.159 & 54057.75 \nl 
8610. & 50.53423 & -15.39811 & 23.22 & 27.310(0.182) & 26.615(0.091) &0.541 & 0.219 & 1.031 & 54064.55 \nl 
6631. & 50.53610 & -15.41348 & 24.81 & 26.988(0.150) & 26.405(0.070) &0.498 & 0.267 & 1.304 & 54060.05 \nl 
6737. & 50.53598 & -15.41296 & 25.45 & 27.336(0.217) & 26.628(0.152) &0.541 & 0.279 & 1.291 & 54060.20 \nl 
34523 & 50.52358 & -15.40722 & 25.52 & 27.211(0.198) & 26.476(0.110) &0.521 & 0.210 & 1.266 & 54061.51 \nl 
44606 & 50.51955 & -15.40855 & 25.84 & 27.082(0.228) & 26.699(0.228) &0.563 & 0.242 & 2.614 & 54059.06 \nl 
48719 & 50.51726 & -15.40686 & 26.70 & 27.142(0.167) & 26.370(0.098) &0.516 & 0.142 & 1.036 & 54045.30 \nl 
55736 & 50.50417 & -15.38446 & 27.18 & 27.440(0.125) & 26.670(0.101) &0.419 & 0.300 & 1.534 & 54069.83 \nl 
54039 & 50.50975 & -15.38449 & 27.64 & 27.223(0.235) & 26.415(0.069) &0.545 & 0.353 & 1.301 & 54057.11 \nl 
41542 & 50.52035 & -15.39916 & 27.71 & 27.075(0.141) & 26.293(0.103) &0.463 & 0.222 & 1.187 & 54051.06 \nl 
12340 & 50.53212 & -15.39489 & 27.84 & 26.987(0.165) & 26.575(0.137) &0.445 & 0.378 & 1.169 & 54063.03 \nl 
52644 & 50.51172 & -15.37853 & 29.17 & 26.875(0.136) & 26.244(0.098) &0.574 & 0.338 & 2.819 & 54064.19 \nl 
2343. & 50.54025 & -15.40458 & 29.61 & 27.084(0.182) & 26.489(0.221) &0.602 & 0.402 & 1.923 & 54068.76 \nl 
23076 & 50.52816 & -15.40843 & 30.66 & 27.288(0.261) & 26.467(0.104) &0.496 & 0.239 & 1.192 & 54061.02 \nl 
85974 & 50.51896 & -15.40298 & 30.86 & 26.900(0.155) & 26.396(0.104) &0.521 & 0.301 & 1.776 & 54047.85 \nl 
7224. & 50.53585 & -15.41538 & 30.90 & 27.390(0.257) & 26.555(0.092) &0.683 & 0.268 & 2.330 & 54050.49 \nl 
7255. & 50.53427 & -15.38705 & 31.15 & 27.265(0.159) & 26.531(0.064) &0.452 & 0.265 & 1.162 & 54045.49 \nl 
50024 & 50.51566 & -15.39516 & 31.69 & 27.299(0.214) & 26.290(0.124) &0.512 & 0.302 & 1.724 & 54048.75 \nl 
59151 & 50.53574 & -15.41413 & 32.61 & 26.772(0.174) & 26.180(0.082) &0.445 & 0.241 & 1.602 & 54065.95 \nl 
60583 & 50.53360 & -15.39877 & 32.95 & 27.064(0.296) & 26.136(0.094) &0.684 & 0.423 & 1.746 & 54054.39 \nl 
4322. & 50.53615 & -15.38579 & 33.25 & 26.793(0.132) & 26.208(0.068) &0.612 & 0.327 & 2.840 & 54067.55 \nl 
30349 & 50.52525 & -15.40856 & 33.51 & 27.043(0.212) & 26.430(0.136) &0.528 & 0.236 & 0.899 & 54066.02 \nl 
3108. & 50.53807 & -15.38939 & 33.71 & 26.989(0.180) & 26.245(0.149) &0.472 & 0.238 & 1.974 & 54067.12 \nl 
9099. & 50.53391 & -15.39736 & 33.74 & 27.050(0.186) & 26.379(0.116) &0.458 & 0.135 & 0.698 & 54057.49 \nl 
30771 & 50.52352 & -15.38006 & 33.74 & 27.297(0.228) & 26.524(0.124) &0.530 & 0.266 & 1.816 & 54069.89 \nl 
14063 & 50.53191 & -15.40485 & 34.07 & 26.754(0.188) & 26.033(0.151) &0.519 & 0.334 & 1.382 & 54068.80 \nl 
59846 & 50.53441 & -15.40029 & 35.12 & 26.831(0.157) & 26.052(0.173) &0.475 & 0.144 & 1.036 & 54065.28 \nl 
31655 & 50.52487 & -15.41096 & 35.82 & 26.606(0.123) & 25.976(0.077) &0.297 & 0.132 & 1.204 & 54049.49 \nl 
7868. & 50.53405 & -15.38828 & 36.17 & 27.244(0.184) & 26.456(0.080) &0.472 & 0.270 & 1.354 & 54069.11 \nl 
69637 & 50.52730 & -15.39672 & 37.23 & 26.591(0.260) & 25.841(0.085) &0.532 & 0.175 & 1.073 & 54074.11 \nl 
41024 & 50.52078 & -15.40182 & 38.62 & 26.958(0.216) & 26.440(0.171) &0.525 & 0.352 & 1.006 & 54043.19 \nl 
34163 & 50.52292 & -15.39251 & 39.34 & 26.687(0.183) & 25.985(0.090) &0.423 & 0.282 & 1.016 & 54079.33 \nl 
7989. & 50.53524 & -15.41099 & 39.42 & 27.391(0.166) & 26.482(0.149) &0.410 & 0.304 & 0.784 & 54055.14 \nl 
44069 & 50.51958 & -15.40478 & 39.90 & 26.868(0.213) & 26.258(0.117) &0.487 & 0.367 & 1.629 & 54072.88 \nl 
27980 & 50.52622 & -15.41000 & 39.92 & 27.127(0.253) & 26.059(0.043) &0.815 & 0.238 & 1.739 & 54050.36 \nl 
52975 & 50.51305 & -15.41224 & 40.52 & 27.301(0.155) & 26.533(0.217) &0.510 & 0.355 & 1.691 & 54067.73 \nl 
7994. & 50.53392 & -15.38694 & 40.69 & 26.936(0.169) & 26.038(0.078) &0.602 & 0.371 & 1.340 & 54059.04 \nl 
1166. & 50.54164 & -15.39645 & 41.11 & 27.386(0.175) & 26.279(0.062) &0.514 & 0.270 & 1.478 & 54076.96 \nl 
76534 & 50.52410 & -15.40276 & 41.86 & 26.645(0.162) & 26.005(0.132) &0.490 & 0.315 & 0.853 & 54073.82 \nl 
22918 & 50.52824 & -15.40865 & 42.03 & 26.707(0.193) & 25.889(0.096) &0.595 & 0.262 & 1.397 & 54071.01 \nl 
2032. & 50.54017 & -15.39411 & 42.53 & 26.497(0.183) & 25.768(0.056) &0.624 & 0.246 & 2.908 & 54053.85 \nl 
48747 & 50.51610 & -15.38609 & 42.68 & 27.234(0.182) & 26.373(0.111) &0.566 & 0.356 & 1.002 & 54079.41 \nl 
67393 & 50.52880 & -15.39772 & 42.74 & 26.533(0.131) & 25.685(0.061) &0.443 & 0.162 & 1.433 & 54068.90 \nl 
58298 & 50.53507 & -15.38553 & 43.52 & 26.694(0.111) & 25.922(0.095) &0.521 & 0.341 & 1.704 & 54083.59 \nl 
7331. & 50.53498 & -15.40072 & 44.58 & 26.382(0.145) & 25.850(0.092) &0.377 & 0.195 & 1.042 & 54080.80 \nl 
2979. & 50.53940 & -15.40977 & 44.90 & 26.467(0.123) & 25.737(0.073) &0.503 & 0.320 & 3.097 & 54076.82 \nl 
1732. & 50.54059 & -15.39462 & 45.00 & 26.252(0.094) & 25.525(0.096) &0.261 & 0.170 & 1.854 & 54082.39 \nl 
19368 & 50.52978 & -15.40833 & 45.25 & 26.566(0.158) & 25.759(0.117) &0.442 & 0.238 & 1.308 & 54065.50 \nl 
49584 & 50.51619 & -15.39832 & 45.67 & 26.838(0.143) & 26.046(0.107) &0.409 & 0.322 & 1.434 & 54082.05 \nl 
16143 & 50.53022 & -15.39054 & 46.74 & 27.014(0.173) & 26.015(0.090) &0.476 & 0.296 & 1.459 & 54058.63 \nl 
15346 & 50.53148 & -15.40689 & 46.85 & 26.746(0.147) & 25.925(0.081) &0.409 & 0.091 & 0.722 & 54068.16 \nl 
54502 & 50.50920 & -15.39192 & 47.13 & 26.996(0.205) & 26.100(0.080) &0.666 & 0.332 & 1.593 & 54075.84 \nl 
59642 & 50.53436 & -15.39644 & 47.39 & 26.428(0.155) & 25.925(0.071) &0.379 & 0.311 & 1.889 & 54059.19 \nl 
52566 & 50.51322 & -15.40390 & 47.41 & 26.641(0.105) & 26.095(0.092) &0.425 & 0.331 & 1.436 & 54088.98 \nl 
52170 & 50.51350 & -15.39881 & 47.99 & 26.915(0.134) & 25.940(0.024) &0.520 & 0.235 & 2.161 & 54058.99 \nl 
4882. & 50.53701 & -15.41209 & 48.91 & 26.808(0.158) & 26.046(0.087) &0.576 & 0.246 & 1.970 & 54084.46 \nl 
68817 & 50.52830 & -15.40526 & 49.93 & 26.578(0.160) & 26.036(0.081) &0.406 & 0.190 & 1.546 & 54086.75 \nl 
15318 & 50.53038 & -15.38675 & 51.46 & 26.562(0.088) & 25.781(0.124) &0.615 & 0.354 & 2.909 & 54055.96 \nl 
71911 & 50.52661 & -15.40578 & 51.99 & 26.207(0.098) & 25.447(0.057) &0.287 & 0.169 & 1.137 & 54059.51 \nl 
28132 & 50.52604 & -15.40768 & 52.24 & 26.820(0.129) & 25.815(0.068) &0.405 & 0.073 & 0.883 & 54084.04 \nl 
6581. & 50.53598 & -15.41154 & 58.98 & 26.810(0.262) & 26.005(0.177) &0.637 & 0.254 & 1.363 & 54073.09 \nl 
25965 & 50.52610 & -15.39321 & 59.02 & 26.562(0.148) & 25.964(0.071) &0.371 & 0.265 & 0.757 & 54100.04 \nl 
6542. & 50.53606 & -15.41233 & 59.12 & 26.575(0.093) & 25.663(0.020) &0.311 & 0.220 & 1.889 & 54096.28 \nl 
53187 & 50.51202 & -15.39909 & 59.75 & 26.757(0.119) & 26.179(0.098) &0.300 & 0.191 & 1.006 & 54113.61 \nl 
69494 & 50.52808 & -15.40923 & 60.17 & 27.077(0.173) & 26.150(0.087) &0.619 & 0.250 & 1.337 & 54069.54 \nl 
2647. & 50.53992 & -15.40867 & 60.67 & 26.278(0.127) & 25.452(0.097) &0.553 & 0.305 & 3.064 & 54115.54 \nl 
17049 & 50.53025 & -15.39867 & 61.24 & 26.272(0.154) & 25.415(0.137) &0.363 & 0.189 & 0.959 & 54082.07 \nl 
19918 & 50.52958 & -15.40892 & 64.94 & 26.241(0.086) & 25.425(0.093) &0.351 & 0.189 & 2.152 & 54082.93 \nl 
64757 & 50.53107 & -15.40794 & 65.03 & 26.796(0.138) & 25.710(0.041) &0.345 & 0.156 & 0.934 & 54107.87 \nl 
13102 & 50.53150 & -15.38948 & 66.34 & 26.331(0.115) & 25.506(0.076) &0.539 & 0.243 & 2.231 & 54113.61 \nl 
45088 & 50.51857 & -15.39463 & 71.40 & 26.699(0.197) & 25.912(0.156) &0.662 & 0.607 & 2.057 & 54128.10 \nl 
3836. & 50.53789 & -15.40643 & 73.27 & 26.390(0.128) & 25.504(0.120) &0.401 & 0.195 & 1.401 & 54066.32 \nl 
7702. & 50.53554 & -15.41410 & 73.76 & 26.473(0.110) & 25.596(0.089) &0.615 & 0.342 & 1.718 & 54094.45 \nl 
49485 & 50.51648 & -15.40236 & 74.19 & 26.071(0.155) & 25.670(0.045) &0.451 & 0.267 & 1.158 & 54099.77 \nl 
23616 & 50.52841 & -15.41752 & 82.13 & 26.481(0.178) & 25.579(0.091) &0.450 & 0.242 & 1.307 & 54107.33 \nl 
19777 & 50.52881 & -15.39393 & 82.38 & 26.246(0.135) & 25.437(0.105) &0.325 & 0.193 & 1.002 & 54128.08 \nl 
65015 & 50.53048 & -15.40066 & 89.03 & 26.625(0.137) & 25.576(0.087) &0.343 & 0.238 & 1.220 & 54084.36 \nl 
4908. & 50.53644 & -15.40175 & 97.89 & 26.201(0.098) & 25.247(0.069) &0.309 & 0.210 & 1.559 & 54115.33 \nl 
\enddata
\end{deluxetable}
\begin{deluxetable}{cccccccccc}
\footnotesize
\tablenum{9}
\tablecaption{Cepheid Candidates in NGC 3370}
\tablehead{\colhead{id}&\colhead{$\alpha$}&\colhead{$\delta$}&\colhead{period}&\colhead{$<V>$}&\colhead{$<I>$}&\colhead{Amp$_V$} &\colhead{Amp$_I$}& \colhead{$L_V$}&\colhead{$t_0$} \\
\colhead{  }&\colhead{(J2000)}&\colhead{(J2000)}&\colhead{(days)}&\colhead{(mag)}&\colhead{(mag)}&\colhead{(mag)} &\colhead{(mag)}& \colhead{ }&\colhead{(2400000+)}} 
\startdata
88045 & 161.77365 & 17.27522 & 14.45 & 27.620(0.262) & 26.950(0.179) &0.526 & 0.228 & 1.523 & 54087.72 \nl 
35527 & 161.76834 & 17.26353 & 15.09 & 27.634(0.188) & 26.544(0.134) &0.450 & 0.493 & 1.255 & 54074.51 \nl 
8849. & 161.75805 & 17.27787 & 15.39 & 27.414(0.215) & 26.445(0.074) &0.615 & 0.318 & 1.662 & 54085.07 \nl 
52153 & 161.78670 & 17.27016 & 15.77 & 27.495(0.188) & 26.804(0.142) &0.557 & 0.293 & 1.502 & 54088.84 \nl 
2204. & 161.75537 & 17.28996 & 16.22 & 27.626(0.135) & 26.723(0.118) &0.479 & 0.273 & 1.749 & 54071.32 \nl 
24497 & 161.76928 & 17.28204 & 16.78 & 27.609(0.256) & 26.737(0.175) &0.556 & 0.438 & 1.757 & 54084.18 \nl 
4668. & 161.75797 & 17.28751 & 17.26 & 27.494(0.251) & 26.660(0.110) &0.563 & 0.277 & 1.310 & 54081.86 \nl 
78990 & 161.76383 & 17.26409 & 17.37 & 27.381(0.182) & 26.474(0.217) &0.498 & 0.297 & 1.892 & 54077.47 \nl 
2638. & 161.75400 & 17.28417 & 17.46 & 27.290(0.179) & 26.530(0.042) &0.586 & 0.459 & 2.549 & 54081.90 \nl 
37156 & 161.76811 & 17.26012 & 17.84 & 27.603(0.138) & 26.792(0.165) &0.551 & 0.313 & 2.003 & 54086.86 \nl 
9842. & 161.75587 & 17.26928 & 18.39 & 27.747(0.321) & 27.056(0.180) &0.593 & 0.224 & 1.623 & 54075.80 \nl 
11908 & 161.76236 & 17.28264 & 18.76 & 27.238(0.173) & 26.209(0.150) &0.400 & 0.126 & 1.463 & 54088.84 \nl 
8367. & 161.75989 & 17.28410 & 18.88 & 27.693(0.250) & 26.772(0.249) &0.607 & 0.282 & 1.741 & 54078.46 \nl 
44450 & 161.77260 & 17.25851 & 19.10 & 27.693(0.248) & 26.843(0.130) &0.556 & 0.350 & 1.411 & 54078.88 \nl 
5394. & 161.75420 & 17.27504 & 19.21 & 27.432(0.181) & 26.475(0.168) &0.544 & 0.272 & 2.211 & 54080.20 \nl 
21444 & 161.76766 & 17.28206 & 19.64 & 27.452(0.222) & 26.525(0.114) &0.477 & 0.271 & 1.670 & 54090.17 \nl 
52086 & 161.78246 & 17.25879 & 19.91 & 27.201(0.122) & 26.347(0.053) &0.500 & 0.418 & 2.413 & 54089.91 \nl 
45559 & 161.78086 & 17.27835 & 20.21 & 27.272(0.172) & 26.524(0.119) &0.567 & 0.414 & 2.188 & 54076.41 \nl 
45280 & 161.77907 & 17.27414 & 20.27 & 27.603(0.215) & 26.766(0.225) &0.570 & 0.259 & 0.854 & 54077.95 \nl 
51454 & 161.78510 & 17.26936 & 20.51 & 27.328(0.166) & 26.400(0.062) &0.434 & 0.176 & 1.151 & 54083.93 \nl 
50670 & 161.77931 & 17.25660 & 20.52 & 27.237(0.131) & 26.399(0.140) &0.479 & 0.241 & 2.157 & 54079.69 \nl 
26546 & 161.76597 & 17.26987 & 21.57 & 27.028(0.206) & 26.160(0.054) &0.472 & 0.068 & 1.625 & 54080.85 \nl 
11992 & 161.75778 & 17.26972 & 21.78 & 27.509(0.192) & 26.547(0.082) &0.587 & 0.319 & 1.590 & 54093.66 \nl 
10540 & 161.76365 & 17.28919 & 21.79 & 27.442(0.243) & 26.607(0.153) &0.578 & 0.245 & 1.771 & 54092.36 \nl 
51357 & 161.78000 & 17.25566 & 23.43 & 27.198(0.144) & 26.349(0.088) &0.517 & 0.349 & 2.023 & 54092.68 \nl 
871.0 & 161.75495 & 17.29607 & 23.66 & 27.329(0.165) & 26.322(0.130) &0.543 & 0.281 & 2.276 & 54091.36 \nl 
8807. & 161.75882 & 17.28007 & 23.72 & 27.484(0.158) & 26.425(0.129) &0.524 & 0.175 & 1.537 & 54085.34 \nl 
47494 & 161.77627 & 17.25957 & 24.43 & 27.215(0.195) & 25.988(0.101) &0.532 & 0.217 & 2.100 & 54073.89 \nl 
23575 & 161.76105 & 17.26054 & 24.46 & 27.310(0.123) & 26.353(0.108) &0.448 & 0.273 & 1.512 & 54099.40 \nl 
47059 & 161.77441 & 17.25595 & 24.49 & 27.332(0.190) & 26.402(0.090) &0.589 & 0.436 & 1.913 & 54098.58 \nl 
53228 & 161.78689 & 17.25998 & 24.73 & 27.490(0.198) & 26.634(0.141) &0.487 & 0.289 & 1.719 & 54085.77 \nl 
61720 & 161.75875 & 17.28389 & 25.43 & 26.630(0.137) & 25.724(0.114) &0.429 & 0.236 & 1.991 & 54082.44 \nl 
21354 & 161.76299 & 17.26917 & 25.59 & 27.131(0.220) & 26.459(0.105) &0.380 & 0.271 & 1.186 & 54097.74 \nl 
23818 & 161.76845 & 17.28078 & 26.33 & 27.098(0.230) & 26.043(0.109) &0.486 & 0.230 & 1.355 & 54095.29 \nl 
22838 & 161.76132 & 17.26232 & 26.38 & 27.246(0.157) & 26.248(0.193) &0.476 & 0.256 & 1.916 & 54085.26 \nl 
81239 & 161.77044 & 17.27853 & 26.42 & 27.013(0.192) & 26.342(0.134) &0.551 & 0.430 & 1.258 & 54082.13 \nl 
39583 & 161.77518 & 17.27540 & 26.87 & 26.958(0.213) & 26.006(0.185) &0.552 & 0.157 & 1.950 & 54086.41 \nl 
18872 & 161.76209 & 17.27022 & 27.02 & 27.024(0.160) & 26.347(0.118) &0.583 & 0.480 & 2.014 & 54076.36 \nl 
22029 & 161.76515 & 17.27416 & 27.32 & 26.772(0.221) & 25.836(0.149) &0.422 & 0.036 & 1.325 & 54084.89 \nl 
48470 & 161.78219 & 17.27266 & 27.35 & 27.344(0.201) & 26.261(0.193) &0.518 & 0.259 & 1.982 & 54079.93 \nl 
5744. & 161.75647 & 17.28052 & 27.74 & 27.299(0.131) & 26.232(0.106) &0.755 & 0.352 & 2.625 & 54090.86 \nl 
51334 & 161.78015 & 17.25611 & 28.79 & 27.496(0.266) & 26.509(0.088) &0.669 & 0.476 & 2.088 & 54100.09 \nl 
49159 & 161.77565 & 17.25210 & 29.07 & 26.890(0.183) & 26.049(0.116) &0.512 & 0.366 & 2.097 & 54107.61 \nl 
4531. & 161.75831 & 17.28893 & 29.23 & 27.124(0.162) & 26.124(0.128) &0.581 & 0.276 & 1.516 & 54092.40 \nl 
62219 & 161.75791 & 17.28025 & 29.43 & 27.323(0.242) & 26.218(0.172) &0.785 & 0.434 & 2.143 & 54088.70 \nl 
46992 & 161.77485 & 17.25741 & 29.60 & 27.119(0.218) & 26.168(0.075) &0.627 & 0.278 & 2.739 & 54101.12 \nl 
4032. & 161.76154 & 17.29942 & 29.78 & 26.756(0.114) & 25.898(0.118) &0.547 & 0.249 & 3.155 & 54095.81 \nl 
51430 & 161.78069 & 17.25719 & 30.43 & 27.408(0.207) & 26.352(0.126) &0.518 & 0.244 & 1.715 & 54101.23 \nl 
27556 & 161.77156 & 17.28401 & 30.69 & 27.202(0.140) & 26.227(0.126) &0.421 & 0.230 & 1.872 & 54094.43 \nl 
19943 & 161.76245 & 17.26963 & 30.80 & 26.582(0.088) & 25.845(0.084) &0.343 & 0.206 & 1.649 & 54083.47 \nl 
35299 & 161.77125 & 17.27173 & 31.20 & 26.483(0.162) & 25.569(0.109) &0.371 & 0.218 & 0.759 & 54100.86 \nl 
13380 & 161.76194 & 17.27869 & 31.74 & 26.477(0.126) & 25.603(0.102) &0.261 & 0.162 & 0.810 & 54080.07 \nl 
13249 & 161.76203 & 17.27921 & 31.78 & 27.010(0.195) & 26.180(0.141) &0.440 & 0.274 & 1.494 & 54102.53 \nl 
32684 & 161.77068 & 17.27447 & 32.04 & 26.620(0.138) & 25.511(0.113) &0.460 & 0.272 & 1.661 & 54104.72 \nl 
15081 & 161.76554 & 17.28584 & 32.56 & 27.510(0.166) & 26.281(0.076) &0.423 & 0.343 & 1.419 & 54086.09 \nl 
4710. & 161.75647 & 17.28320 & 32.62 & 27.231(0.150) & 26.150(0.108) &0.470 & 0.207 & 1.513 & 54108.24 \nl 
18200 & 161.76118 & 17.26863 & 32.87 & 27.001(0.150) & 25.894(0.104) &0.463 & 0.305 & 2.597 & 54090.74 \nl 
26416 & 161.76487 & 17.26700 & 33.40 & 27.044(0.178) & 25.851(0.084) &0.495 & 0.182 & 1.799 & 54097.26 \nl 
52428 & 161.78414 & 17.26088 & 33.48 & 26.998(0.131) & 25.962(0.128) &0.483 & 0.214 & 2.033 & 54098.93 \nl 
31439 & 161.77252 & 17.28137 & 33.49 & 26.970(0.170) & 25.953(0.104) &0.428 & 0.139 & 1.968 & 54092.83 \nl 
17886 & 161.76336 & 17.27517 & 33.56 & 26.800(0.203) & 25.557(0.043) &0.688 & 0.289 & 2.100 & 54101.75 \nl 
49211 & 161.78007 & 17.26419 & 33.57 & 27.125(0.175) & 26.169(0.110) &0.484 & 0.259 & 1.834 & 54104.43 \nl 
52279 & 161.78547 & 17.26580 & 33.69 & 27.022(0.167) & 25.987(0.131) &0.445 & 0.359 & 2.101 & 54094.93 \nl 
33812 & 161.77245 & 17.27762 & 34.07 & 26.916(0.138) & 26.034(0.111) &0.488 & 0.380 & 1.673 & 54105.37 \nl 
4345. & 161.75620 & 17.28353 & 34.07 & 27.348(0.144) & 26.141(0.116) &0.739 & 0.360 & 2.373 & 54089.34 \nl 
17595 & 161.76259 & 17.27355 & 34.22 & 27.176(0.229) & 26.021(0.152) &0.741 & 0.375 & 2.019 & 54095.37 \nl 
1320. & 161.75212 & 17.28576 & 34.58 & 27.551(0.156) & 26.487(0.097) &0.414 & 0.196 & 1.104 & 54084.23 \nl 
17969 & 161.76144 & 17.26969 & 34.71 & 26.941(0.206) & 25.778(0.161) &0.673 & 0.328 & 2.096 & 54108.94 \nl 
10677 & 161.76072 & 17.28069 & 35.24 & 27.152(0.175) & 26.027(0.077) &0.645 & 0.381 & 2.083 & 54090.46 \nl 
40369 & 161.77544 & 17.27471 & 35.41 & 27.592(0.225) & 26.062(0.069) &0.581 & 0.208 & 1.189 & 54098.72 \nl 
19618 & 161.76333 & 17.27244 & 35.66 & 26.758(0.222) & 25.544(0.184) &0.509 & 0.209 & 1.731 & 54109.86 \nl 
22097 & 161.76764 & 17.28103 & 36.02 & 26.679(0.119) & 25.681(0.133) &0.480 & 0.246 & 1.666 & 54093.05 \nl 
50582 & 161.78077 & 17.26097 & 36.69 & 26.958(0.167) & 25.899(0.062) &0.489 & 0.343 & 1.736 & 54105.54 \nl 
59919 & 161.75710 & 17.28387 & 36.99 & 26.703(0.165) & 25.819(0.162) &0.506 & 0.377 & 2.357 & 54108.29 \nl 
45614 & 161.77520 & 17.26245 & 37.02 & 26.788(0.144) & 25.958(0.103) &0.526 & 0.171 & 2.055 & 54090.95 \nl 
21445 & 161.76923 & 17.28640 & 37.10 & 27.064(0.150) & 25.975(0.077) &0.618 & 0.284 & 1.887 & 54090.02 \nl 
16214 & 161.75886 & 17.26535 & 37.21 & 27.394(0.171) & 26.191(0.127) &0.513 & 0.189 & 1.208 & 54110.28 \nl 
30965 & 161.77252 & 17.28201 & 37.28 & 26.549(0.097) & 25.594(0.105) &0.606 & 0.322 & 3.412 & 54094.59 \nl 
20306 & 161.76295 & 17.27049 & 37.28 & 26.834(0.131) & 26.072(0.106) &0.257 & 0.214 & 0.871 & 54082.04 \nl 
21506 & 161.76844 & 17.28412 & 38.54 & 26.163(0.098) & 25.409(0.067) &0.480 & 0.339 & 2.652 & 54103.44 \nl 
31251 & 161.77276 & 17.28228 & 39.30 & 26.314(0.094) & 25.373(0.071) &0.337 & 0.179 & 2.062 & 54118.69 \nl 
47492 & 161.77587 & 17.25844 & 39.41 & 27.278(0.185) & 26.133(0.059) &0.396 & 0.157 & 1.264 & 54084.59 \nl 
7613. & 161.75941 & 17.28456 & 39.50 & 27.209(0.196) & 26.166(0.167) &0.442 & 0.239 & 1.149 & 54120.64 \nl 
18990 & 161.76910 & 17.28952 & 40.01 & 26.664(0.136) & 25.855(0.037) &0.522 & 0.325 & 3.251 & 54097.25 \nl 
13355 & 161.76178 & 17.27834 & 41.09 & 26.420(0.123) & 25.492(0.126) &0.412 & 0.168 & 1.564 & 54116.06 \nl 
20732 & 161.76774 & 17.28324 & 41.55 & 26.881(0.124) & 25.715(0.042) &0.389 & 0.169 & 1.612 & 54099.06 \nl 
29982 & 161.76730 & 17.26882 & 42.75 & 26.440(0.149) & 25.523(0.055) &0.386 & 0.250 & 1.278 & 54107.93 \nl 
9431. & 161.76075 & 17.28388 & 43.10 & 26.126(0.109) & 25.287(0.044) &0.447 & 0.310 & 2.588 & 54087.71 \nl 
6440. & 161.75761 & 17.28213 & 43.94 & 26.483(0.135) & 25.311(0.065) &0.632 & 0.248 & 3.243 & 54087.45 \nl 
85483 & 161.77245 & 17.27666 & 44.57 & 26.444(0.136) & 25.310(0.107) &0.362 & 0.116 & 1.106 & 54088.24 \nl 
25870 & 161.76779 & 17.27589 & 44.81 & 26.379(0.155) & 25.410(0.139) &0.523 & 0.336 & 1.405 & 54099.75 \nl 
37212 & 161.77170 & 17.26998 & 44.87 & 26.705(0.175) & 25.701(0.141) &0.529 & 0.225 & 2.175 & 54121.50 \nl 
28504 & 161.76514 & 17.26491 & 44.97 & 26.387(0.101) & 25.564(0.073) &0.378 & 0.287 & 1.768 & 54109.23 \nl 
9014. & 161.75980 & 17.28234 & 45.10 & 27.106(0.190) & 25.985(0.160) &0.563 & 0.234 & 2.191 & 54101.07 \nl 
40168 & 161.77694 & 17.27923 & 45.40 & 26.626(0.095) & 25.606(0.076) &0.552 & 0.288 & 2.503 & 54122.22 \nl 
5439. & 161.75713 & 17.28309 & 45.82 & 26.409(0.124) & 25.424(0.052) &0.576 & 0.258 & 2.602 & 54089.11 \nl 
46830 & 161.77864 & 17.26849 & 45.88 & 27.114(0.232) & 26.032(0.132) &0.602 & 0.355 & 1.969 & 54104.19 \nl 
34313 & 161.77155 & 17.27429 & 47.19 & 26.634(0.204) & 25.659(0.049) &0.451 & 0.204 & 2.193 & 54099.56 \nl 
20949 & 161.77120 & 17.29254 & 47.37 & 26.415(0.120) & 25.453(0.074) &0.431 & 0.216 & 2.543 & 54098.79 \nl 
2074. & 161.75695 & 17.29511 & 49.85 & 26.699(0.143) & 25.738(0.079) &0.480 & 0.244 & 2.455 & 54129.77 \nl 
46035 & 161.78023 & 17.27523 & 50.13 & 26.580(0.110) & 25.497(0.106) &0.636 & 0.286 & 3.021 & 54112.43 \nl 
28129 & 161.76619 & 17.26830 & 50.57 & 26.245(0.122) & 25.118(0.092) &0.465 & 0.233 & 2.670 & 54127.73 \nl 
5361. & 161.75695 & 17.28274 & 50.60 & 25.948(0.115) & 25.052(0.086) &0.394 & 0.255 & 1.797 & 54115.05 \nl 
28534 & 161.77050 & 17.27974 & 51.15 & 26.929(0.146) & 25.944(0.062) &0.462 & 0.333 & 1.986 & 54095.76 \nl 
48903 & 161.77827 & 17.26024 & 51.69 & 26.717(0.161) & 25.620(0.139) &0.668 & 0.323 & 2.414 & 54136.82 \nl 
15864 & 161.76420 & 17.28080 & 52.41 & 26.144(0.067) & 25.109(0.042) &0.357 & 0.243 & 1.478 & 54109.82 \nl 
4367. & 161.75525 & 17.28084 & 52.72 & 26.941(0.148) & 25.723(0.064) &0.312 & 0.181 & 1.130 & 54088.61 \nl 
13303 & 161.76500 & 17.28732 & 52.74 & 26.716(0.122) & 25.561(0.043) &0.539 & 0.245 & 2.246 & 52817.50 \nl 
1528. & 161.75194 & 17.28407 & 60.68 & 25.982(0.076) & 25.156(0.043) &0.350 & 0.217 & 2.805 & 54113.52 \nl 
5501. & 161.75677 & 17.28193 & 62.71 & 26.093(0.053) & 24.833(0.025) &0.176 & 0.050 & 0.853 & 54142.71 \nl 
6706. & 161.75874 & 17.28466 & 64.79 & 26.238(0.096) & 25.267(0.072) &0.559 & 0.224 & 2.216 & 54148.62 \nl 
7014. & 161.75977 & 17.28685 & 66.71 & 25.781(0.104) & 24.700(0.070) &0.320 & 0.198 & 2.104 & 54126.33 \nl 
33669 & 161.76845 & 17.26671 & 67.23 & 26.670(0.154) & 25.282(0.126) &0.408 & 0.184 & 1.565 & 54115.81 \nl 
9063. & 161.75862 & 17.27891 & 68.90 & 25.687(0.045) & 24.875(0.038) &0.201 & 0.136 & 1.571 & 54135.47 \nl 
22612 & 161.76869 & 17.28313 & 69.35 & 25.746(0.069) & 24.703(0.073) &0.391 & 0.212 & 2.692 & 54095.88 \nl 
29662 & 161.76544 & 17.26415 & 71.53 & 26.272(0.119) & 25.366(0.110) &0.484 & 0.259 & 1.869 & 54152.55 \nl 
22718 & 161.76805 & 17.28118 & 73.36 & 25.673(0.108) & 24.763(0.042) &0.430 & 0.207 & 1.547 & 54112.11 \nl 
1454. & 161.75252 & 17.28609 & 79.26 & 26.521(0.092) & 25.188(0.094) &0.674 & 0.374 & 2.578 & 54159.30 \nl 
33346 & 161.76861 & 17.26770 & 80.85 & 26.002(0.121) & 25.105(0.064) &0.407 & 0.216 & 0.979 & 54171.59 \nl 
33195 & 161.77345 & 17.28137 & 81.04 & 25.895(0.072) & 24.884(0.048) &0.281 & 0.164 & 1.654 & 54177.57 \nl 
4471. & 161.75697 & 17.28539 & 83.28 & 27.010(0.093) & 25.584(0.095) &0.351 & 0.213 & 1.314 & 54158.04 \nl 
22098 & 161.76770 & 17.28115 & 86.33 & 25.892(0.116) & 24.822(0.129) &0.247 & 0.112 & 1.918 & 54146.68 \nl 
3205. & 161.75284 & 17.27851 & 88.25 & 26.112(0.120) & 25.097(0.083) &0.356 & 0.186 & 1.005 & 54172.21 \nl 
31067 & 161.76816 & 17.26970 & 88.54 & 25.667(0.100) & 24.516(0.046) &0.330 & 0.131 & 1.147 & 54130.16 \nl 
48741 & 161.77799 & 17.26004 & 96.49 & 25.522(0.051) & 24.561(0.028) &0.248 & 0.146 & 2.139 & 54131.89 \nl 
8038. & 161.76096 & 17.28786 & 96.82 & 26.286(0.101) & 25.194(0.080) &0.336 & 0.182 & 2.181 & 54135.30 \nl 
17501 & 161.76206 & 17.27215 & 98.72 & 26.034(0.131) & 24.952(0.051) &0.295 & 0.208 & 2.189 & 54158.29 \nl 
\enddata
\end{deluxetable}

\subsection{Long-Period Cepheids}

The initial {\it HST} imaging campaigns of the SN hosts were too brief
to reliably identify Cepheids with $P > 60$~d as their duration was
shorter than a full pulsation cycle.  While long-period Cepheids are
rarer than their older cousins, their greater brightness and
contrast makes them easier to detect and they are very valuable for
extending the range of Cepheids as distance indicators.  At
$P > 100$~d, the \PL relations flatten (e.g., Freedman et al. 1992)
and such Cepheids require the use of different relations as discussed
by \citet{bird09}.  While we have as yet no Cepheids with $P>100$ days in our sample, in the future such objects could be useful for extending the range to which SNe Ia may be calibrated.

For the three SN hosts exclusively observed with ACS (NGC 3370, 1309,
and 3021), we have identified a total of 39 and 18 Cepheids with
periods greater than 60 and 75~d, respectively.  For NGC 4258 we
identified additional Cepheids using the new epochs beyond those
analyzed by \citet{macri06}, 149 in all including 68 with $P > 20$~d,
6 with $P > 60$~d, and 3 with $P > 75$~d.  The optical Cepheid data
for NGC 4258 are given by \citet{macri09}.  In Table 10 we indicate the
number of Cepheids with long periods found in the previously discussed
hosts.

\begin{table}[h]
\begin{small}
\begin{center}
\vspace{0.4cm}
\begin{tabular}{cccc}
\multicolumn{4}{c}{Table 10: Long-Period Cepheids} \\
\hline
\hline
Host & SN Ia &  $P > 60$~d & $P > 75$~d \\
\hline
NGC 4536 & SN 1981B &  5   &   4  \\
NGC 4639 & SN 1990N &   2   &   1   \\
NGC 3982 & SN 1998aq & 5   &   3   \\
NGC 3370 & SN 1994ae & 19  &  10  \\
NGC 3021 & SN 1995al &  6   &   4  \\
NGC 1309 & SN 2002fk &  14    &   4 \\
NGC 4258 & ----------- &  6   &   3   \\
\hline
 \hline
\end{tabular}
\end{center}
\end{small} 
\end{table}

The new epochs of imaging of the three hosts initially observed with
WFPC2 (NGC 3982, NGC 4536, and NGC 4639) also enabled the discovery of
new Cepheids at $P > 60$~d not previously identified.  However, the
fractional coverage of these hosts is less, limited by the smaller
field of WFPC2.  To look for such objects we retrieved the WFPC2 data
which originated from programs GO-5427, GO-5981, and GO-8100 (SN~Ia
{\it HST} Calibration Program, P.I. Sandage) to combine with the Cycle
15 data from our SHOES (Supernovae and $H_0$ for the Equation
of State)    program.  These data were processed in the same
manner as the ACS data described in the previous section while making
use of the latest WFPC2 CTE corrections.  The search for Cepheids
in these hosts netted a dozen Cepheids with $P > 60$~d to augment
those previously detected in the original analyses of Saha et al.
(1996, 1997, 2001), \citet{gibson00} and \citet{sg01} .\footnote{We rediscovered
$\sim$90\% the Cepheids presented in the previous analyses.  For these
Cepheids we found a negligible difference (0.3~d) in their mean
period, with a dispersion about the difference of 1.3~d.}  Additional
imaging might reveal so-called ``ultra long-period Cepheids''
\citep[$P > 100$~d; e.g.,][]{bird09} of future value.

\subsection{Cepheid Homogeneity}

The reliable use of Cepheids along the distance ladder relies on their homogeneity.  To test this we constructed
composite light curves in the well-sampled $V$ band, as shown in
Figure 9.  We limited inclusion to long period Cepheids, i.e.,  those with $P > 10$~d, resulting in averages of 20
to 40~d in the hosts.  We find that these mean light curves in each galaxy to be quite
homogeneous.  Table 11 gives the mean half-amplitude for the SN hosts, each of which is measured to 1\% to 2\% precision and is consistent with the sample average of 0.48, including
that of the inner field of NGC 4258.  The exception may be the outer
field of NGC 4258 whose mean is 13\% lower than the sample mean, though this difference
is not significant due to the small number of long-period Cepheids in
this field.  {\it Individual} half-amplitudes for the Cepheids range from
0.15 to 0.80 mag.

There are pertinent reasons why the {\it mean} shape or amplitude of
the light curve might vary from galaxy to galaxy.  A difference in
blending would alter the apparent amplitudes, reducing them in the
presence of greater blending.  Chemical composition may also affect
Cepheid amplitudes.  \citet{paczynski00} found that the mean amplitude
of Cepheids in the Galaxy is 7\% greater than in the LMC, and the mean
amplitude of Cepheids in the Small Magellanic Cloud is 25\% smaller
than in the LMC; they suggested that a natural explanation for the
difference is their relative metal content.  This difference would be
in the same direction as the low metallicity of the outer region of
NGC 4258, though more data would be required to see if this is a
significant difference.  Pulsation models also indicate a dependence
between amplitude and chemical composition
\citep{bono99,marconi05}. Finally, Cepheid amplitudes also vary with
temperature or color, resulting in the amplitude-color relations
\citep[e.g.,][]{kanbur06}.

The 2\% limit on the difference in the mean of the half amplitude for
NGC 4258 (inner field) and the SN hosts constrains the {\it
differential blending} between NGC 4258 and the SN hosts to $<5$\% of
the mean Cepheid flux. The uniformity of the observed amplitudes is also consistent with the finding in \S 3 that the metallicities near the Cepheids are
homogeneous.

In Figure 10 we show the composite light curves of all Cepheids with
$P > 60$~d.  These have lower mean amplitudes, as expected, but the
characteristic sawtooth shape clearly demonstrates their
authenticity.


\begin{table}[h]
\begin{small}
\begin{center}
\vspace{0.4cm}
\begin{tabular}{cccc}
\multicolumn{4}{c}{Table 11: $V$-Band Half-Amplitudes} \\
\hline
\hline
Host & Mean $V$ $(\sigma$) &  $\sigma$ from sample mean &  No. $P > 10$~d \\
\hline
NGC 3370   &    0.486(0.011)   &  0.56   &      130 \\
NGC 1309    &   0.474(0.013)    & $-$0.48    &      86 \\
NGC 3021     &  0.464(0.020)    & $-$0.78     &     33 \\
NGC 4536     &  0.467(0.024)    & $-$0.54      &    34 \\
NGC 3982     &  0.490(0.019)     & 0.56       &   37 \\
NGC 4639     &  0.483(0.017)     & 0.14       &   30 \\
NGC 4258i     &  0.472(0.011)    & $-$0.72      &   144 \\
NGC 4258o     &  0.418(0.045)     & $-$1.39       &    7 \\
\hline
 \hline
\end{tabular}
\end{center}
\end{small} 
\end{table}

\section{Cepheid Metallicities and Pulsation Relations}

Past work \citep{kochanek97,kennicutt98,sakai04,macri06} has
demonstrated a significant dependence between the apparent magnitudes
of Cepheids at a fixed period and the metallicity in the environment
of the Cepheid.  This dependence and its uncertainty propagates as one
of the largest sources of systematic error in the Hubble constant measured via the LMC,
$\sim$4\% \citep{freedman01}. The SHOES program was designed to
mitigate the sensitivity of the Hubble constant measurement to
metallicity by 1) utilizing Cepheids in a narrow range of the metallicity
parameter [O/H], and by 2) measuring Cepheids in the near-infrared where
the metallicity dependence is diminished
\citep{alibert99,persson04,marconi05,gieren08}.

Nevertheless, to account for even a modest metallicity dependence and
its uncertainty, we measured the [O/H] abundance from 93 H~II regions
in the vicinity of the Cepheids in all of the galaxies in Table 2
using slit masks with the Low Resolution Imaging Spectrometer on the
Keck I telescope \citep{oke95}.  Our analysis methods are described in
\S 2.5 of \citet{riess05} and follow the calibration from
\citet{zaritsky94}, for which [O/H]$_{\rm solar} = 7.9 \times 10^{-4}$
and the solar abundance is 12 + log[O/H] = 8.9.  The result is the
measurement of a gradient in 12 + log[O/H] for each galaxy across the
deprojected radii occupied by the Cepheids (Table 12), as shown in
Figure 11.  The intercepts and gradients were used to estimate the
metallicity at the deprojected radius of each individual Cepheid.

\begin{table}[h]
\begin{small}
\begin{center}
\vspace{0.4cm}
\begin{tabular}{cccccc}
\multicolumn{5}{c}{Table 12: Metallicity (12 + log[O/H]) of SHOES Hosts} \\
\hline
\hline
Host & at $r = 30''$ & change per $10''$  & avg. at Cepheid positions & dispersion \\
\hline
NGC1309 & 9.013 & $-$0.098  & 8.90 & 0.19 \\
NGC3021 & 9.018 & $-$0.224  & 8.94 & 0.25 \\
NGC4536 & 9.104 & $-$0.025  & 8.79  & 0.12 \\
NGC4639 & 9.130 & $-$0.086  & 8.96  & 0.14 \\
NGC4258$^a$ & 9.015 & $-$0.006  & 8.94 & 0.05  \\
NGC3370 & 9.030 & $-$0.090  & 8.82  & 0.19 \\
NGC3982 & 8.998 & $-$0.152  & 8.74  & 0.25 \\
\hline
\hline
\multicolumn{5}{l}{$^a$For inner field of Macri et al. (2006); outer field 
average = 8.72, dispersion = 0.03.} \\
\end{tabular}
\end{center}
\end{small} 
\end{table}


The mean value of 12 + log[O/H] at the positions of the Cepheids in the SN hosts is quite similar to that in the {\it inner field} of NGC 4258, with a difference that is less than the dispersion of the means of the SN hosts.
This statement is independent of the normalization of the metallicity scale
as it depends on the difference in metallicity.
Thus, a correction to the distance scale is
unwarranted when comparing these Cepheids.\footnote{However, even using the \citet{sakai04}
zeropoint relation of $\delta (m-M) \delta {\rm [O/H]} = -0.24 \pm
0.05$ and these Cepheids to calibrate the distance scale would result
in a modest 1\% correction to the Hubble constant with a systematic
error of 0.2\%, a substantially reduced sensitivity from past use of
LMC Cepheids and their values of 12 + log[O/H] = 8.5.}
The exception is the outer field of NGC
4258 whose values are determined to be 8.72 $\pm$ 0.03. 

Interestingly, the metallicities of the Cepheids in the SN hosts and
the inner field of NGC 4258 are very similar to the solar
neighborhood value of 8.9 and the value of 8.81 measured for 68 Galactic Cepheids
\citep{andrievsky02,andrievsky04}.  Thus, in principle, Galactic
Cepheids could provide a suitable calibration of the luminosities of
our Cepheid sample in the SN hosts independent of NGC 4258.  In
practice, the calibration of Galactic Cepheid luminosities is compromised
by the precision and accuracy of their distance estimates, their large
extinction, and the inhomogeneity of their photometry.

\citet{sandage08} contend that the {\it slope} of the Cepheid \PL
relation is sensitive to chemical composition and that
solar-metallicity Cepheids used in the distance scale should be
calibrated with Galactic Cepheids.  This conclusion could
have important consequences for the determination of the distance scale via the LMC. \citet{tammann03} used a mixture
of Baade-Becker-Wesselink and cluster-based distance estimates
to Galactic Cepheids to calibrate the $V$-band and $I$-band \PL
relations which should then be applicable to the solar-metallicity
Cepheids in SN~Ia hosts.  Corrections for the extinction of Cepheids
in SN hosts are subsequently made by the use of two colors and a
Galactic reddening law \citep{saha06}. This is equivalent to the use
of a ``Wesenheit reddening-free'' mean magnitude, $m_w$, defined by
\citet{madore82} as

\bq m_w = m_V - R(m_V - m_I) = a_w\, {\rm log} P + b_w, \eq 

\noindent
where $R \equiv A_V/(A_V - A_I)$, and $a_w$ and $b_w$ are the
slope and intercept of this $P-w$ relation, respectively.

The \citet{tammann03} Cepheid analysis provides a slope for the $P-w$
relation in equation (3) of $-3.75 \pm 0.09$ mag, which is used by
\citet{saha06} to determine distances to the SN~Ia hosts.  This slope
is the steepest estimate of the Galactic relation to date
\citep{tammann03} and is much steeper than the LMC $P-w$ slope of $-3.2$
to $-3.3$ mag \citep{udalski99}.  Because the mean period for Cepheids
seen in SN hosts, $<P>=30-35$ days,
is longer than that in the LMC, $<P>=5$ days, this steeper slope
results in the bulk of the 15\% increase in the \citet{sandage06}
determination of the distance scale from that of \citet{freedman01}.

As seen in Table 12, the measurements here provide the largest sample
of Cepheids to date with solar metallicity (mean 12 + log[O/H] = 8.9),
long periods \citep[beyond the break in the LMC relation at $P =
10$~d]{kanbur04}, and uniform measurements.  Here we use them to
measure the slope of the $P-w$ relation, whose value is vital to measurements of the Hubble constant.

For the Cepheids in each host we fit equation (3) to determine the
slope and zeropoint.  We used an iterative clipping of $\pm$3$\sigma$
from the mean to remove outliers and limited the fit to Cepheids with
$10 < P < 100$~d to avoid the possibility of a break at $P = 10$~d and a change of slope at $P>100$ days (Bird, Stanek, and Prieto 2009).  In the SN
hosts we also limited the Cepheids to $20 < P$ ~d to mitigate selection as discussed in \S 2.2.  We did not include the outer
field of NGC 4258 because it has too few Cepheids (7 with
$P > 10$~d, 2 with $P > 20$~d) to yield a reliable result.  The values
and uncertainties of the slopes based on the 445 Cepheids are given in
Table 13, and the results are plotted versus the host metallicity in
Figure 12.  The mean slope for all Cepheids was 
$-2.98 \pm 0.07$ mag.
Changing the lower period cutoff to $P > 10$~d for the Cepheids in the
SN hosts gives a mean of $-3.00 \pm 0.07$ mag.  Setting the lower
period cutoff for each galaxy individually based on the completeness
boundaries like those shown in Figures 6 to 8 yields a mean slope of $-2.98 \pm
0.08$ mag.  Overall, we found the mean slope to be insensitive to the
period range, with a moderately shallower slope at the longest periods
as expected \citep{bird09}.  This finding is consistent with the analysis by \citep{madfre09} who show that a color tilt in the monochromatic $P-L$ slope due to metallicity is diminished in the $P-w$ due to dereddening.

We find no evidence that the slope is
steeper than the LMC slope, as ours is consistent with the LMC though
1$\sigma$ to 2$\sigma$ {\it shallower}.  The mean slope is significantly lower and inconsistent with the results from
\citet{tammann03}.   (Even the poorly determined mean slope of the 3 hosts observed with WFPC2 yield -3.26 $\pm 0.22$ which is $>2$ $\sigma$ shallower than the Tammann et al. slope).  We discuss likely origins of this difference in 
\S 5.

\begin{table}[h]
\begin{small}
\begin{center}
\vspace{0.4cm}
\begin{tabular}{ccc}
\multicolumn{3}{c}{Table 13: Cepheid $P-w$ slopes} \\
\hline
\hline
Host & slope (mag) &  $\sigma$ \\
\hline
NGC 3370  &   $-$2.94  &   0.14 \\
NGC 1309   &  $-$2.82  &   0.21 \\
NGC 3021    & $-$2.60  &   0.24 \\
NGC 3982   &  $-$3.15  &   0.42 \\
NGC 4639   &  $-$3.07  &   0.55 \\
NGC 4536   &   $-$3.38  &   0.30 \\
NGC 4258i   &  $-$3.05   &  0.10 \\
\hline
 \hline
\end{tabular}
\end{center}
\end{small} 
\end{table}

\section{Light Curves of SN 1995al and SN 2002fk}

Use of the Cepheid data in the previous sections and the flux-calibrated light curves of the new SNe presented in this section provide the means to determine their luminosity.

Both SN 1995al (NGC 3021) and SN 2002fk (NGC 1309) were
spectroscopically normal SNe~Ia (Wei et al. 1995; Ayani \& Yamaoka
2002).

SN 2002fk was extensively monitored by the 0.76~m Katzman Automatic
Imaging Telescope \citep[KAIT;][]{li00,filippenko01} commencing 13~d
before $B$-band maximum.  The SN photometry was measured with the
benefit of galaxy subtraction and PSF fitting relative to stars in the
field of the SN.  These field stars were later calibrated on five
photometric nights with KAIT and the Nickel 1 m telescope at the Lick
Observatory using Landolt (1992) standards.

The mean magnitudes of the field stars are given in Table 14 and their
positions are shown in Figure 13.  The light curves of the SN are
shown in Figure 14 compared to their model fits using the MLCS2k2
\citep{jha07} algorithm.   The fits are
consistent with those obtained for well-observed SNe~Ia and are
sufficient to constrain the relative distance modulus to a precision
of 0.068 mag or 0.105 mag, with the intrinsic contribution
of 0.08 mag \citep{jha07} added in quadrature.

For SN 1995al, photometric monitoring was conducted
starting 5~d before $B$-band maximum with the FLWO 1.2~m telescope
equipped with a thick, front-illuminated Loral CCD (``Andycam")
and a set of Johnson $UBV$ and Kron-Cousins $RI$ filters.  The
observations were initially presented by \citet{riess99} as a member
of a set of 22 SNe~Ia and we direct the reader there for additional
details.  The photometry presented by Riess et al. lacked the
benefit of galaxy template subtraction and multiple zeropoint
calibrations.  These steps can improve the precision of SN Ia
photometry and a reanalysis is warranted for including this SN in the
small calibration set.  We have now undertaken the galaxy subtraction
and obtained zeropoint calibrations on 5 independent nights using the
\citet{landolt92} standard stars for the fundamental calibration.

The mean magnitudes of the field stars are given in Table 15 and their
positions are shown in Figure 13.  The photometric differences with
the version from \citet{riess99} are a few hundredths of a magnitude in
all bands except $I$, which differed in the mean by 0.1 mag.
The light curves of the SNe are shown in Figure 14 compared to their
model fits using the MLCS2k2 \citep{jha07} algorithm. 

Photometry of the two SNe is given in
Tables 16 and 17 and their photometric parameters in Table 18.

\begin{deluxetable}{ccccccccccc}
\tablenum{14}
\footnotesize
\tablecaption{Comparison Stars for SN 2002fk}
\tablehead{\colhead{Star}&\colhead{$U$}&\colhead{N}&\colhead{$B$}&\colhead{N}&\colhead{$V$}&\colhead{N}&\colhead{$R$}&\colhead{N}&\colhead{$I$}&\colhead{N}}
\startdata
1&----&0&16.255(0.007)&5&15.748(0.013)&5&15.441(0.006)&3&15.034(0.010)&3\nl 
2&----&0&16.415(0.006)&5&15.784(0.012)&3&15.405(0.015)&3&15.002(0.002)&2\nl 
3&----&0&16.331(0.009)&5&15.787(0.007)&5&15.444(0.012)&4&15.047(0.015)&4\nl 
4&----&0&17.513(0.010)&3&16.954(0.012)&5&16.611(0.014)&2&16.176(0.009)&2\nl 
\enddata
\end{deluxetable}

\begin{deluxetable}{ccccccccccc}
\tablenum{15}
\footnotesize
\tablecaption{Comparison Stars for SN 1995al}
\tablehead{\colhead{Star}&\colhead{$U$}&\colhead{N}&\colhead{$B$}&\colhead{N}&\colhead{$V$}&\colhead{N}&\colhead{$R$}&\colhead{N}&\colhead{$I$}&\colhead{N}}
\startdata
0&15.078(0.019)&3&14.619(0.004)&4&13.795(0.004)&4&13.307(0.004)&3&12.836(0.008)&3\nl 
1&19.027(0.079)&2&18.050(0.051)&2&16.865(0.007)&2&16.067(0.002)&2&15.350(0.011)&2\nl 
2&13.633(0.021)&2&13.686(0.009)&2&13.246(0.015)&2&12.977(0.013)&2&12.692(0.004)&2\nl 
3&14.586(0.025)&3&14.628(0.006)&3&14.132(0.008)&3&13.842(0.008)&3&13.551(0.009)&3\nl 
4&15.390(0.018)&3&14.870(0.002)&4&14.050(0.006)&4&13.620(0.005)&3&13.244(0.006)&3\nl 
5&16.536(0.012)&3&16.744(0.002)&4&16.322(0.007)&4&16.042(0.009)&4&15.737(0.012)&4\nl 
6&18.499(0.007)&2&17.451(0.013)&4&16.425(0.005)&4&15.822(0.012)&4&15.326(0.014)&4\nl 
7&16.894(0.027)&2&16.667(0.004)&4&15.990(0.004)&4&15.609(0.008)&4&15.250(0.010)&4\nl 
8&17.893(0.012)&3&17.538(0.006)&4&16.757(0.006)&4&16.295(0.009)&4&15.881(0.017)&4\nl 
9&17.558(0.019)&2&16.878(0.009)&4&15.967(0.011)&4&15.429(0.011)&4&14.928(0.015)&4\nl 
10&99.999(9.999)&0&17.371(0.015)&4&15.879(0.003)&4&14.820(0.005)&3&13.629(0.016)&3\nl 
11&15.432(0.021)&3&15.291(0.009)&4&14.654(0.006)&4&14.296(0.005)&4&13.951(0.009)&4\nl 
12&17.086(0.006)&3&16.860(0.012)&4&16.132(0.015)&4&15.693(0.013)&4&15.318(0.008)&4\nl 
\enddata
\end{deluxetable}

\begin{deluxetable}{ccccc}
\tablenum{16}
\footnotesize
\tablecaption{Photometry of SN 2002fk}
\tablehead{\colhead{JD-2.4e6}&\colhead{$B$}&\colhead{$V$}&\colhead{$R$}&\colhead{$I$}}
\startdata
52535.99&15.061(0.010)&15.152(0.032)&15.090(0.049)&14.879(0.025) \nl
52537.00&14.667(0.011)&14.760(0.010)&14.721(0.038)&14.518(0.020) \nl
52537.97&14.347(0.010)&14.450(0.012)&14.396(0.033)&14.238(0.011) \nl
52541.99&13.600(0.010)&13.714(0.011)&13.693(0.042)&13.603(0.015) \nl
52544.00&13.472(0.010)&13.592(0.028)&13.529(0.016)&13.557(0.020) \nl
52544.96&13.386(0.010)&13.508(0.020)&13.462(0.010)&13.510(0.011) \nl
52548.99&13.327(0.010)&13.366(0.011)&13.382(0.011)&13.573(0.010) \nl
52549.98&13.311(0.010)&13.363(0.010)&13.380(0.010)&13.589(0.011) \nl
52550.95&13.349(0.011)&13.400(0.010)&13.380(0.010)&13.654(0.012) \nl
52551.97&13.383(0.011)&13.424(0.010)&13.397(0.013)&13.691(0.010) \nl
52553.95&13.519(0.011)&13.468(0.011)&13.481(0.013)&13.784(0.012) \nl
52555.94&13.658(0.010)&13.549(0.010)&13.647(0.036)&13.953(0.010) \nl
52559.95&13.973(0.013)&13.766(0.010)&13.941(0.035)&14.199(0.012) \nl
52562.89&14.298(0.011)&13.957(0.010)&14.111(0.033)&14.292(0.010) \nl
52565.90&14.669(0.010)&14.127(0.013)&14.187(0.049)&14.296(0.034) \nl
52570.95&----&14.372(0.025)&14.191(0.030)&14.179(0.039) \nl
52574.87&15.549(0.013)&14.569(0.022)&14.269(0.040)&14.023(0.011) \nl
52576.89&15.757(0.042)&14.684(0.011)&14.378(0.038)&14.044(0.010) \nl
52579.90&15.977(0.048)&14.859(0.016)&14.488(0.010)&14.121(0.011) \nl
52582.87&16.109(0.031)&15.055(0.026)&14.690(0.010)&14.293(0.010) \nl
52588.93&16.363(0.038)&15.367(0.025)&15.024(0.034)&---- \nl
52594.84&16.513(0.018)&15.538(0.021)&15.263(0.017)&15.003(0.028) \nl
52604.83&16.650(0.017)&15.819(0.012)&15.635(0.011)&15.490(0.026) \nl
52607.80&16.751(0.027)&15.884(0.014)&15.705(0.049)&15.600(0.018) \nl
52610.85&16.812(0.034)&15.996(0.019)&15.826(0.024)&15.755(0.026) \nl
52613.81&16.799(0.017)&16.037(0.020)&15.955(0.014)&15.917(0.032) \nl
52616.79&16.873(0.038)&16.161(0.027)&16.011(0.033)&16.097(0.060) \nl
52619.77&16.889(0.020)&16.223(0.037)&16.147(0.053)&16.090(0.035) \nl
52631.79&17.153(0.028)&16.547(0.026)&16.584(0.063)&16.595(0.069) \nl
52644.73&17.323(0.048)&16.830(0.018)&16.977(0.020)&17.103(0.074) \nl
52657.72&17.301(0.084)&17.148(0.036)&17.309(0.050)&17.792(0.135) \nl
52673.63&17.723(0.101)&17.551(0.035)&17.903(0.050)&18.125(0.139) \nl
\hline
\enddata
\end{deluxetable}

\begin{deluxetable}{cccccc}
\tablenum{17}
\footnotesize
\tablecaption{Photometry of SN 1995al}
\tablehead{\colhead{JD-2.4e6}&\colhead{$U$}&\colhead{$B$}&\colhead{$V$}&\colhead{$R$}&\colhead{$I$}}
\startdata
       50024.990&----&13.45(0.02)&13.43(0.02)&13.32(0.06)&13.64(0.02)\nl
       50026.030&13.01(0.02)&13.40(0.02)&----&13.32(0.02)&13.61(0.02)\nl
       50030.010&13.23(0.04)&13.28(0.02)&13.25(0.04)&13.24(0.02)&13.66(0.02)\nl
       50032.000&13.35(0.02)&13.32(0.02)&13.28(0.02)&13.24(0.02)&13.70(0.02)\nl
       50035.020&13.48(0.04)&13.53(0.02)&13.33(0.02)&13.34(0.02)&13.82(0.02)\nl
       50037.020&----&13.69(0.02)&13.43(0.02)&13.32(0.10)&13.87(0.02)\nl
       50038.020&13.60(0.04)&13.70(0.02)&13.43(0.02)&13.48(0.02)&13.96(0.02)\nl
       50040.000&----&13.85(0.02)&13.55(0.02)&13.62(0.02)&13.99(0.02)\nl
       50042.000&14.16(0.06)&14.08(0.02)&13.71(0.02)&13.82(0.03)&14.20(0.07)\nl
       50047.980&----&14.55(0.02)&13.82(0.02)&13.81(0.02)&13.94(0.03)\nl
       50051.020&15.18(0.03)&15.00(0.02)&14.01(0.03)&13.88(0.03)&13.89(0.04)\nl
       50067.000&16.38(0.04)&16.06(0.02)&14.88(0.02)&14.50(0.02)&14.22(0.02)\nl
       50070.860&----&16.19(0.02)&15.07(0.02)&14.74(0.02)&14.47(0.02)\nl
       50078.950&----&16.38(0.06)&15.35(0.02)&14.99(0.06)&14.90(0.02)\nl
       50086.980&----&16.54(0.04)&15.55(0.03)&15.32(0.06)&15.25(0.05)\nl
       50088.900&----&16.51(0.06)&15.58(0.03)&15.33(0.05)&15.25(0.05)\nl
       50103.820&----&16.76(0.08)&16.02(0.04)&15.86(0.03)&15.91(0.05)\nl
       50136.900&----&17.31(0.12)&16.86(0.08)&16.90(0.12)&17.21(0.17)\nl
       50161.730&----&17.81(0.21)&17.53(0.16)&17.85(0.21)&17.83(0.45)\nl
\enddata
\end{deluxetable}

It is important to insure that the photometry of the other 4 SNe Ia in Table 2 is also reliable.  For SN 1990N, Lira et al. (1999) undertook a comprehensive recalibration of SN 1990N.  SN 1994ae was recalibrated using galaxy subtraction in Riess et al. (2005) which also contained the calibration of SN 1998aq using the same techniques.

We have also verified the photometric calibration of SN 1981B
presented by \citet{buta83}.  For the Buta \& Turner stars A, B, and
C, we found respective differences in the $B$-band of 0.02, -0.01, and
0.02 mag (an average of 0.01 mag), where a positive difference
indicates our photometry is brighter. For the $V$ band we found
differences of 0.01, 0.00, and 0.00 mag, respectively.

\begin{deluxetable}{ccccccc}
\tablecaption{Table 18: SN Observables$^a$}
\tablenum{18}
\tablehead{\colhead{SN}&\colhead{$U_{\rm max}$ (mag)}&\colhead{$B_{\rm
max}~$ (mag)}&\colhead{$V_{\rm max}$~(mag)}&\colhead{$R_{\rm
max}$~(mag)}&\colhead{$I_{\rm max}$~(mag)}&\colhead{$\Delta m_{15}(B)$
}}
\startdata
SN 2002fk &  ------ & 13.32(0.02) & 13.38(0.02) & 13.39(0.03) & 13.50(0.04) &  1.08(0.03)  \\
SN 1995al &  13.01(0.05) & 13.31(0.02) & 13.27(0.02) & 13.23(0.02) & 13.62(0.03) &  1.00(0.05)   \\
\enddata
\tablenotetext{a}{The uncertainty is given in parentheses.}

\end{deluxetable}

\section{Discussion}

In a companion paper \citep{riess09}, we present IR measurements from NICMOS on {\it HST} of
the Cepheids analyzed here.  Although the
new, optical photometry of SNe Ia and Cepheids presented here already
addresses some of the largest sources of systematic error in the
determination of the Hubble constant, the {\it remaining} errors are further reduced from observations of these Cepheids at
longer wavelengths.

 One of the biggest of these remaining uncertainties results from the
corrections applied for Cepheid reddening.  Systematic errors in the
apparent color excess arise from differences in the photometric system
used to measure Cepheid colors in the SN hosts and the anchor galaxy or from intrinsic differences in color resulting from those in metallicity.  Errors in the optical color excess are further amplified by
use of a $V-I$ reddening ratio, $R={A_V \over E_{V-I}} \approx 2.5$.  Another source of
error arises from differences in the value of $R$ from host to
host or sight line to sight line for which Galactic variations in $R$ are $\sim$ 0.2 (Valencic, Clayton and Gordon 2004).  Reobserving the Cepheids with a single instrument (to negate
photometric system differences) and at redder wavelengths (reducing
the scale of $R$ and its variations) would mitigate these uncertainties.  We therefore
defer our full analysis of the Hubble constant resulting from the data
presented here to Riess et al. (2009), where we present the
long-wavelength observations from NICMOS.

However, we note here that the two new SNe Ia, SN 1995al and SN 2002fk, and the Cepheids in their hosts yield estimates of the SN Ia luminosity (corrected to the fiducial of the luminosity-light curve shape relation) which are quite consistent with the other SNe in Table 2.  Due to the relationship between luminosity and light curve shape, the value of the fiducial luminosity depends on which light curve shape within a family of light curves is chosen to be the fiducial 
.  Based on the maser distance to NGC 4258 and its Cepheids, Riess et al. (2009) find for the MLCS2k2 light curve family (Jha, Riess, Kirshner 2005) the fiducial, dereddened
absolute magnitude, $M^0_V$ (defined at the time of $B$ max) of SN 1995al and SN 2002fk is $-18.99 \pm 0.14$ and
$-19.16 \pm 0.12$ mag, respectively, in good agreement with the
average of $-19.05 \pm 0.07$ mag for the mean of the previous four (also from Riess et al. 2009).  Thus the average for all 6 is $M^0_V=-19.06 \pm 0.05$.

From our analysis it appears that the slopes for \PL and $P-w$ derived by Tammann et al. (2003) from Galactic Cepheids are
inaccurate.  Even limiting the analysis of the Tammann et al. Cepheids to the 27 with $P
> 10$~d, in order to remove more rapid ones not represented in our sample, we find a slope of $-4.37 \pm 0.18$.  The Galactic Cepheids have the  {\it same metallicity} as those presented here yet the Tammann et al.  slope is 7$\sigma$ greater than the mean of $-2.98 \pm 0.07$ for the 7 hosts in Table 13.  While one might invoke an unusual helium abundance as a possible explanation for the conflict with a single host (e.g., for NGC 4258 when Macri et al. 2006 showed the same discrepancy), this would be unlikely to explain why each of the 7 hosts in Table 13 are inconsistent with the Tammann et al. slope for $P>10$~d Cepheids.

We note that it is far more difficult to reliably measure the $P-w$ (or \PL)
slope of Galactic Cepheids (from our vantage point within the Galaxy)
than for extragalactic Cepheids.  While Cepheids in an external host
can be treated as coincident in distance, the estimate of the Galactic Cepheid
slope suffers from the need for many individual, accurate distances.
The cluster-based distances used for the longest period Galactic Cepheids
are not very reliable as they frequently utilize tenuous stellar
associations rather than cluster membership.  In addition, past distance
estimates to Galactic Cepheids using the Baade-Becker-Wesselink method
did not consider a period dependence of the projection factor which
may be significant \citep{gieren05,fouque07} and could lead to an inaccurate
slope.  

Geometric distance measurements via parallax are the "gold standard" for estimating distances.  They are much more robust than the previous methods.  The direct parallaxes of 10 Galactic Cepheids were measured by \citet{benedict07} using the Fine Guidance Sensor on {\it HST}.  These 10 Cepheids alone or combined with Hipparcos
parallaxes yield a $P-w$ slope of $-3.29 \pm 0.15$ \citep{leeuwen07},
consistent with the SN hosts ($-2.98 \pm 0.07$) and the LMC ($-3.2$
to $-3.3$).

The mean extinction of the Tammann et al Galactic Cepheids is high,
with a mean of 1.8 visual magnitudes, and is even higher for the longest
period Cepheids.  A possible error in the assumed value of the reddening
parameter (e.g., $\sigma_R \approx 0.5$) for these Cepheids would bias
the pulsation relations by 0.3 mag if each had the same extinction.
Thus, without more precise knowledge of this ratio of optical
absorption to reddening, it does not seem possible to use the
Galactic Cepheids with high extinction to determine $H_0$ to better than $\sim$ 15\%.
Because younger and higher mass Cepheids have more than the average
extinction, the {\it slope} of the inferred Galactic \PL can also be
biased by an error in $R$.  The mean of the Benedict et al Cepheids is a more modest 0.36 visual magnitudes which also adds to their reliability.

The Galactic measurement at the long-period end also suffers from
limited statistics.  The \citet{tammann03} sample has a mean period of
12~d with only 7 Cepheids at $P > 30$~d and only 1 at $P > 50$~d.  In
our SN hosts, over 200 Cepheids (more than half of the sample) have $P
>30$~d.  A Galactic sample of just 7 Cepheids at this range and with
the aforementioned concerns would appear insufficient, even if their
luminosities were accurate, to support a very precise measurement of
the Hubble constant.  However, the product of the 0.5 mag steeper slope from that of the Benedict et al
sample or the SN host sample 
and difference in the mean period between the Galactic and SN host samples of
$\Delta log P =0.4-0.5$ causes a decrease in $H_0$ of 10\%-12\% in Sandage et al.

Comparing results for the sum of the standardized peak magnitudes of nearby SNe Ia and the Hubble diagram intercept, 
$m_{v,i}^0 +5a_v$, a quantity which is invariant to the method of standardization,
the difference of the weighted mean for the four (out of six)
SNe~Ia (SN 1981B, SN 1990N, SN 1994ae, and SN 1998aq) that are in common
with \citet{sandage06} is $0.02 \pm 0.04$ mag, showing excellent agreement.   \citet{sandage06} also use
SNe~Ia data obtained from the photographic era (SN 1937C, 1960F, and
1974G) which R05 do not.  There are clear reasons for
not using the photographic data: they are not measured in the same way
as for the Hubble-flow set, they have been shown to be unreliable, and
they disagree with modern data \citep{riess05}. Even within the Sandage et al. analysis, the
the absolute magnitude inferred from the
photographic subset ($-19.68 \pm 0.10$) differs from the modern set ($-19.42
\pm 0.05$) by 2.6$\sigma$, making the average brighter by 0.05 mag and decreasing $H_0$ by 2.5\%.

Little of the difference in $H_0$ results from the determination of the absolute distance scale.
 \citet{sandage06} use an LMC distance modulus
$\mu_0 = 18.54$ mag as one route to their value.  
The implied LMC distance based on NGC 4258 and Galactic Cepheids  \citep{macri06,benedict07}  is $\mu_0 \sim 18.42$~mag, and Freedman et al. (2001) and \citet{riess05}adopted $mu_0 \equiv 18.50$~mag.

\section{Summary and Conclusions}

(1) Using ACS we have observed a total of 237 Cepheids (216 with $P >
20$~d) in three recent SN~Ia hosts: NGC 1309 (SN 2002fk), NGC 3021 (SN
1995al), and NGC 3370 (SN 1994ae).  We also present the multi-band
light curves of SN 1995al and SN 2002fk.

(2) We reobserved the hosts of 6 reliable SNe~Ia and the ``maser
galaxy'' NGC 4258 in {\it HST} Cycle 15 (following the
initial contiguous cycle of {\it HST} discovery observations) to
identify longer period Cepheids.  We found 57 with $60 < P < 100$~d and
29 with $75 < P < 100$~d which can aid in extending Cepheid
measurements to greater distances.

(3) We have measured the metallicity parameter, 12 + log[O/H], in H~II
regions to infer the metallicity of the Cepheid sample.  We find the
values for the Cepheids in the SN hosts and the inner region of NGC
4258 to be very homogeneous, all consistent with the solar value of
8.9.

(4) Based on 445 Cepheids across 7 hosts of solar metallicity we find
the mean slope of the $P-w$ relation using $V$-band and $I$-band
measurements, the most commonly used for distance-scale work, to be
$-2.98 \pm 0.07$.  The 7 individual slopes are all consistent with the
mean.  The observed slope is fairly consistent with the slope of LMC
Cepheids and consistent with the slope from Galactic Cepheid parallax distances.  It is inconsistent with the slope of Galactic Cepheid
distances from \citet{tammann03} using Baade-Becker-Wesselink and
cluster distances.  Using the slope derived here increases the value
of $H_0$ from that measured by \citet{sandage06} and \citet{tammann08}
by $\sim$ 15\% and constitutes the bulk of the difference in $H_0$
with \citet{freedman01} and \citet{riess05}.  A companion paper
\citep{riess09} provides the addition of $H$-band measurements of the
Cepheids in the 7 hosts and takes advantage of the full 
reduction in systematic errors afforded by our refurbished distance ladder to provide a new determination of $H_0$.

\bigskip 
\medskip 

We are grateful to Peter Stetson for making his DAOPHOT/ALLFRAME
software available. Financial support for this work was provided by
NASA through programs GO-9352, GO-9728, GO-10189, GO-10339, GO-10497 and
GO-10802 from the Space Telescope Science Institute, which is operated
by AURA, Inc., under NASA contract NAS 5-26555. A.V.F.'s supernova
group at U.C. Berkeley is also supported by NSF grant AST--0607485 and
by the TABASGO Foundation.  KAIT was constructed and supported by
donations from Sun Microsystems, Inc., the Hewlett-Packard Company,
AutoScope Corporation, Lick Observatory, the US National Science
Foundation (NSF), the University of California, the Sylvia \& Jim
Katzman Foundation, and the TABASGO Foundation. Some of the data
presented herein were obtained with the W. M. Keck Observatory, which
is operated as a scientific partnership among the California Institute
of Technology, the University of California, and NASA; the observatory
was made possible by the generous financial support of the W. M. Keck
Foundation.  We wish to extend special thanks to those of Hawaiian
ancestry on whose sacred mountain we are privileged to be guests.





\vfill 
\eject

{\bf Figure Captions}

Figure 1: Image of a $2' \times 2'$ region of the field of NGC 1309
from ACS WFC $F555W$. North is up and east to the left.  The locations
of the Cepheids identified are indicated by circles.

Figure 2:  Same as Figure 1 but for NGC 3021.

Figure 3: $F555W$ and $F814W$ light curves of Cepheids in NGC
3021. For each of the Cepheid candidates listed by identification
number in Table 8, the photometry is fit to the period-specific
light-curve template.

Figure 4: Same as Figure 3 for NGC 1309 (see Table 7).

Figure 5: $V$-band, $I$-band, and Wesenheit dereddened [$V -
  2.45(V-I)$] \PL relations for Cepheids identified in NGC 1309. The
\PL relation shown is from the LMC and is only fit to the filled symbols (see Figure 12 for a slope fit to the individual relations).
An approximate
instability-strip width of 0.35 mag is also given for comparison.
Open symbols indicate short-period Cepheids in the period range which
appears to suffer an incompleteness bias (preferentially bright or
significantly blended with a blue source).

Figure 6:  Same as Figure 5 but for NGC 3021.

Figure 7:  Same as Figure 5 but for NGC 3370.

Figure 8: Dereddened distance modulus from the $P-w$ relation as a
function of the period cutoff.  Although there is some evidence of an
incompleteness bias at the short-period end of the $V$-band and
$I$-band \PL relations, there is no significant change in the
dereddened distance modulus with period cutoff.

Figure 9: Composite $V$-band Cepheid light curves.  For the Cepheids
with $P > 10$~d in each host, the fitted mean magnitude was subtracted
from the observations which were then plotted as a function of their
phase, as inferred from the fitted period.  As seen and shown in Table
11, for each host the mean half-amplitude is consistent with 0.48 mag.
This consistency indicates a lack of significant variation in blending
from host to host and also provides a more direct probe of metallicity
which can cause amplitude differences in the mean.

Figure 10: Composite $V$-band Cepheid light curves of the longer
period Cepheids, $P > 60$~d, whose discovery was enabled by
reobservations of each host {\it HST} cycles subsequent to the
discovery cycle.  Although Cepheids at $P > 60$~d have reduced
amplitudes, they exhibit the characteristic sawtooth light curve of a
classical Cepheid and serve to extend the use of the \PL to greater
distances.

Figure 11: Radial gradients of the oxygen-to-hydrogen ratios [O/H] of
H~II regions in the Cepheid hosts. Following the method of Zaritsky et
al. (1994), emission-line intensity ratios of [O~II]
$\lambda\lambda$3726, 3729, [O~III] $\lambda\lambda$4959, 5007, and
H$\beta$ were used to derive values of 12 + log[O/H] for H~II
regions. The fitted linear gradient was used to determine the value of
the metallicity correction at the individual radii of the observed
Cepheids.  The variance in the data is dominated by spatial variations, not measurement errors.

Figure 12: The measured slopes of the dereddened (Wesenheit) $P-w$
relations as a function of the mean metallicity parameter.  The 7
hosts in Table 2 (including the inner region of NGC 4258) have
solar-like metallicities and slopes which are mutually consistent with
their mean of $-2.98 \pm 0.07$.  This mean is consistent with the
slope inferred from parallax distances to Galactic Cepheids (Leeuwen
et al. 2007) indicated by the X, in good agreement with the LMC
slope (square) and inconsistent with the Galactic slope (diamond)
inferred from non-geometric distance measures (Tammann et al. 2003).

Figure 13: Ground-based comparison stars in the field of NGC 1309 (SN
2002fk; {\it left}) and NGC 3021 (SN 1995al; {\it right}) used to
calibrate the light curves of the respective SNe~Ia.

Figure 14: MLCS2k2 fits to the multi-band data for SN 1995al and SN
2002fk.

\begin{figure}[ht]
\vspace*{140mm}
\figurenum{1}
\includegraphics{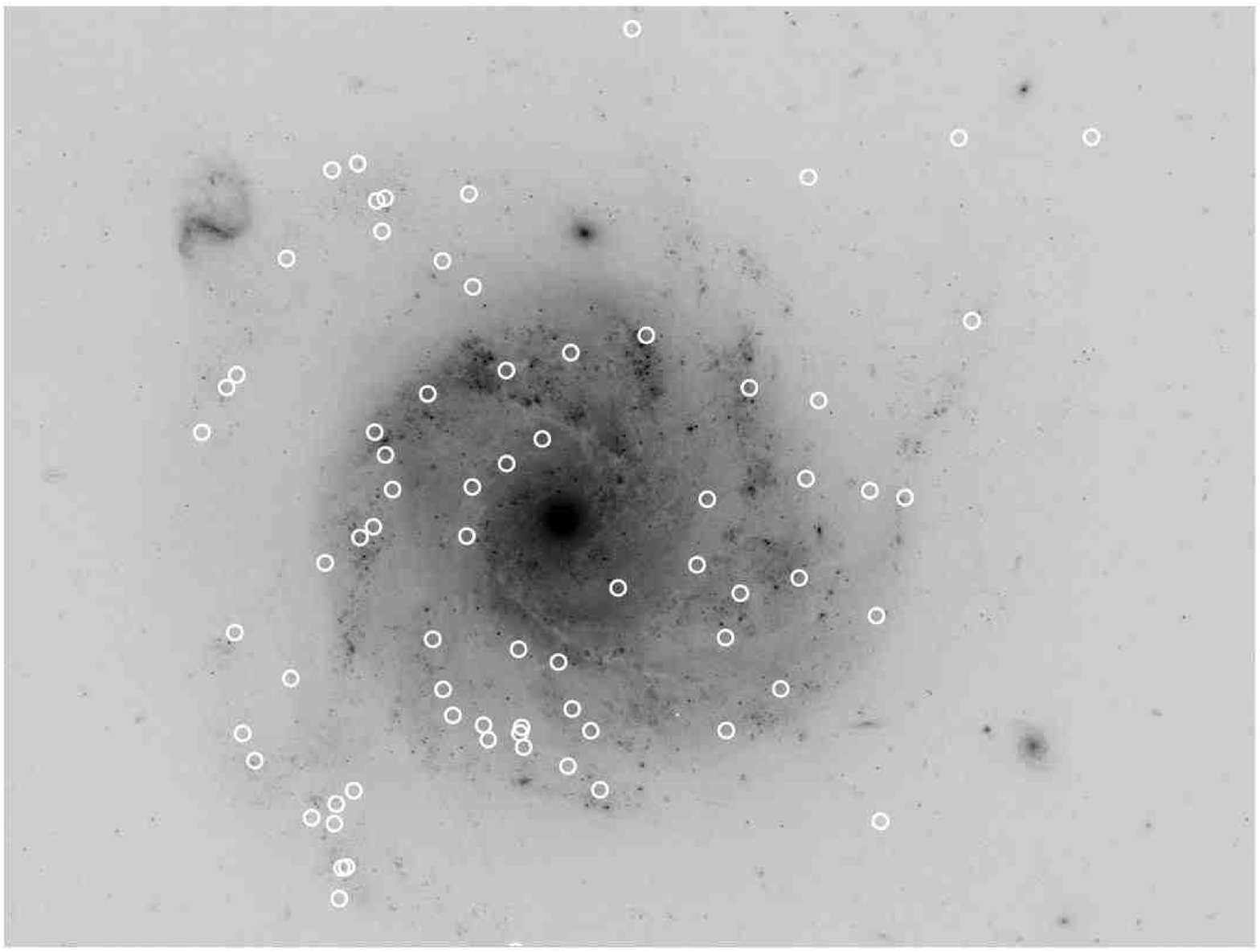}
\caption { }
\end{figure}

\begin{figure}[ht]
\vspace*{140mm}
\figurenum{2}
\includegraphics{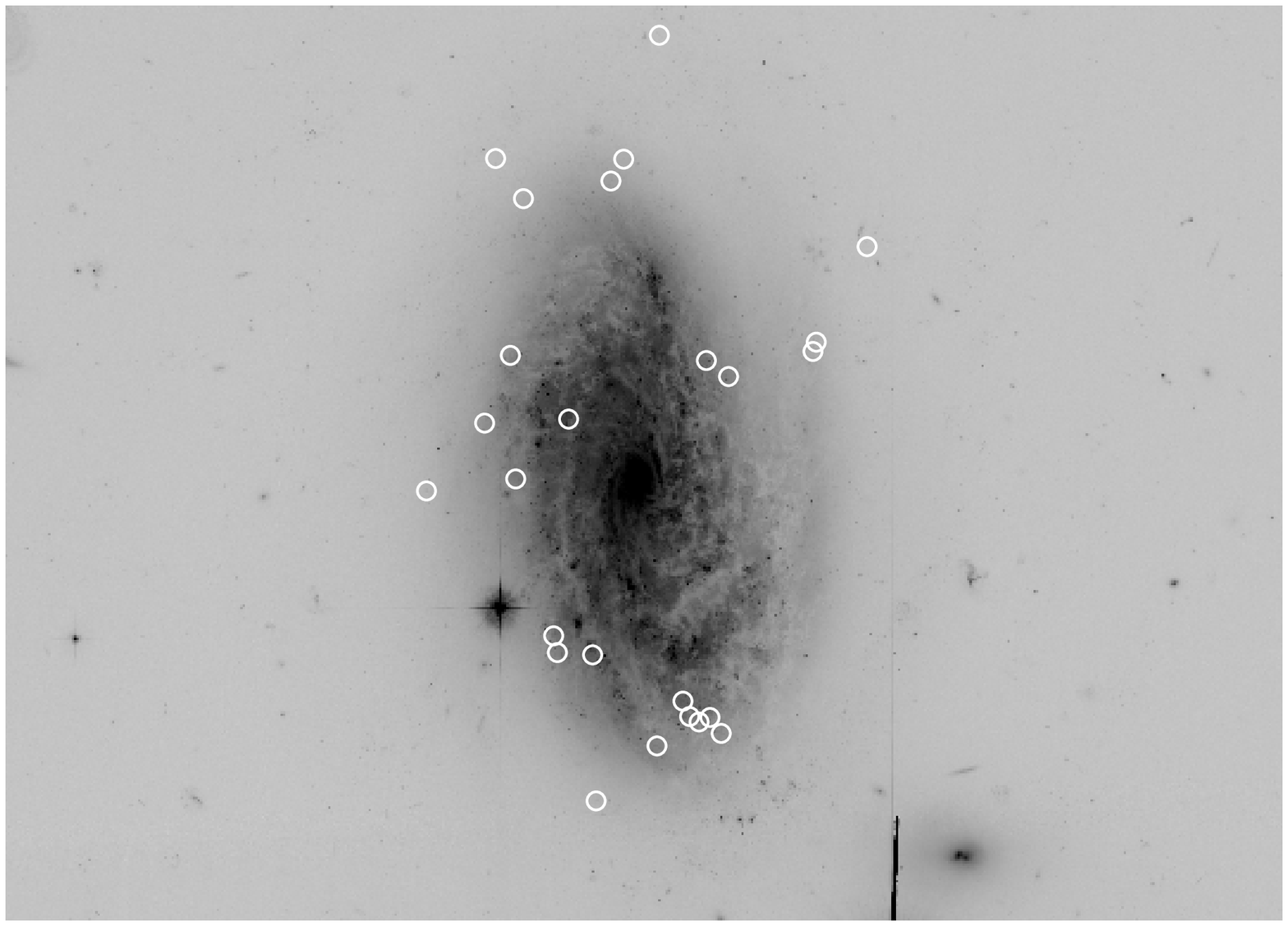}
\caption { }
\end{figure}

\begin{figure}[ht]
\vspace*{140mm}
\figurenum{3}
\includegraphics{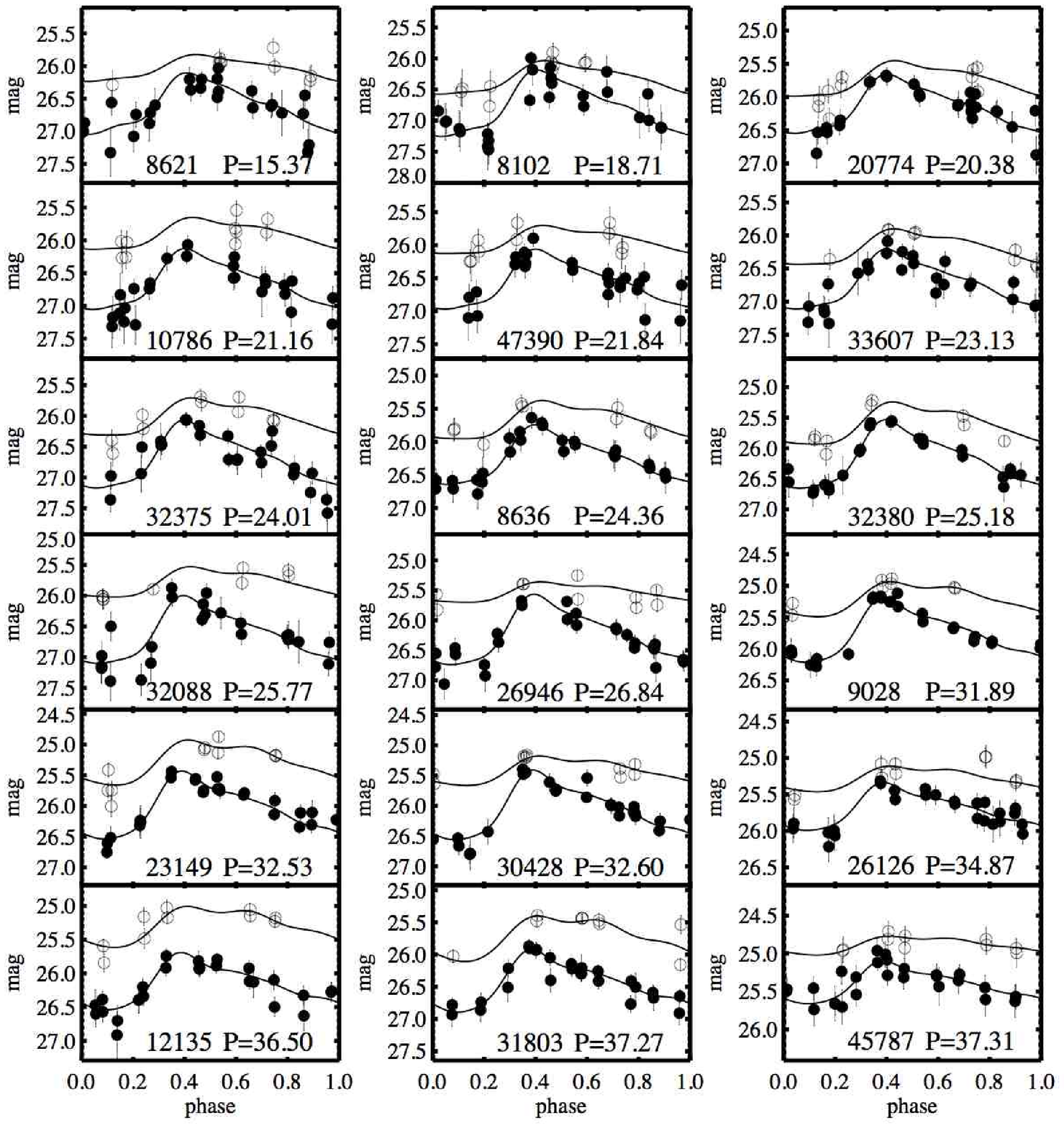}
\caption { Page 1 of 2 3021}
\end{figure}

\begin{figure}[ht]
\vspace*{140mm}
\figurenum{3}
\includegraphics{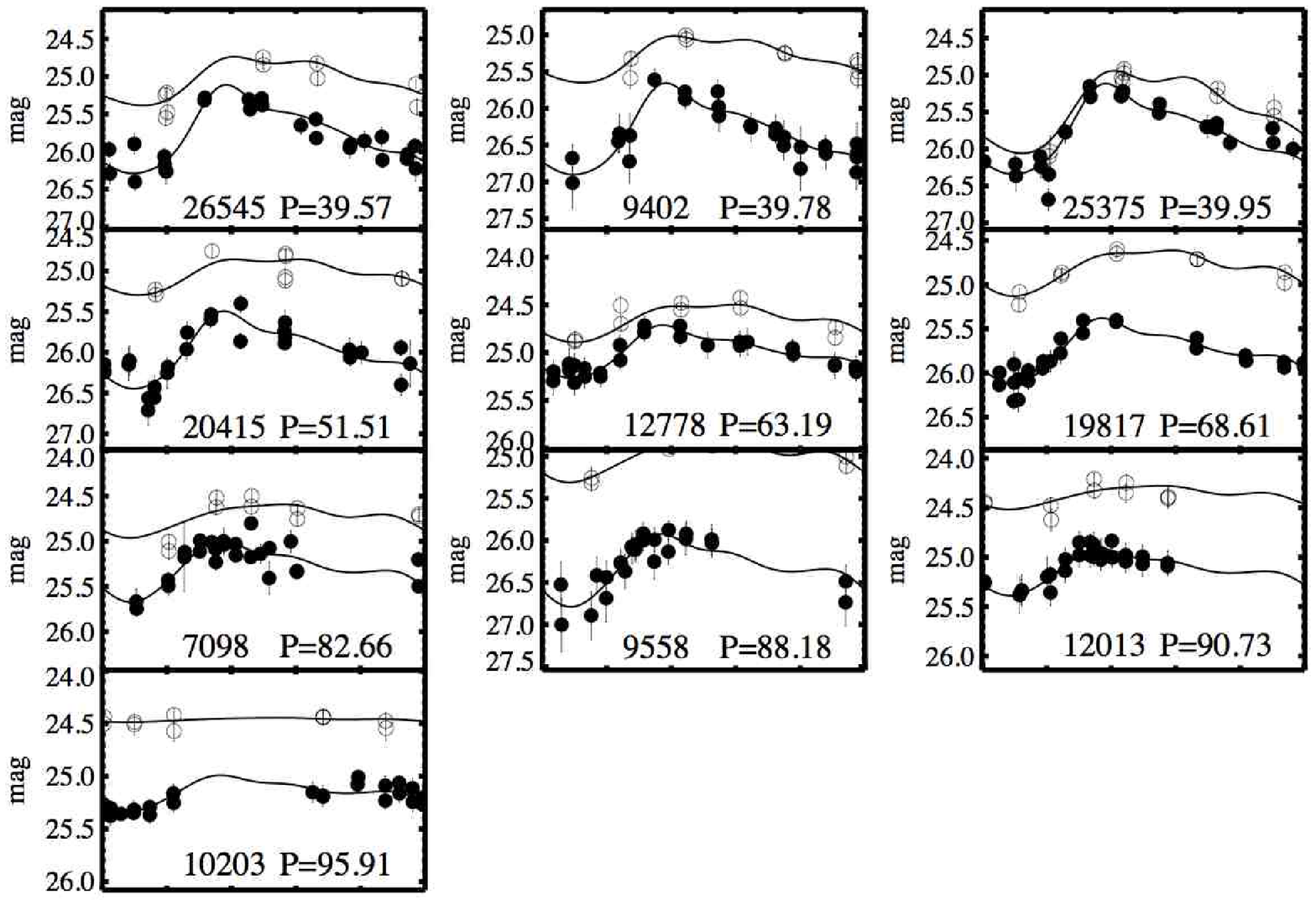}
\caption { Page 2 of 2 3021}
\end{figure}

\begin{figure}[ht]
\vspace*{140mm}
\figurenum{4}
\includegraphics{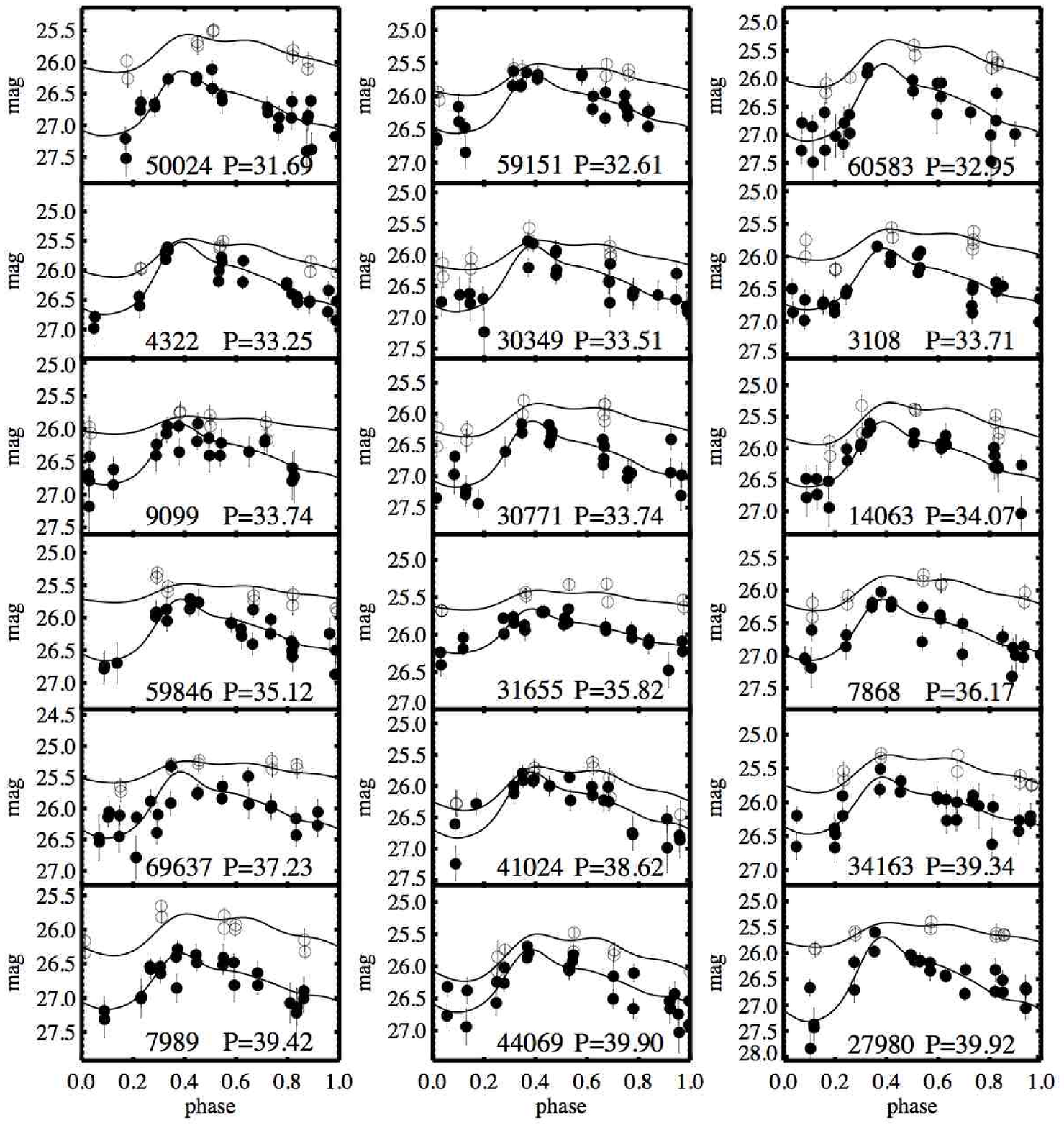}
\caption {Page 1 of 5 1309 }
\end{figure}

\begin{figure}[ht]
\vspace*{140mm}
\figurenum{4}
\includegraphics{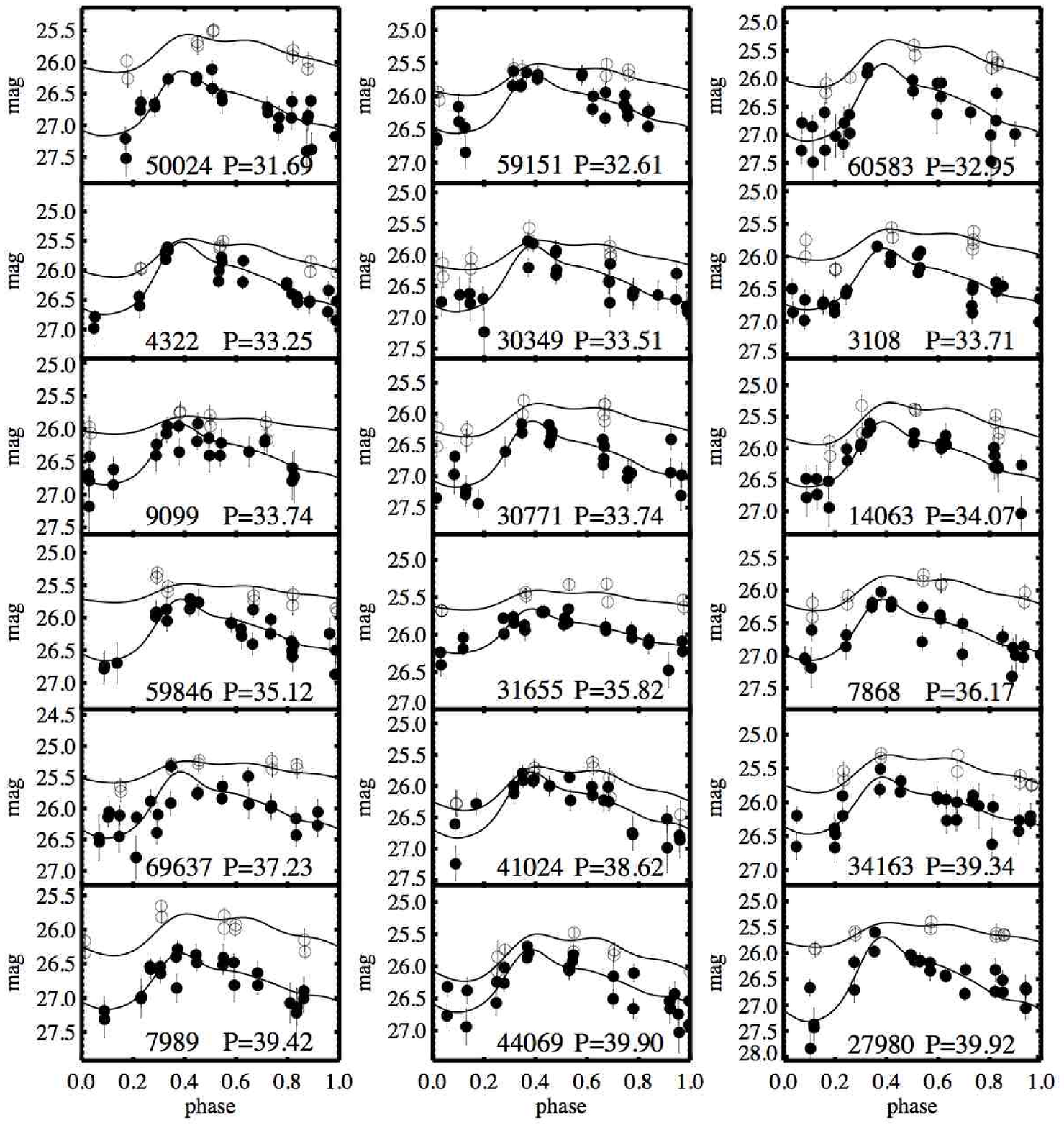}
\caption {Page 2 of 5 1309 }
\end{figure}

\begin{figure}[ht]
\vspace*{140mm}
\figurenum{4}
\includegraphics{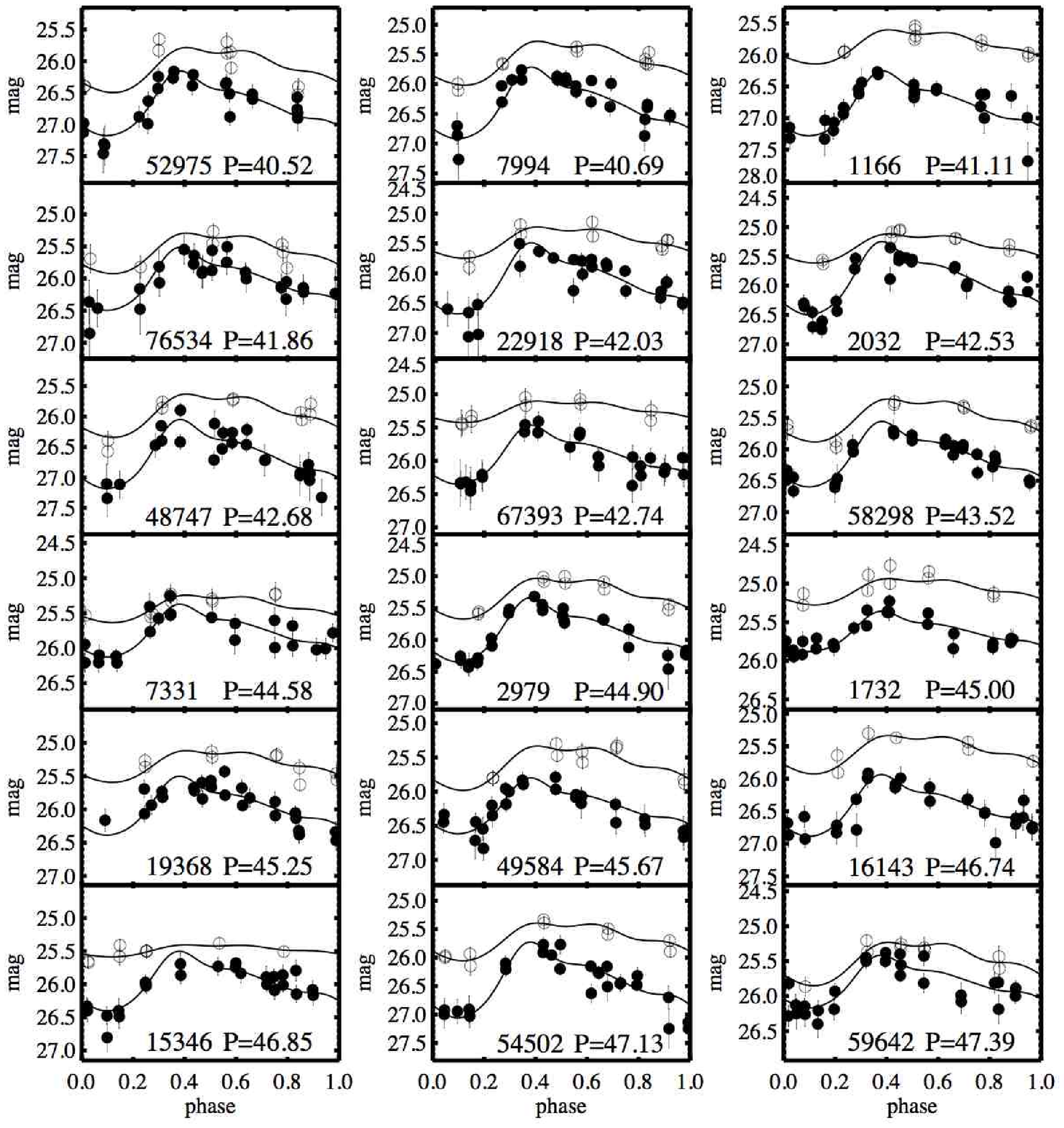}
\caption {Page 3 of 5 1309 }
\end{figure}

\begin{figure}[ht]
\vspace*{140mm}
\figurenum{4}
\includegraphics{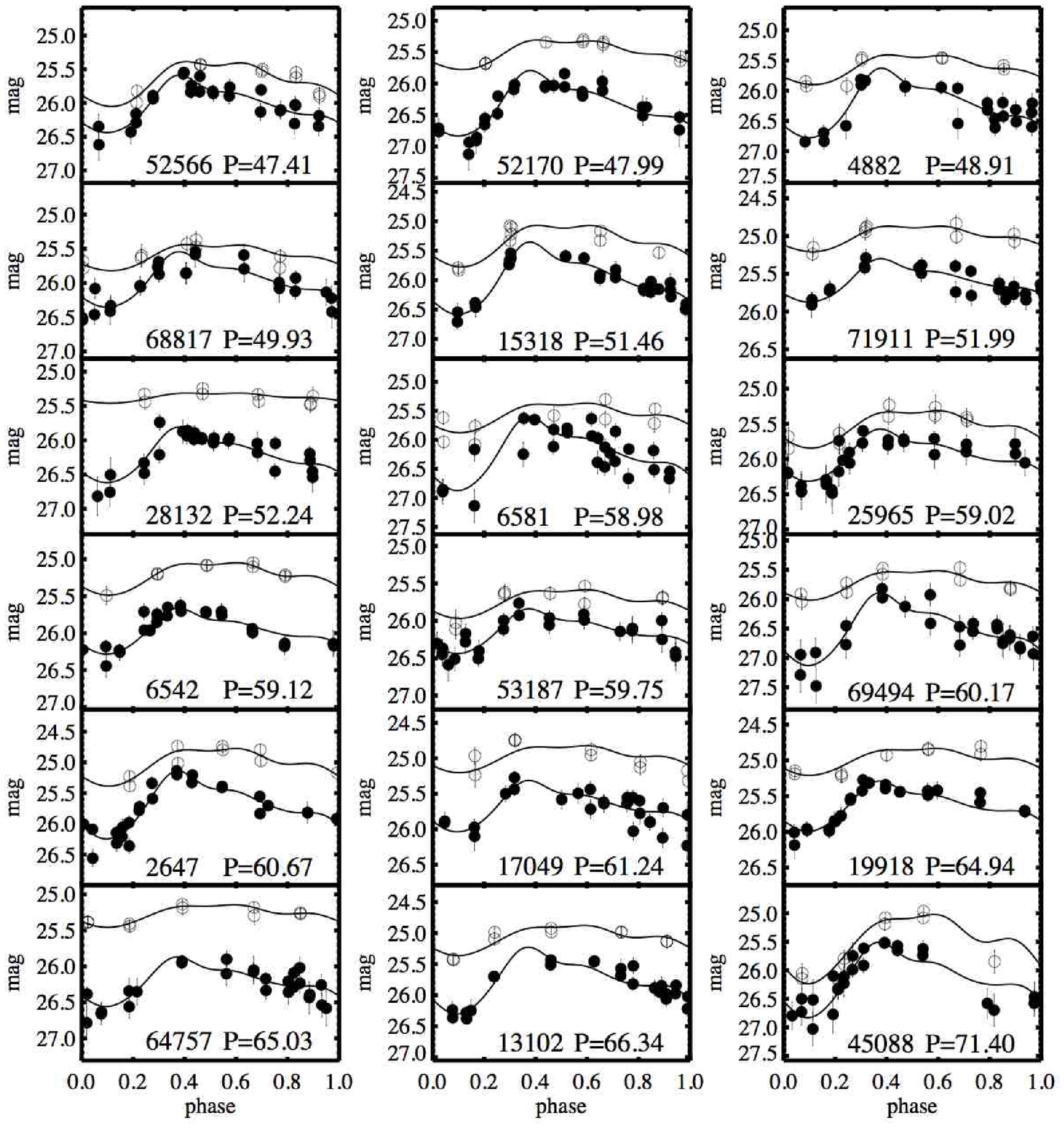}
\caption {Page 4 of 5 1309 }
\end{figure}

\begin{figure}[ht]
\vspace*{140mm}
\figurenum{5}
\includegraphics{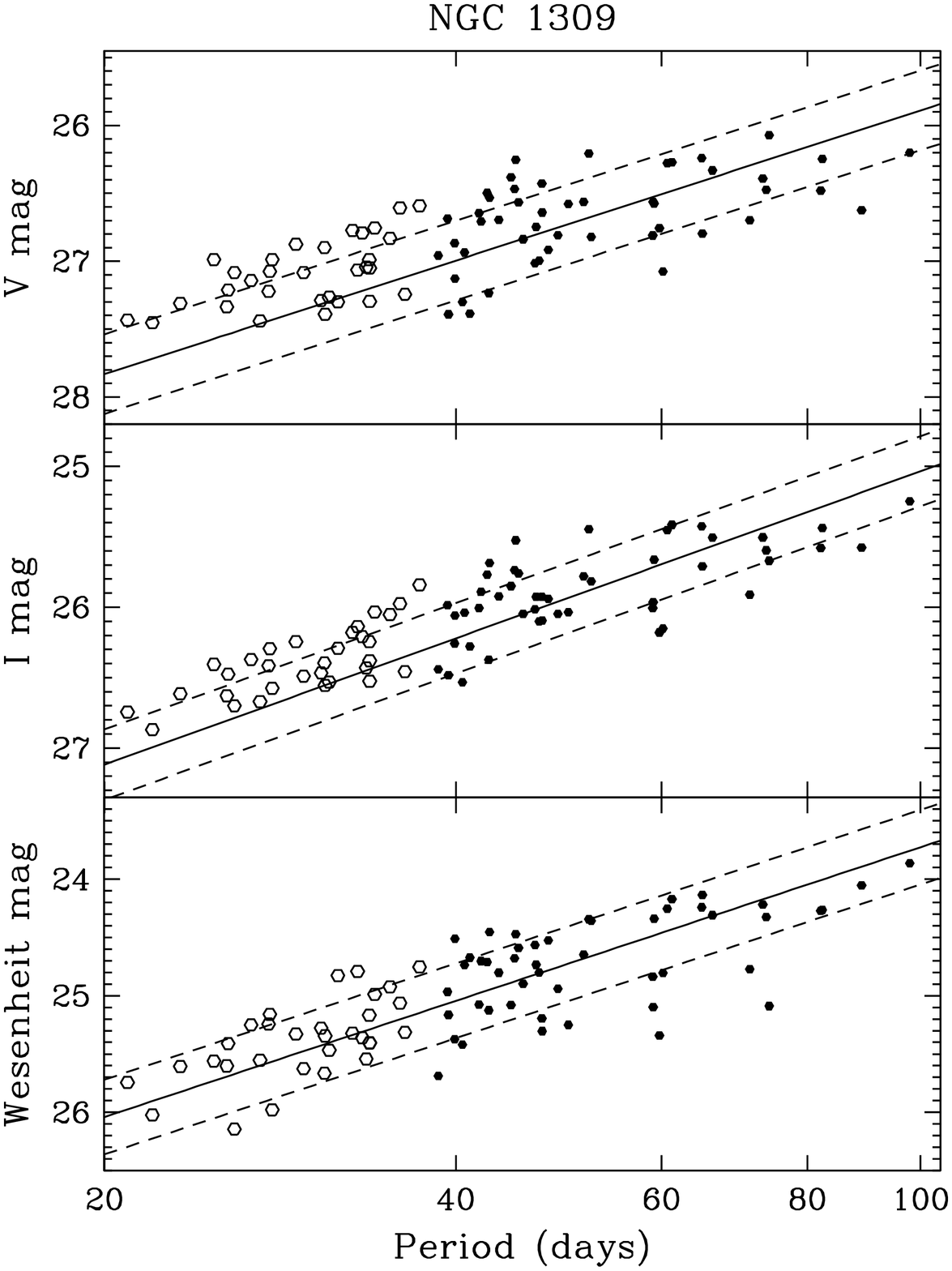}
\caption { }
\end{figure}

\begin{figure}[ht]
\vspace*{140mm}
\figurenum{6}
\includegraphics{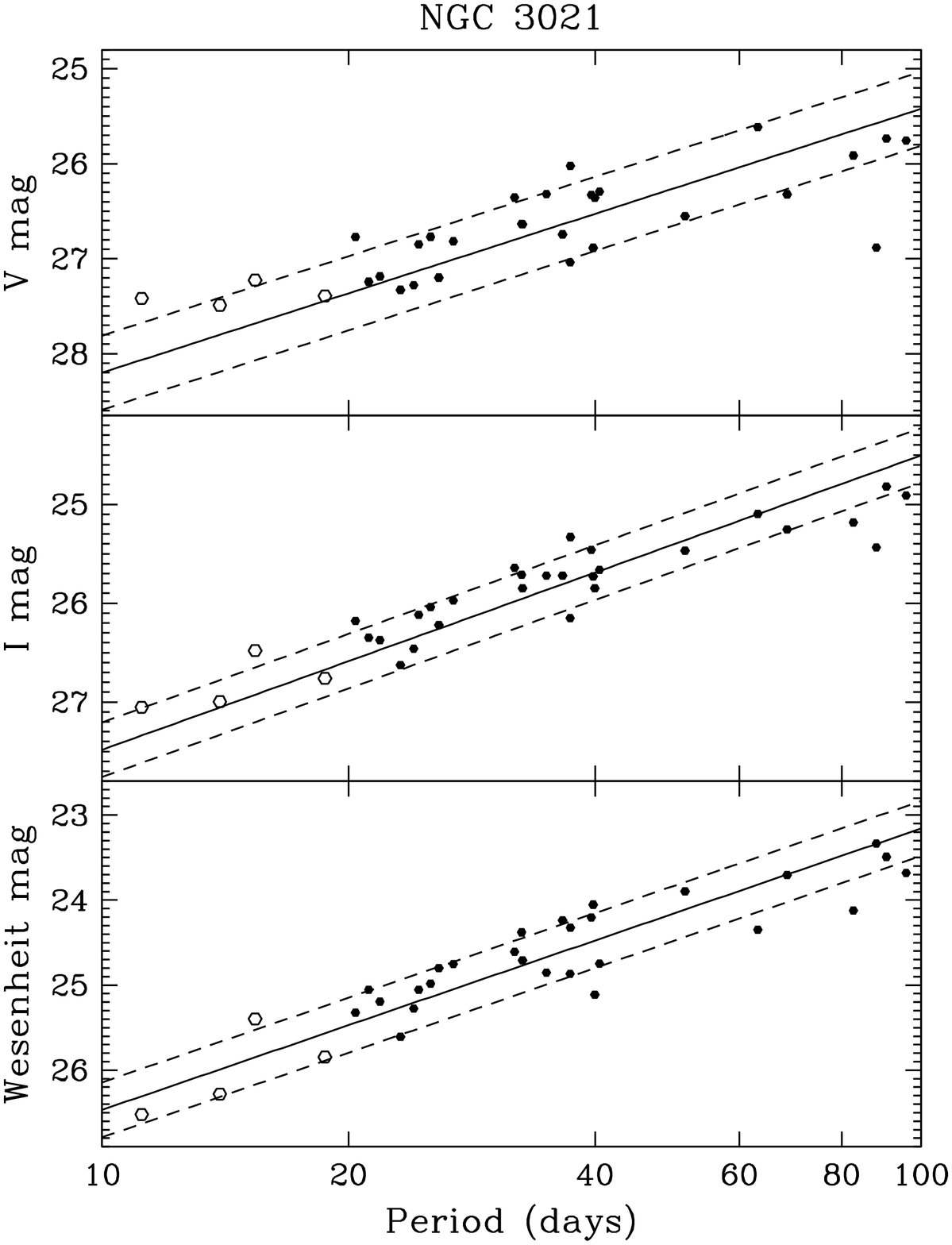}
\caption { }
\end{figure}

\begin{figure}[ht]
\vspace*{140mm}
\figurenum{7}
\includegraphics{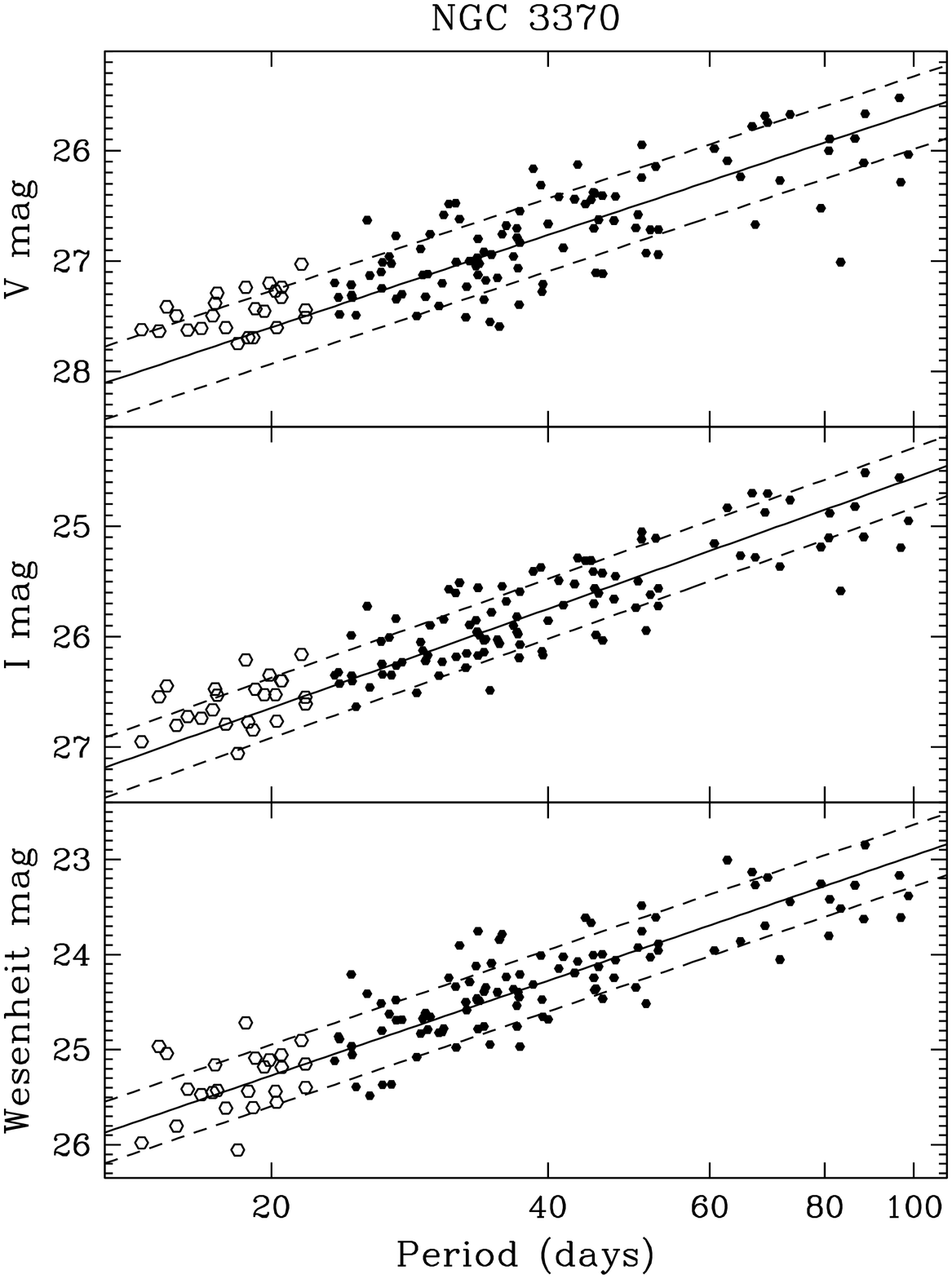}
\caption { }
\end{figure}

\begin{figure}[ht]
\vspace*{140mm}
\figurenum{8}
\includegraphics{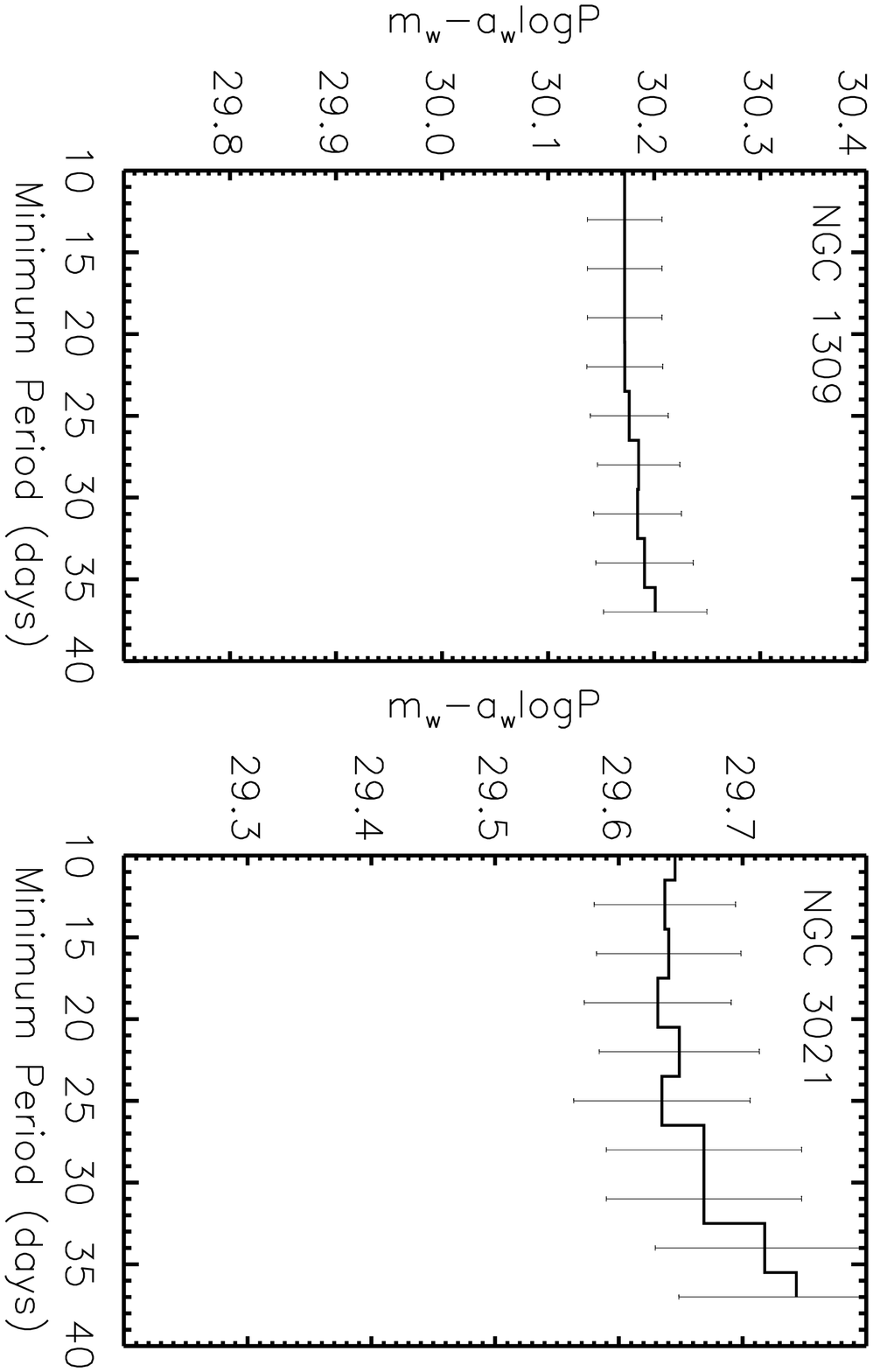}
\caption { }
\end{figure} 

\begin{figure}[ht]
\vspace*{140mm}
\figurenum{9}
\includegraphics{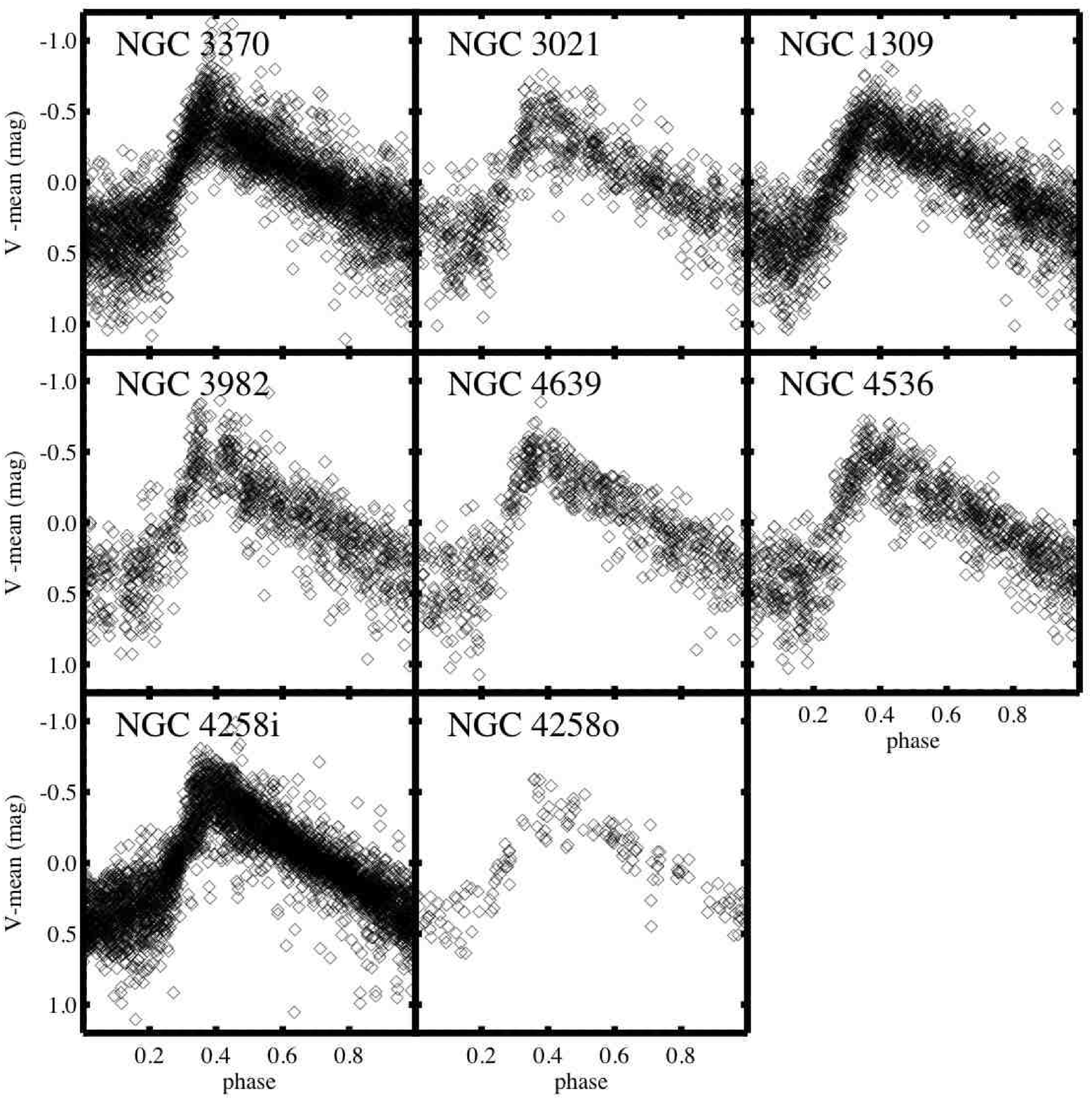}
\caption { }
\end{figure}

\begin{figure}[ht]
\vspace*{140mm}
\figurenum{10}
\includegraphics{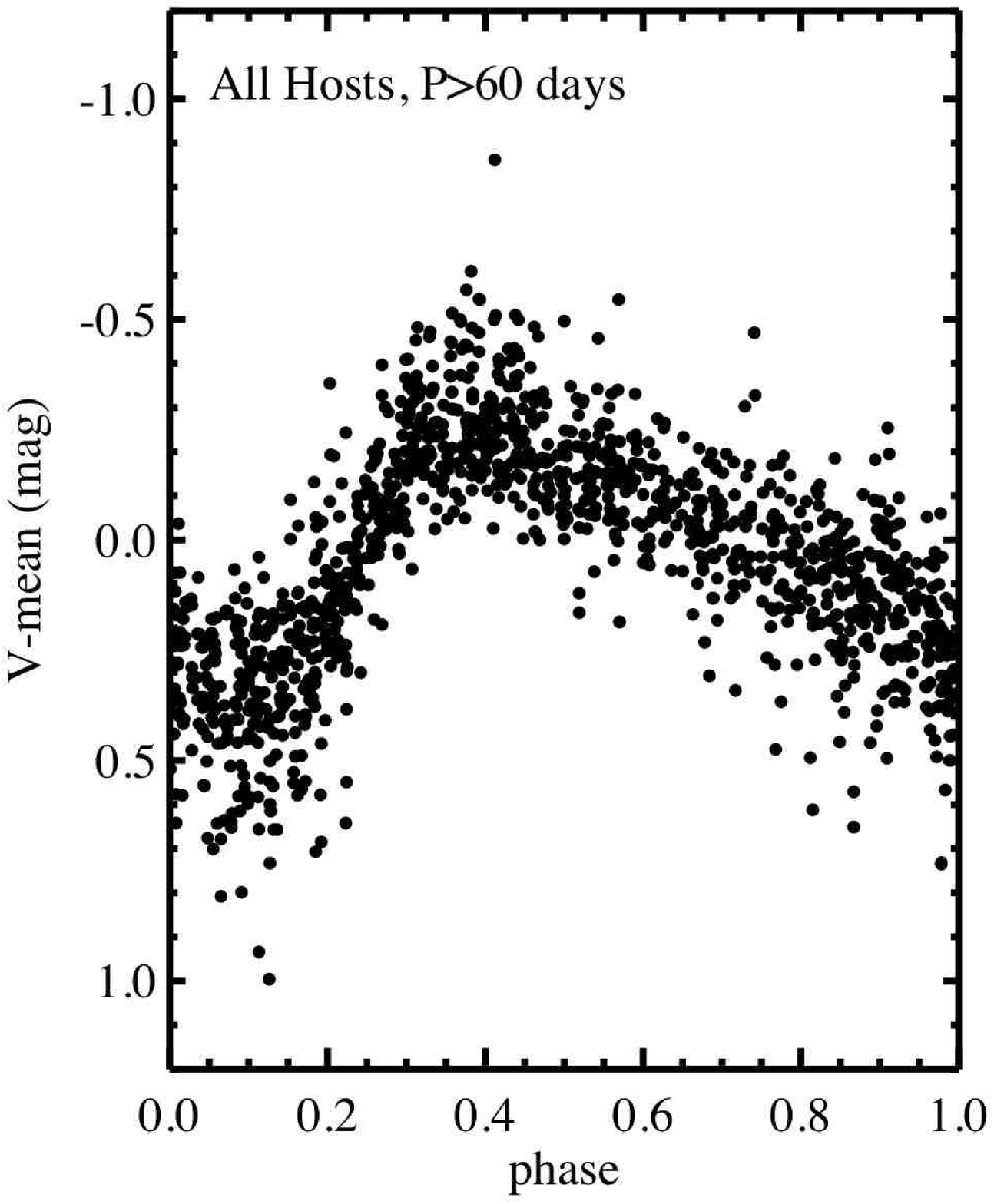}
\caption { }
\end{figure}

\begin{figure}[ht]
\vspace*{140mm}
\figurenum{11}
\includegraphics{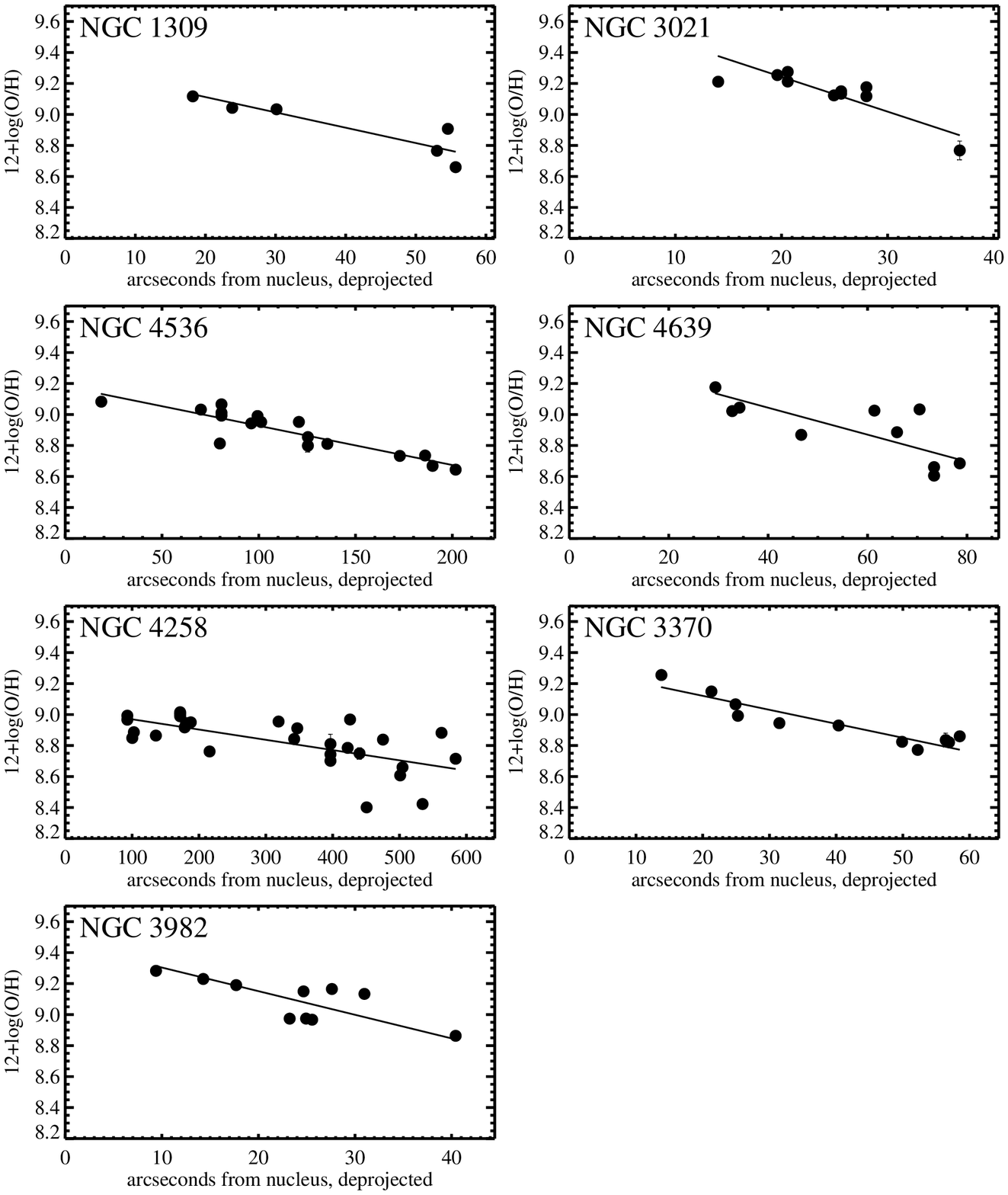}
\caption { }
\end{figure}

\begin{figure}[ht]
\vspace*{140mm}
\figurenum{12}
\includegraphics{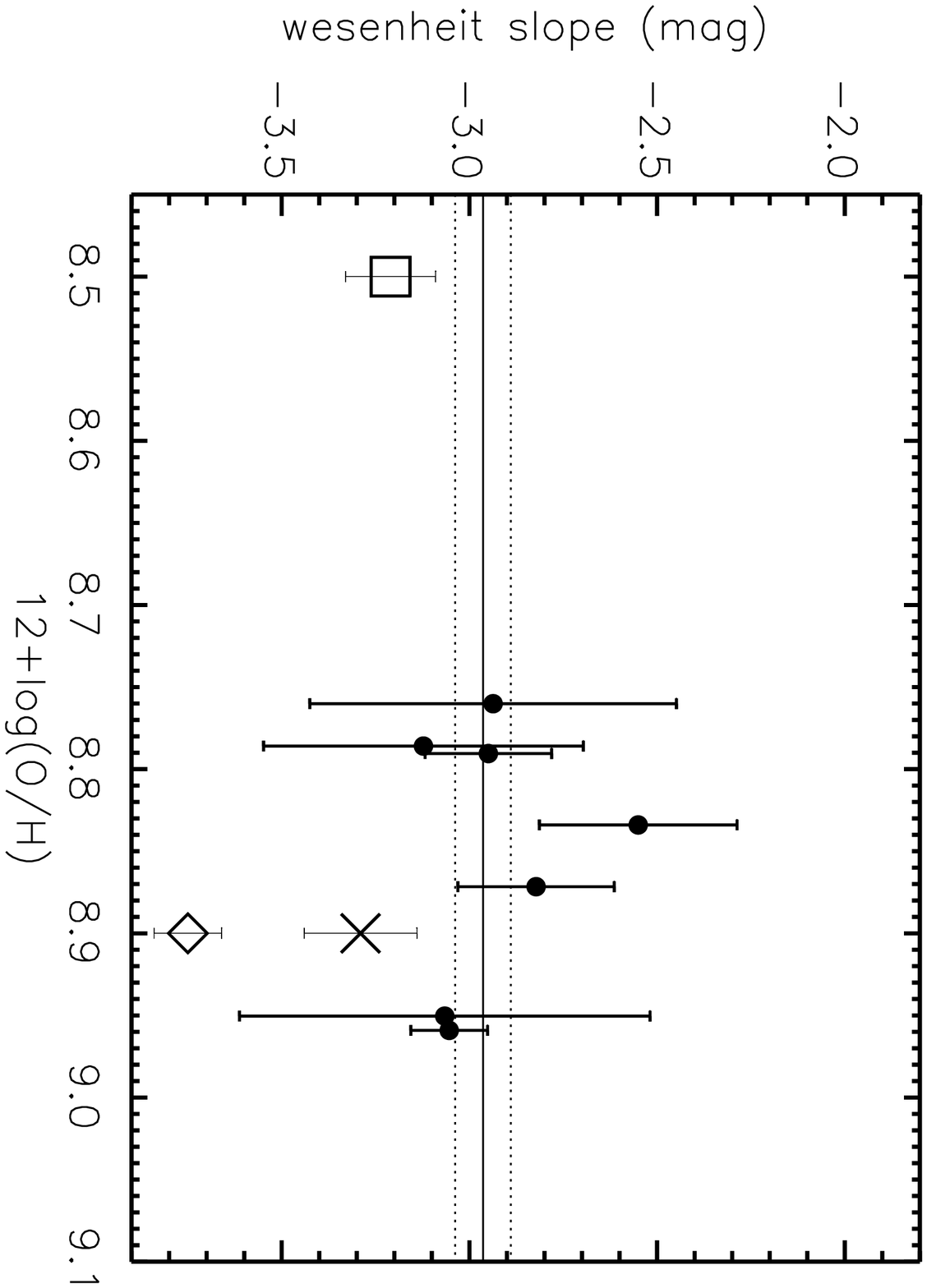}
\caption { }
\end{figure}

\begin{figure}[ht]
\vspace*{90mm}
\figurenum{13}
\includegraphics{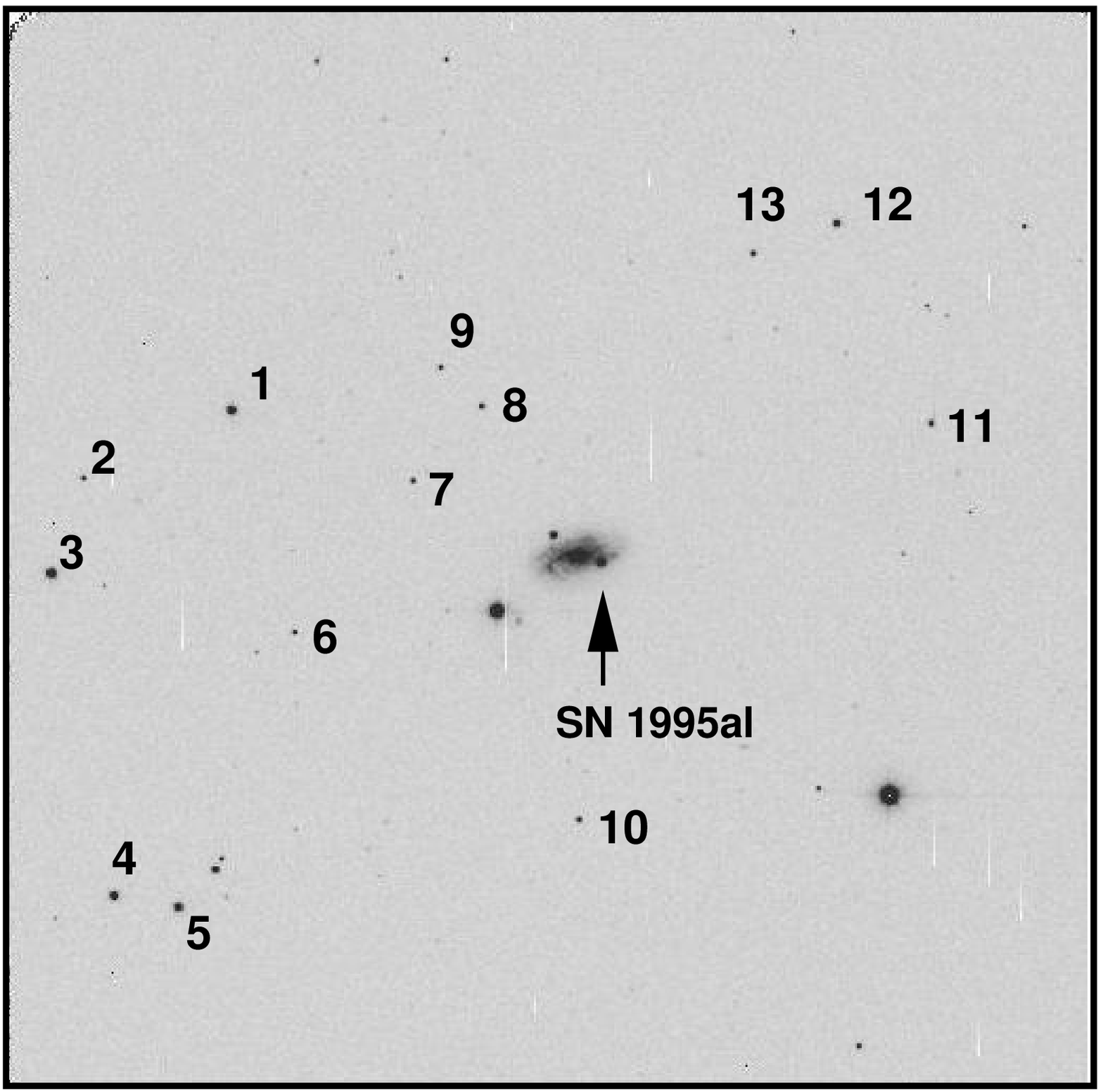}
\includegraphics{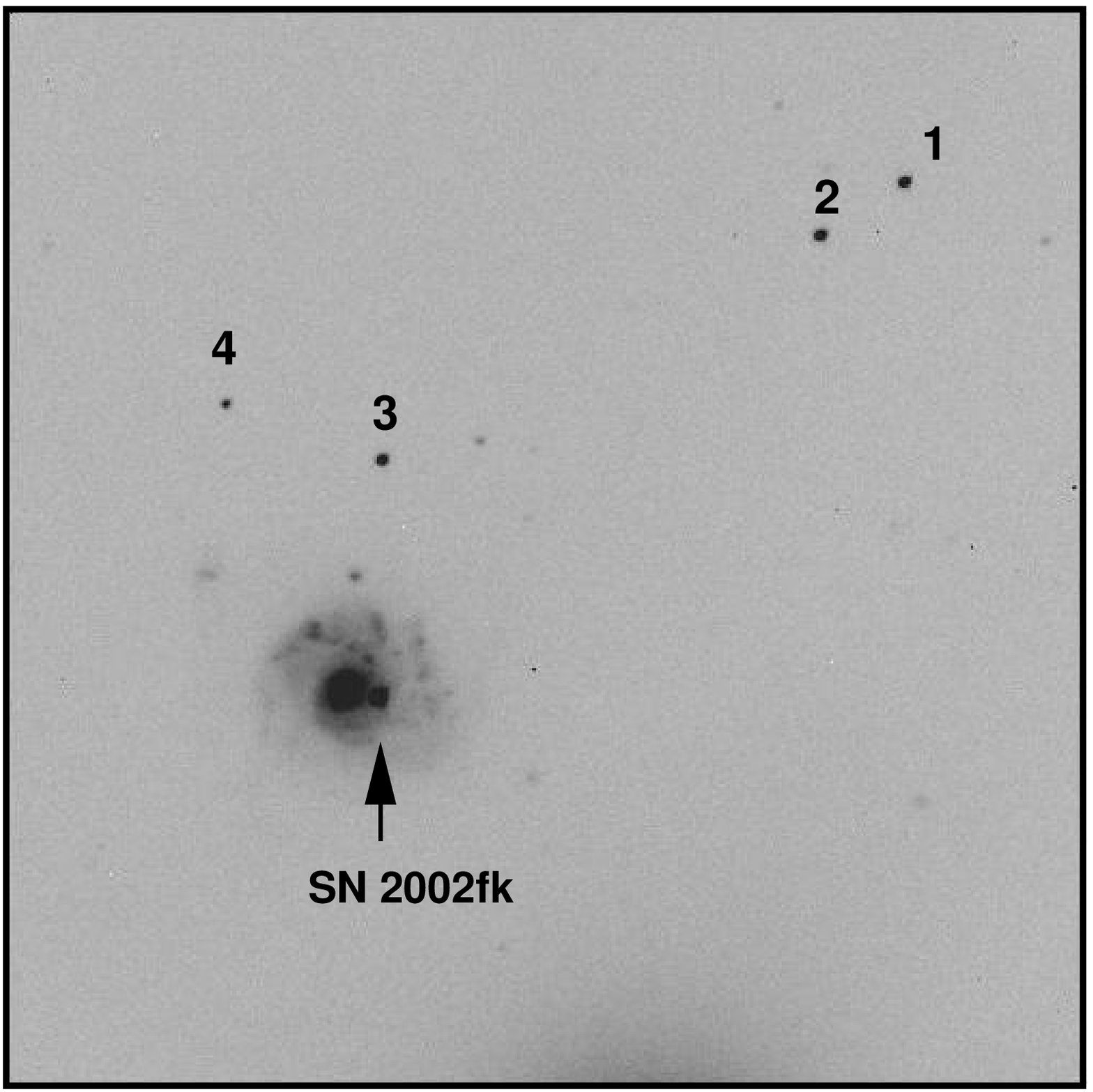}
\caption { Comparison Star Finders for SNe }
\end{figure}

\begin{figure}[ht]
\vspace*{100mm}
\figurenum{14}
\includegraphics{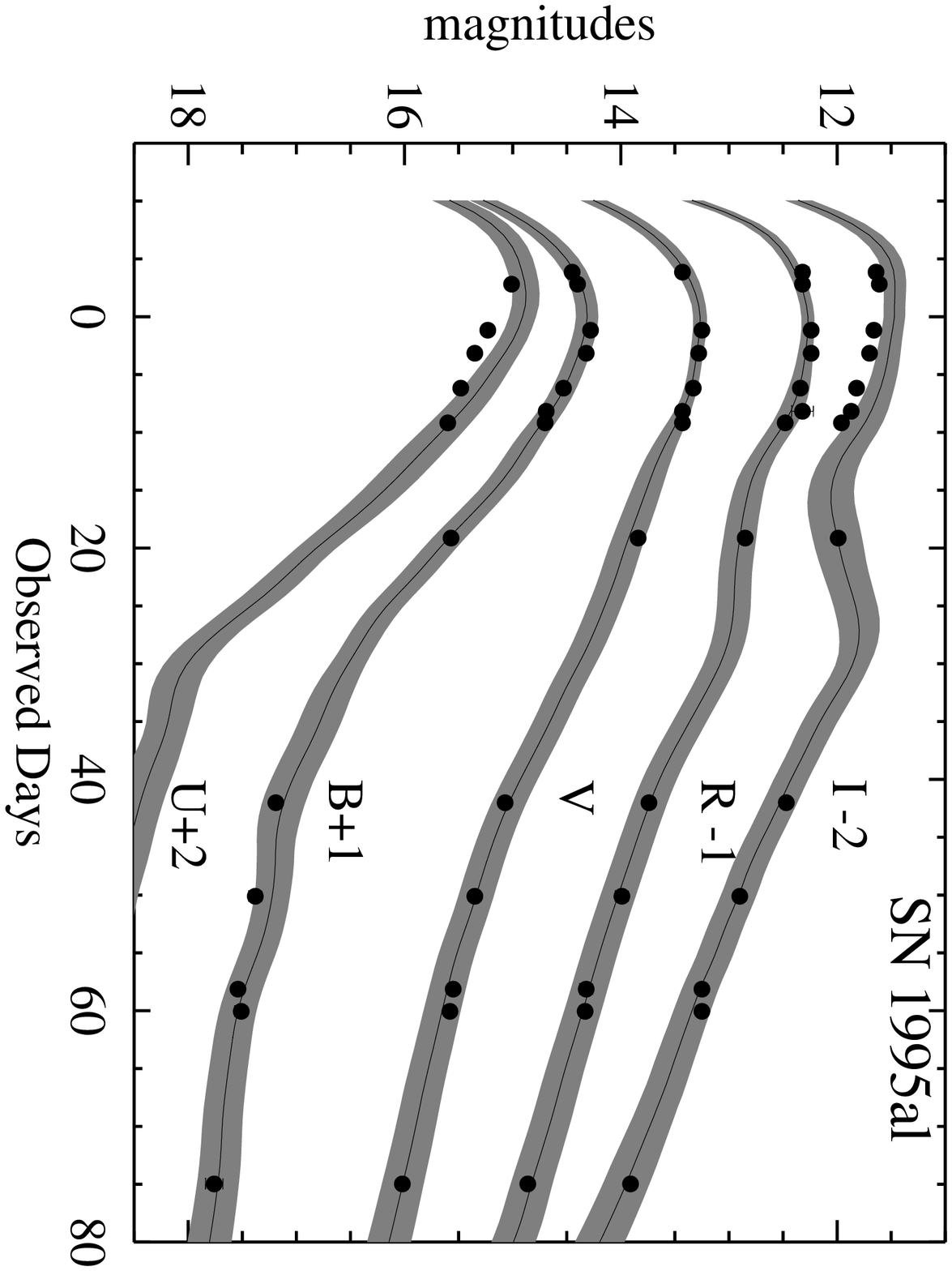}
\includegraphics{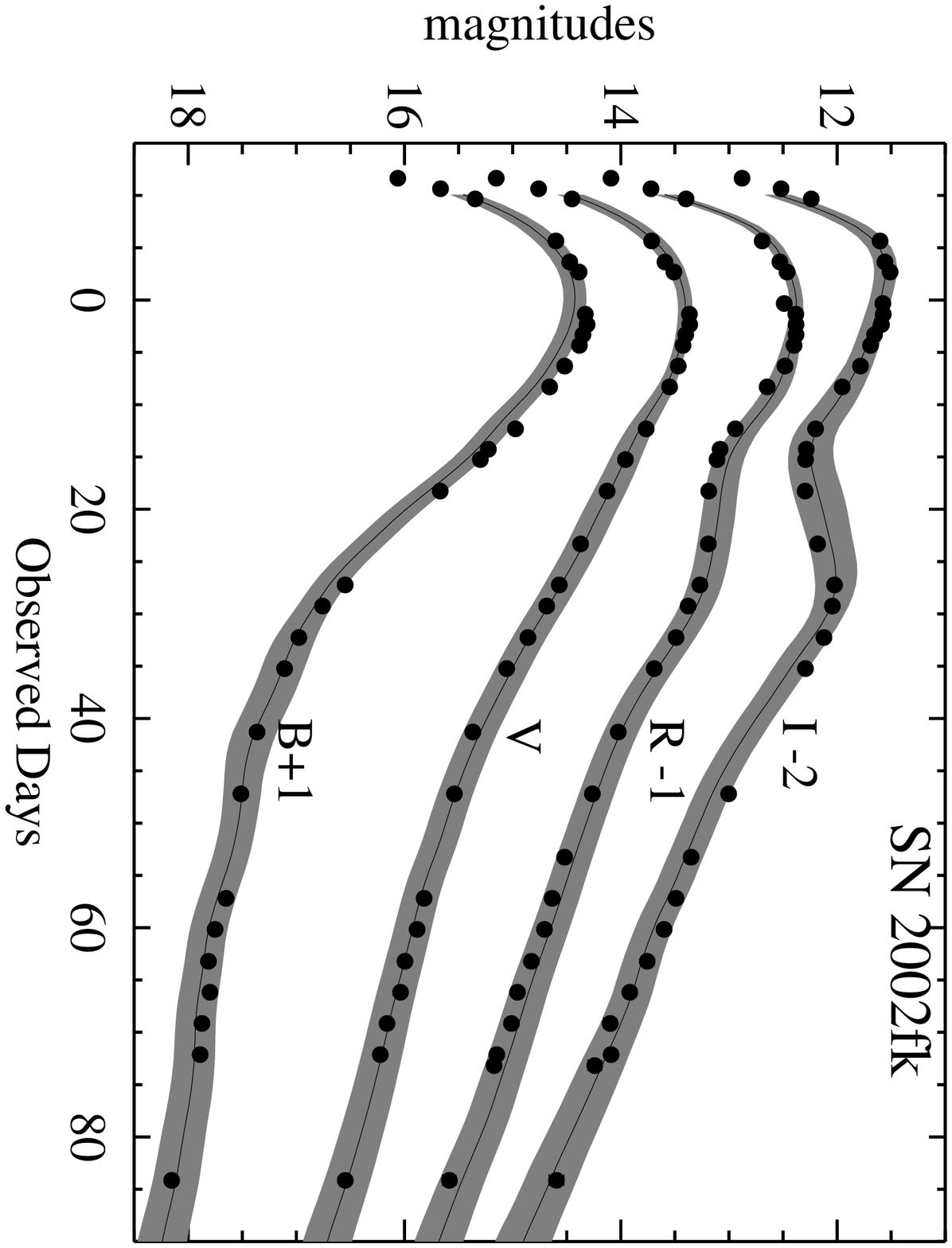}
\caption {Light Curves and Fits }
\end{figure}


\begin{thebibliography}{66}
\expandafter\ifx\csname natexlab\endcsname\relax\def\natexlab#1{#1}\fi

\bibitem[Alibert et al.(1999)]{alibert99} Alibert, Y., Baraffe, I.,
Hauschildt, P., \& Allard, F.\ 1999, \aap, 344, 551

\bibitem[Andrievsky et al.(2002)]{andrievsky02} Andrievsky, S.~M.,
Kovtyukh, V.~V., Luck, R.~E., L{\'e}pine, J.~R.~D., Maciel, W.~J., \&
Beletsky, Y.~V.\ 2002, \aap, 392, 491

\bibitem[Andrievsky et al.(2004)]{andrievsky04} Andrievsky, S.~M.,
Luck, R.~E., Martin, P., \& L{\'e}pine, J.~R.~D.\ 2004, \aap, 413, 159

\bibitem[Ayani \& Yamaoka(2002)]{ayani02} Ayani, K., \& Yamaoka, H.
  2002, IAU Circ. No. 7976

\bibitem[Benedict et al.(2007)]{benedict07} Benedict, G.~F., et al.\
2007, \aj, 133, 1810

\bibitem[Bird et al.(2009)]{bird09} Bird, J.~C., Stanek, K.~Z., \&
Prieto, J.~L.\ 2009, \apj, in press (arXiv:0807.4933)

\bibitem[Bono et al.(1999)]{bono99} Bono, G., Caputo, F., Castellani,
V., \& Marconi, M.\ 1999, \apj, 512, 711

\bibitem[Buta \& Turner(1983)]{buta83} Buta, R.~J., \& Turner, A.\
1983, \pasp, 95, 72

\bibitem[Chiaberge et al. (2009)]{chiaberge09} Chiaberge, M., et al.\
2009, HST ACS Instrument Science Report

\bibitem[Ferrarese et al.(2000)]{ferrarese00} Ferrarese, L.,
Silbermann, N.~A., Mould, J.~R., Stetson, P.~B., Saha, A., Freedman,
W.~L., \& Kennicutt, R.~C., Jr.\ 2000, \pasp, 112, 177

\bibitem[Filippenko et al.(2001)]{filippenko01} Filippenko, A.~V., Li,
W.~D., Treffers, R.~R., \& Modjaz, M.\ 2001, in Small
Telescope Astronomy on Global Scales, ed.
W. P. Chen, C. Lemme, \& B. Paczy\'{n}ski (San Francisco: ASP), 121

\bibitem[Ford et al.(2003)]{ford03} Ford, H.~C., et al.\ 2003,
\procspie, 4854, 81

\bibitem[Fouqu{\'e} et al.(2007)]{fouque07} Fouqu{\'e}, P., et al.\
2007, \aap, 476, 73

\bibitem[Freedman et al.(1992)]{freedman92} Freedman, W. L., Madore, 
B. F., Hawley, S. L., Horowitz, I. K., Mould, J., Navarrete, M., \&
Sallmen, S. 1992, \apj, 396, 80

\bibitem[{{Freedman} {et~al.}(2001){Freedman}, {Madore}, {Gibson},
{Ferrarese}, {Kelson}, {Sakai}, {Mould}, {Kennicutt}, {Ford},
{Graham}, {Huchra}, {Hughes}, {Illingworth}, {Macri}, \&
{Stetson}}]{freedman01}{Freedman}, W.~L., et al.\@ 2001, \apj, 553, 47

\bibitem[{{Gibson} {et~al.}(2000){Gibson}, {Stetson}, {Freedman},
{Mould}, {Kennicutt}, {Huchra}, {Sakai}, {Graham}, {Fassett},
{Kelson}, {Ferrarese}, {Hughes}, {Illingworth}, {Macri}, {Madore},
{Sebo}, \& {Silbermann}}]{gibson00}{Gibson}, B.~K., et al.\@ 2000,
\apj, 529, 723

\bibitem[{{Gieren} {et~al.}(2005){Gieren}, {Storm}, {Barnes},
{Fouqu{\' e}}, {Pietrzy{\' n}ski}, \& {Kienzle}}]{gieren05}{Gieren},
W., et al.\@ 2005, \apj, 627, 224

\bibitem[Gieren et al.(2008)]{gieren08} Gieren, W., et al.\ 2008,
\apj, 683, 611

\bibitem[Guy et al.(2005)]{guy05} Guy, J., Astier, P., Nobili, S.,
Regnault, N., \& Pain, R.\ 2005, \aap, 443, 781

\bibitem[Hamuy et al.(1995)]{hamuy95} Hamuy, M., Phillips, M.~M.,
Maza, J., Suntzeff, N.~B., Schommer, R.~A., \& Aviles, R.\ 1995, \aj,
109, 1

\bibitem[Hamuy et al.(1996)]{hamuy96} Hamuy, M., Phillips, M.~M.,
Suntzeff, N.~B., Schommer, R.~A., Maza, J., Smith, R.~C., Lira, P., \&
Aviles, R.\ 1996, \aj, 112, 2438

\bibitem[Holtzman et al.(1995)]{holtzman95} Holtzman, J.~A., Burrows,
C.~J., Casertano, S., Hester, J.~J., Trauger, J.~T., Watson, A.~M., \&
Worthey, G.\ 1995, \pasp, 107, 1065

\bibitem[Jha et al.(2007)]{jha07} Jha, S., Riess, A. G., \&
Kirshner, R. P. 2007, \apj, 659, 122

\bibitem[Jha et al.(2006)]{jha06} Jha, S., et al.\ 2006, \aj, 131, 527

\bibitem[{{Kanbur} \& {Ngeow}(2004)}]{kanbur04}{Kanbur}, S.~M., \&
{Ngeow}, C.-C. 2004, \mnras, 350, 962

\bibitem[Kanbur \& Ngeow(2006)]{kanbur06} Kanbur, S.~M., \& Ngeow,
C.-C.\ 2006, \mnras, 369, 705

\bibitem[{{Kennicutt} {et~al.}(1998){Kennicutt}, {Stetson}, {Saha},
{Kelson}, {Rawson}, {Sakai}, {Madore}, {Mould}, {Freedman},
{Bresolin}, {Ferrarese}, {Ford}, {Gibson}, {Graham}, {Han}, {Harding},
{Hoessel}, {Huchra}, {Hughes}, {Illingworth}, {Macri}, {Phelps},
{Silbermann}, {Turner}, \& {Wood}}]{kennicutt98}{Kennicutt}, R.~C., et
al.\@ 1998, \apj, 498, 181

\bibitem[{{Kochanek}(1997)}]{kochanek97}{Kochanek}, C.~S. 1997, \apj,
491, 13

\bibitem[Landolt(1992)]{landolt92} Landolt, A.~U.\ 1992, \aj, 104, 340

\bibitem[{{Leonard} {et~al.}(2003){Leonard}, {Kanbur}, {Ngeow}, \&
{Tanvir}}]{leonard03}{Leonard}, D.~C., et al.\@ 2003, \apj, 594, 247

\bibitem[Li et al.(2000)]{li00} Li, W.~D., et al.\ 2000, in Cosmic 
Explosions, ed. S. S. Holt \& W. W. Zhang (New York: AIP), 103

\bibitem[Macri et al.(2001)]{macri01} Macri, L.~M., Stetson, P.~B.,
Bothun, G.~D., Freedman, W.~L., Garnavich, P.~M., Jha, S., Madore,
B.~F., \& Richmond, M.~W.\ 2001, \apj, 559, 243



\bibitem[Macri et al.(2006)]{macri06} Macri, L.~M., Stanek, K.~Z.,
Bersier, D., Greenhill, L.~J., \& Reid, M.~J.\ 2006, \apj, 652, 1133

\bibitem[Macri et al.(2009)]{macri09} Macri, L.~M., et al.\ 2009,
submitted

\bibitem[Madore(1982)]{madore82} Madore, B.~F.\ 1982, \apj, 253, 575

\bibitem[Madore \& Freedman (2009)]{madfre09} Madore, B.~F. \& Freedman, W. L., 2009, \apj in press

\bibitem[Marconi et al.(2005)]{marconi05} Marconi, M., Musella, I., \&
Fiorentino, G.\ 2005, \apj, 632, 590

\bibitem[Oke et al.(1995)]{oke95} Oke, J.~B., et al.\ 1995, \pasp,
107, 375

\bibitem[Paczy{\'n}ski \& Pindor(2000)]{paczynski00} Paczy{\'n}ski,
B., \& Pindor, B.\ 2000, \apjl, 533, L103

\bibitem[Perlmutter et al.(1997)]{perlmutter97} Perlmutter, S., et
al.\ 1997, \apj, 483, 565

\bibitem[Persson et al.(2004)]{persson04} Persson, S.~E., Madore,
B.~F., Krzemi{\'n}ski, W., Freedman, W.~L., Roth, M., \& Murphy,
D.~C.\ 2004, \aj, 128, 2239

\bibitem[Phillips et al.(1999)]{phillips99} Phillips, M.~M., Lira, P.,
Suntzeff, N.~B., Schommer, R.~A., Hamuy, M., \& Maza, J.\ 1999, \aj,
118, 1766

\bibitem[Phillips(1993)]{phillips93} Phillips, M.~M.\ 1993, \apjl,
413, L105

\bibitem[Riess et al.(1995)]{riess95} Riess, A.~G., Press, W.~H., \&
Kirshner, R.~P.\ 1995, \apjl, 438, L17

\bibitem[Riess et al.(1996a)]{riess96a} Riess, A.~G., Press, W.~H., \&
Kirshner, R.~P.\ 1996, \apj, 473, 88

\bibitem[Riess et al.(1996b)]{riess96b} Riess, A.~G., Press, W.~H., \&
Kirshner, R.~P.\ 1996, \apj, 473, 588

\bibitem[Riess et al.(1998)]{riess98} Riess, A.~G., et al.\ 1998, \aj,
116, 1009

\bibitem[Riess et al.(1999)]{riess99} Riess, A.~G., et al.\ 1999, \aj,
117, 707

\bibitem[{{Riess} {et~al.}(2005){Riess}, {Li}, {Stetson},
{Filippenko}, {Jha}, {Kirshner}, {Challis}, {Garnavich}, \&
{Chornock}}]{riess05}{Riess}, A.~G., et al.\@ 2005, \apj, 627, 579

\bibitem[{{Riess} {et~al.}(2009){Riess}, {Macri}, {Casertano},
{Ferguson}, {Filippenko} {et al.}}]{riess09}{Riess}, A.~G., et al.\@
2009, submitted

\bibitem[Saha et al.(1997)]{saha97} Saha, A., Sandage, A., Labhardt,
L., Tammann, G.~A., Macchetto, F.~D., \& Panagia, N.\ 1997, \apj, 486,
1

\bibitem[{{Saha} {et~al.}(1996){Saha}, {Sandage}, {Labhardt},
{Tammann}, {Macchetto}, \& {Panagia}}]{saha96}{Saha}, A., et al.\@
1996, \apj, 466, 55

\bibitem[{{Saha} {et~al.}(2001){Saha}, {Sandage}, {Tammann},
{Dolphin}, {Christensen}, {Panagia}, \& {Macchetto}}]{saha01}{Saha},
A., et al.\@ 2001, \apj, 562, 314

\bibitem[{{Saha} {et~al.}(2006){Saha}, {Thim}, {Tammann}, \&
{Reindl}}]{saha06}{Saha}, A., et al.\@ 2006, \apjs, 165, 108

\bibitem[{{Sakai} {et~al.}(2004){Sakai}, {Ferrarese}, {Kennicutt}, \&
{Saha}}]{sakai04}{Sakai}, S., et al.\@ 2004, \apj, 608, 42

\bibitem[Sandage \& Tammann(2008)]{sandage08} Sandage, A., \& Tammann,
G.~A.\ 2008, \apj, 686, 779

\bibitem[{{Sandage} {et~al.}(2006){Sandage}, {Tammann}, {Saha},
{Reindl}, {Macchetto}, \& {Panagia}}]{sandage06}{Sandage}, A., et
al.\@ 2006, \apj, 653, 843

\bibitem[{{Sirianni} {et~al.}(2005){Sirianni}, {Jee},
{Ben{\'{\i}}tez}, {Blakeslee}, {Martel}, {Meurer}, {Clampin}, {De
Marchi}, {Ford}, {Gilliland}, {Hartig}, {Illingworth}, {Mack}, \&
{McCann}}]{sirianni05}{Sirianni}, M., et al.\@ 2003, \apjs, 148, 175

\bibitem[{{Stetson}(1987)}]{stetson87}{Stetson}, P.~B. \@ 1987, \pasp,
99, 191

\bibitem[{{Stetson}(1994)}]{stetson94}Stetson, P. B. \@ 1994, \pasp, 106, 250

\bibitem[{{Stetson}(1996)}]{stetson96}Stetson, P. B. \@ 1996, \pasp, 108, 851

\bibitem[Stetson(2000)]{stetson00} Stetson, P.~B.\ 2000, \pasp, 112,
925

\bibitem[Stetson \& Gibson (2001)]{sg01}Stetson, P. B., \& Gibson, B. K. \@ 2001, \mnras, 328, L1

\bibitem[{{Stetson} {et~al.}(1998){Stetson}, {Saha}, {Ferrarese},
{Rawson}, {Ford}, {Freedman}, {Gibson}, {Graham}, {Harding}, {Han},
{Hill}, {Hoessel}, {Huchra}, {Hughes}, {Illingworth}, {Kelson},
{Kennicutt}, {Madore}, {Mould}, {Phelps}, {Sakai}, {Silbermann}, \&
{Turner}}]{stetson98}{Stetson}, P.~B., et al.\@ 1998, \apj, 508, 491

\bibitem[Tammann et al.(2008)]{tammann08} Tammann, G.~A., Sandage, A.,
\& Reindl, B.\ 2008, \apj, 679, 52

\bibitem[{{Tammann} {et~al.}(2003){Tammann}, {Sandage}, \&
{Reindl}}]{tammann03}{Tammann}, G.~A., {Sandage}, A., \& {Reindl},
B. \@ 2003, \aap, 404, 423

\bibitem[{{Udalski} {et~al.}(1999){Udalski}, {Soszynski}, {Szymanski},
{Kubiak}, {Pietrzynski}, {Wozniak}, \& {Zebrun}}]{udalski99}{Udalski},
A., et al.\@ 1999, Acta Astronomica, 49, 223

\bibitem[van Leeuwen et al.(2007)]{leeuwen07} van Leeuwen, F., Feast,
M.~W., Whitelock, P.~A., \& Laney, C.~D.\ 2007, \mnras, 379, 723

\bibitem[Valencic, Clayton, \& Gordon 2004]{val04} Valencic, Clayton, \& Gordon, 2004, \apj, 616, 912

\bibitem[Wang et al.(2009)]{wang09} Wang, X., et al.\ 2009, \apj,
submitted (arXiv:0811.1205)

\bibitem[Wei et al.(1995)]{wei95} Wei, J.-Y., et al.\ IAU Circ. No.
6256

\bibitem[{{Zaritsky} {et~al.}(1994){Zaritsky}, {Kennicutt}, \&
{Huchra}}]{zaritsky94}{Zaritsky}, D., {Kennicutt}, R.~C., \& {Huchra},
J.~P. \@ 1994, \apj, 420, 87

\end{thebibliography}
\end{document}